\documentclass[11pt,reqno]{amsart}
\usepackage[foot]{amsaddr}
\usepackage[utf8]{inputenc}
\usepackage[T1]{fontenc}
\usepackage[font={scriptsize}]{caption}
\usepackage[caption=false]{subfig}
\usepackage{enumitem}
\usepackage{hyperref}
\hypersetup{
    colorlinks=true,
    linkcolor=blue,
    filecolor=blue,
    urlcolor=blue,
    citecolor=blue
}

\makeatletter
\def\paragraph{\@startsection{paragraph}{4}%
  \z@\z@{-\fontdimen2\font}%
  {\normalfont\bfseries}}
\makeatother

\usepackage{mathtools}
\mathtoolsset{showonlyrefs=true}
\usepackage{longtable, array}
\usepackage{tabularx}
\usepackage{multirow}
\usepackage[normalem]{ulem}
\usepackage{mathrsfs}
\usepackage{environ}

\usepackage[dvipsnames]{xcolor}
\usepackage{xfrac}
\usepackage{amsmath}
\usepackage{amsthm}
\usepackage{amssymb}
\usepackage{graphicx}
\usepackage{geometry}
\usepackage{verbatim}
\usepackage{float}
\usepackage{algorithmic}
\usepackage{algorithm}
\usepackage{cancel}

\usepackage{tikz}
\usepackage{tikz-cd}
\usetikzlibrary{positioning,arrows.meta}

\usepackage[sort&compress,numbers]{natbib}

\DeclareMathOperator*{\argmax}{arg\,max}

\numberwithin{equation}{section}

\newtheorem{lemma}{Lemma}[section]

\theoremstyle{remark}
\newtheorem{remark}{Remark}[section]

\makeatletter
\newenvironment{breakablealgorithm}
  {%
   \begin{center}
     \refstepcounter{algorithm}%
     \hrule height.8pt depth0pt \kern2pt%
     \renewcommand{\caption}[2][\relax]{%
       {\raggedright\textbf{\fname@algorithm~\thealgorithm} ##2\par}%
       \ifx\relax##1\relax
         \addcontentsline{loa}{algorithm}{\protect\numberline{\thealgorithm}##2}%
       \else
         \addcontentsline{loa}{algorithm}{\protect\numberline{\thealgorithm}##1}%
       \fi
       \kern2pt\hrule\kern2pt
     }
  }{%
     \kern2pt\hrule\relax
   \end{center}
  }
\makeatother

\title[ED Patient Flow Optimization with an Alternative Care Threshold Policy]{Emergency department patient flow optimization with an alternative care threshold policy}
\author{Sahba Baniasadi, Paul M. Griffin, and Prakash Chakraborty}
\email{sqb6360@psu.edu, pmg14@psu.edu, and prakashc@psu.edu}
\address{Marcus Department of Industrial and Manufacturing Engineering, The Pennsylvania State University}

\begin{document}
\begin{abstract}

Emergency department (ED) overcrowding and patient boarding represent critical systemic challenges that compromise care quality. We propose a threshold-based admission policy that redirects non-urgent patients to alternative care pathways, such as telemedicine, during peak congestion. The ED is modeled as a two-class $M/M/c$ preemptive-priority queuing system, where high-acuity patients are prioritized and low-acuity patients are subject to state-dependent redirection. Analyzed via a level-dependent Quasi-Birth-Death (QBD) process, the model determines the optimal threshold by maximizing a long-run time-averaged objective function comprising redirection-affected revenue and costs associated with patient balking and system occupancy. Structural analysis establishes monotone comparative statics relating the optimal threshold to all model parameters analytically. Numerical analysis using national healthcare data reveals that optimal policies are highly context-dependent. While rural EDs generally optimize at lower redirection thresholds, urban EDs exhibit performance peaks at moderate thresholds. Results indicate that our optimal policy yields significant performance gains of up to {$4.84\%$ in rural settings} and {$5.90\%$ in urban environments}. This research provides a mathematically rigorous framework for balancing clinical priority with operational efficiency across diverse ED settings.  

\end{abstract}
\maketitle
\section{Introduction}

The overcrowding of emergency departments (EDs) in the United States (US) is a persistent and complex problem driven by a range of systemic problems. EDs often serve as a safety net for uninsured or under-insured individuals who may lack access to primary care, leading to a surge in non-urgent visits that further strain resources \cite{cowling2013jama, yun2025ha}. The mismatch between patient demand and hospital capacity can lead to inefficient care delivery and compromised patient outcomes \cite{jones2022emj, nyce2021jpe}.

When demand for emergency services exceeds available resources, system congestion drives longer wait times, increased rates of patients leaving without being seen (LWBS), ambulance diversion, diminished patient satisfaction, and worse clinical outcomes. According to the American College of Emergency Physicians, overcrowding is defined as \emph{a situation that occurs when the identified need for emergency services exceeds available resources for patient care in ED, hospital, or both} \cite{savioli2022jpm, american2008emergency}. 

Various approaches have been developed to address ED overcrowding.  These include assigning physicians or advance-practice clinicians to triage to assist in evaluation and initiate treatment, \emph{fast-tracking} low-severity patients (e.g., minor injuries, sore throats) to a separate area that does not use ED beds for treatment, offering remote assessment such as telehealth to low-acuity patients, using a flow coordinator to facilitate patient flow, and steering patients that do not require ED care to other settings. While these interventions have demonstrated measurable benefits, they are often implemented as static structural changes rather than as occupancy-triggered operational policies that respond to real-time system states. State-dependent admission and routing policies have been studied in queuing control settings \cite{stidham1985, lippman1975}, however, their application to the specific context of occupancy-triggered alternative-care redirection with explicit alternative-care revenue has received limited attention in the ED literature. This limits their ability to adapt to transient surges in demand, staffing fluctuations, or seasonal variability in patient mix.

We propose a threshold-based admission policy to address ED overcrowding that offers an alternative service to non-severe patients when the system occupancy exceeds a specific threshold. For this case, patients have been assessed at triage into severe and non-severe categories and offered the alternative once the assessment is completed.  The alternative service depends on the type of setting and could include telemedicine in rural settings and affiliated clinics for hospital networks in urban settings. The threshold is based on a hospital-based profit function. 

The key contribution of this work is to formalize this operational intuition as an analytically tractable and easily implementable threshold-based control policy embedded within a multi-class stochastic service system.
We model the ED as a two-class $M/M/c$ preemptive-priority queuing system analyzed via a level-dependent Quasi-Birth-Death (QBD) process. This framework considers both revenue and costs (e.g., waiting and balking costs) from the hospital perspective to determine the optimal threshold by maximizing a long-run time-averaged objective function.
 We establish structural properties of the optimal threshold, including monotone comparative statics with respect to all model parameters, that analytically characterize how $\theta^*$ responds to changes in the operational environment.
The approach is applied across a variety of hospital settings including rural and urban-based hospitals to quantify the potential benefit by setting. Our numerical results indicate that this optimal policy yields significant performance gains of up to 4.84\% in rural settings and 5.90\% in urban environments. We conduct sensitivity analysis on several of the model parameters and on the arrival and service distributions to test robustness of the approach. We also perform a discrete-event
simulation study to compare preemptive and
non-preemptive priority with referral and no-referral policies.

Unlike previous work that focuses primarily on performance metrics such as waiting time or length of stay, our model explicitly integrates financial, operational, and behavioral components into a unified long-run objective function. This allows hospital decision-makers to evaluate not only clinical flow improvements, but also the economic and access-related implications of redirecting non-urgent demand.

While the literature discusses a rich set of ED operational features
including boarding, fast-tracking, and multi-stage triage, our model
deliberately focuses on a specific, tractable intervention: a
threshold-based redirection policy for non-urgent patients. This
design choice enables exact analytical characterization of the
steady-state performance and optimal policy, which would not be
feasible in a model incorporating all operational complexities
simultaneously. Incorporating richer features such as nonstationary
arrivals, multi-stage triage, and boarding dynamics represents an
important direction for future work.

The layout of this paper is as follows. Section~\ref{sec:lit_review} reviews the literature on ED overcrowding interventions, multi-class priority queuing models, quantitative methods for ED patient flow optimization, and admission-control and threshold redirection policies.  Section~\ref{sec:descr-prob} describes the model and formulates the threshold optimization problem. Section~\ref{sec:struct-props} establishes the structural properties of the threshold policy, including the stochastic orderings and the monotone comparative statics of the optimal threshold. Section~\ref{sec:methodology} develops the QBD solution method and reports the model parametrization and base computational results for representative rural and urban settings. Section~\ref{sec:sensitivity} presents the sensitivity and robustness analysis, including a discrete-event simulation study. Section~\ref{sec:capacity} addresses capacity allocation, and Section~\ref{sec:conclusion} concludes with a discussion of limitations and future work.

\section{Literature Review}\label{sec:lit_review}

This section reviews two key areas that shape the work presented in the paper.  This includes a discussion of operational interventions to address ED overcrowding followed by a discussion of queuing-based methods suitable for priority-based patient flow.

\subsection{Approaches for Reducing ED Overcrowding}

Overcrowding in the ED stems from an imbalance between hospital capacity and demand for services, and has been identified as a primary barrier to providing safe and efficient patient care \cite{morley2018po, savioli2022jpm}.  In addition to expanding physical capacity, interventions fall into three main categories: i) input-level interventions, ii) throughput-level interventions, and iii) output-level interventions.  

Input-level interventions focus on improving the triage process and reducing non-urgent arrivals. Assigning a physician or an advanced-practice clinician to triage can streamline decision-making and initiate treatment earlier. In a case-controlled study \cite{han2010jem} found that physician-based triage decreased LWBS rates but had only a modest effect of patient ED waiting times to be seen (WT) and length of stay (LOS) times.  However, a meta-analysis of 30 studies on the impact of physician-based triage on ED LOS found that median time decreased by 15.3 minutes compared to traditional nurse led triage \cite{jeyaraman2022bmj}. \cite{kamali2019ms} developed a steady-state, many-server fluid approximation to show that physician-based triage can outperform nurse-based triage when arrival rates are high.  \cite{zayas2019iise} modeled triage and treatment as a two-stage stochastic system developing a heuristic-based threshold to prioritize triage versus treatment for low-acuity patients.

Reducing non-urgent arrivals can also be achieved by introducing an alternative service such as telemedicine.  Two systematic reviews \cite{tsou2021jmir, scott2025jtt} showed that in rural areas this approach can help reduce ED WT and LOS without compromising patient outcomes.  Using a large data set from New York state, \cite{sun2020isr} also found that telemedicine in the ED significantly reduces LOS.

Throughput-level interventions focus on improving flow within the ED.  One approach is \emph{fast-tracking} low-acuity patients (based on the emergency severity index (ESI)) to a dedicated area that does not utilize ED beds.  Several studies show
this reduces overall WT and LOS without negatively impacting high-acuity patients \cite{considine2008emj, chrusciel2019bmj}.  However, \cite{ferrand2018jom} used simulation to compare fast-track to a dynamic priority queue (DPQ) approach, which prioritizes patients based on ESI and accumulated wait times. In their model, DPQ dominated fast-track in several performance measures. Other studies have also supported the use of DPQ \cite{hou2020jma, vaezi2022esa}.

Output-level interventions address the efficient disposition of patients, specifically focusing on the \emph{boarding} phenomenon where admitted patients await transfer to inpatient beds. Boarding diminishes ED capacity and is a critical determinant of ambulance diversion \cite{asplin2003}. The detrimental effects are substantial; research demonstrates a dose-dependent association between
ED stays exceeding five hours and increased 30-day mortality, with standardized mortality ratios rising by 10\% for patients boarding 8-12 hours \cite{jones2022emj}. 

\subsection{Applicable Multi-Class Queuing Models}
An ED can be modeled as a stochastic service system characterized by unscheduled arrivals, finite service capacity determined by beds and clinical staff, heterogeneity in patient acuity, and real-time decisions regarding routing, prioritization, and redirection. These structural features naturally align with multi-class queuing models, particularly those incorporating priority mechanisms.

The methodological foundation for analyzing block-structured stochastic systems lies in the matrix geometric methods pioneered by \cite{neuts1994} and further systematized by \cite{ramaswami1999}. Building upon this general theory, a foundational framework for priority-based multi-class queues was established by \cite{kao1991}, who analyzed a multiprocessor system with preemptive priorities and introduced a tractable method for computing steady-state distributions. We utilize their specific formulation because it explicitly captures the complex level-dependent transitions inherent in multi-server systems with preemption, which are not easily tractable via standard birth-death approximations. Although motivated by computer-processing environments, the underlying mechanics are directly applicable to ED operations, where high-acuity (urgent) patients preempt service for lower-acuity individuals. This approach allows for the exact calculation of steady-state probabilities in systems with complex, level-dependent transitions that are difficult to analyze via standard birth-death processes.

Subsequent research has adapted such priority-driven frameworks to healthcare settings. Stochastic population models for ED crowding show how time-varying arrival rates, length-of-stay dynamics, and capacity constraints collectively shape fluctuations in ED occupancy and crowding levels \cite{parnass2023}. Although not explicitly formulated as multi-class queues, these models capture key sources of heterogeneity. 

The importance of class-based competition for scarce ED resources is also reflected in the broader patient-flow and scheduling literature. A review of queuing and optimization models, \cite{saghafian2015edreview}, highlights how acuity differences and resource needs shape operational performance.  Beyond the ED, multi-class queuing formulations have examined routing and staffing across networks. For example, a multi-class–multiserver model with workload-dependent service times showed that variability in workload and class-dependent priorities significantly affect system delays \cite{nambiar2023}. 

Queuing theory has also shaped dynamic routing and admission-control policies, often modeled by Markov decision processes (MDPs). While Markov Decision Processes (MDPs) are often used to derive optimal routing policies, they can suffer from the \emph{curse of dimensionality} in large state spaces. In contrast, our work utilizes the structural properties of threshold-based policies to formulate a tractable QBD process.

In summary, the literature demonstrates that multi-class queuing systems incorporating priority, preemption, and dynamic routing provide a rigorous analytical foundation for modeling ED operations. The methodological insights from matrix-analytic frameworks for priority-based queues \cite{kao1991}, combined with ED-specific queuing and patient-flow models \cite{parnass2023, saghafian2015edreview, nambiar2023}, motivate the formulation of our preemptive-priority queue with a threshold-based redirection mechanism for non-urgent patients. 

\subsection{Quantitative Methods for ED Patient Flow Optimization}\label{sec:lit-review-quant-methods}
Several quantitative methods have been applied to the ED patient flow
optimization. Simulation-based approaches build computational models
to test scenarios and interventions: \cite{saghafian2015edreview} provide a comprehensive review of operations research contributions
to ED flow optimization, and \cite{terning2022} develop a hybrid
agent-based and discrete-event simulation model for ED patient flow
under pandemic conditions. Markov decision processes (MDPs) formulate
flow control as optimization problems: \cite{nunes2009} apply an MDP to elective admissions control using value iteration, and
\cite{clissold2015} formulate a finite-horizon MDP for hospital bed
occupancy management using dynamic programming.
Machine learning methods have
been used to predict patient volumes and classify acuity:
\cite{elbouri2021} review these applications and note the limited
generalizability across institutions and a lack of methods tailored specifically to operational flow management. Exact analytical queuing
models derive steady-state performance measures directly without
asymptotic assumptions; the matrix-geometric framework of
\cite{neuts1994} and its systematization by \cite{ramaswami1999}
form the methodological foundation for this family.
 
Since ED care is governed by patient acuity, multi-class priority queue
models are a natural fit. Simulation-based approaches have been used
to compare priority schemes: \cite{laskowski2009} develop a multi-priority agent-based framework to forecast wait times across
interacting EDs, and \cite{auyeung2006} develop a multi-class
Markovian queuing network model of ED patient flow, using real
patient timing data to compare priority schemes and response time
distributions. On the exact analytical side, \cite{siddharthan1996}
propose a priority queuing model to reduce average waiting times by
separating non-emergency patients, and \cite{lin2013} develop an
exact framework incorporating class-specific waiting time targets for
patients requiring both ED treatment and inpatient admission.
\cite{wang2025} construct preemptive and non-preemptive queuing models for urgent and routine patients and embed exact performance expressions into a mixed-integer programming model, minimizing total costs of waiting, patient rejection, beds, and servers.

\subsection{Admission Control, Threshold Policies, and Redirection}
 
Admission control has been studied as a distinct but closely related
policy problem, and a consistent finding is that threshold policies
emerge as the natural and often provably optimal structure for controlling patient flows in multi-class service systems.
\cite{pehlivan2022} establish the existence of an optimal threshold
admission policy for multi-class loss networks with parallel multiserver stations. \cite{zayascaban2020} prove via dynamic programming that an optimal threshold policy exists for a two-class
multi-server loss system under both stationary and non-stationary
conditions, motivated in part by emergency departments, showing that simpler policies ignoring non-stationarity incur significant reward
losses. \cite{helm2011} characterize an optimal admission threshold policy using an MDP model controlling both scheduled and expedited patient gateways, demonstrating through simulation on historical hospital data that threshold-based admission control reduces emergency blockages and cancellations. \cite{prodel2014} address ED hospitalization admission control using an MDP with class-specific length of stay, confirming that MDP-derived threshold policies improve overall performance across different facility types.
 
Referral and redirection of patients to alternative care settings have
received growing attention as a mechanism for managing ED congestion.
\cite{weinick2010} estimate that a substantial fraction of ED visits
could be managed at retail clinics or urgent care centers, with
significant potential cost savings. \cite{tsou2021jmir} and
\cite{sun2020isr} find that telemedicine reduces ED length of stay without compromising clinical outcomes, particularly in rural settings.
\cite{zayascaban2023} study server allocation in a two-stage tandem
queuing system with abandonment and develop threshold policies that
switch between service phases once total congestion exceeds a
threshold.
 
What this body of work shows collectively is that threshold policies are a natural and well-supported structure for admission and redirection control in multi-class
service systems, and that exact analytical models provide reliable performance characterizations without the approximation errors that accompany asymptotic methods.
To the best of our knowledge, no existing study embeds a state-dependent redirection threshold within an exact priority queue model. We addressed this gap by deriving an optimal threshold control policy within a two-class $M/M/c$ priority queue.

\section{Model Description and Problem Formulation}\label{sec:descr-prob}

\subsection{System Overview and Control Objective}
We address ED congestion by identifying an optimal threshold policy for {offering} {alternative care (such as telemedicine or affiliated urgent care clinics)} to non-urgent patients when system demand is high. The objective is to maximize revenue penalized by waiting and balking costs all from the hospital perspective. We model the ED as a two-class $M/M/c$ priority queuing system with distinct service time distributions for urgent and non-urgent patients. This framework is designed to provide hospital administrators with an implementable decision tool for efficient ED management.  
\paragraph*{Patient Flow and Priority Structure.}
Patients arrive either via ambulance and are considered as \emph{urgent} patients or arrive as walk-ins and are triaged into \emph{urgent} (walk-in triage levels 1-2) or \emph{non-urgent} (triage levels 3-5) classes.
Urgent patients are prioritized for immediate service, while non-urgent patients are assigned lower priority.
Non-urgent patients may either enter the ED queue or be offered an external alternative-care option depending on a congestion threshold. This mechanism optimizes resource allocation by directing lower-acuity patients to suitable external care pathways. The ED is thus modeled as a two-class preemptive-priority $M/M/c$ system with state-dependent threshold-based admission of non-urgent arrivals.
\paragraph*{Alternative Care.}
{In this study, alternative care refers to any lower-acuity treatment pathway available to non-urgent patients outside the main emergency department. These options may include telemedicine consultations, affiliated urgent care clinics, or hospital-based outpatient services that can deliver appropriate treatment at a lower operational cost and shorter wait time. From a modeling perspective, the alternative care pathway is associated with its own revenue rate (e.g. $r^{\mathrm{Alt}}$ for in-person alternatives) and potential balking or acceptance costs. The acceptance probability $p_a$ denotes the probability that a non-urgent patient accepts the offered alternative care option, regardless of its specific form.}

\subsection{Queuing Model with Threshold-based Admission and Balking}\label{sec:queue-model}

We now explain in detail our model of the ED as a two-class $M/M/c$ priority queuing system. We assume all patients arrive according to a Poisson process at a total rate $\lambda$. Furthermore, \emph{urgent} patients arrive with probability $p_{u}$, and \emph{non-urgent} patients with probability $p_n =1 - p_{u}$. This yields effective arrival rates $\lambda_u = \lambda  p_u$ for urgent patients and $\lambda_n = \lambda  p_n$ for non-urgent patients, each with exponentially distributed service times at rates $\mu_u$ and $\mu_n$, respectively.
The ED is equipped with $c_u$ urgent beds and $c_n$ non-urgent beds, for a total capacity of $c = c_u + c_n$ beds.
Urgent patients occupy urgent beds first, with preemptive priority over non-urgent patients if this capacity is exceeded, while non-urgent patients are restricted to non-urgent beds. 
That is, urgent patients wait only for other urgent patients if all $c$ beds are occupied by urgent patients, whereas non-urgent patients wait behind all urgent patients (both in service and queue) in addition to preceding non-urgent patients. 

Denote the total number of patients in the system at time $t$ by $N(t) = N_u(t) + N_n(t)$, where $N_u(t)$ and $N_n(t)$ denote the numbers of urgent and non-urgent patients, respectively.
Let $N_u^s = \min\{c, N_u\}$ and $N_n^s=\min\{c - N_u^s, N_n, c_n\}$ denote respectively the number of urgent and non-urgent patients in service. 
When $N(t)$ reaches or exceeds a balking threshold $k$, newly arriving non-urgent patients \emph{balk}, that is, leave without entering the system. If $N(t) < k$ non-urgent arrivals enter the system. However, when the system occupancy lies between a lower threshold $\theta$ and $k$, non-urgent arrivals may be routed to alternative care with probability $p_a$. Formally:
 \begin{enumerate}[wide, labelwidth=!, labelindent=0pt, label=(\roman*)]
\item If $\theta \leq N(t) < k$, an incoming non-urgent patient is offered alternative care. This offer is accepted with probability $p_a$ or declined with probability $(1-p_a)$. Patients who decline the offer remain in the ED for treatment.
\item  If $N(t) < \theta$, a non-urgent patient stays in the ED for treatment. 
\end{enumerate}
The two thresholds reflect fundamentally different mechanisms. The threshold $\theta$ is an \emph{ED-controlled} decision variable: the ED monitors real-time occupancy and proactively offers alternative care to arriving non-urgent patients whenever $ \theta \leq N(t) < k$. This is an operational protocol entirely within the hospital's discretion, and it is the quantity optimized in this paper. The balking threshold $k$, by contrast, is a \emph{patient-driven} behavioral parameter: patients who arrive to find
$N(t)\geq k$ self-select out based on their own assessment of expected waiting time and leave without entering the system. 
$k$ is treated as an exogenous input and the optimization is conducted over $\theta$ for a fixed $k$.

Since the number of non-urgent patients is constrained by the balking threshold $k$, overall system stability is ensured solely by the stability of the urgent queue.  Consequently, the ED system is stable when $\rho_u = \frac{\lambda_u}{c \mu_u} < 1$.

\paragraph*{Model Justification.}
The $M/M/c$ priority queuing framework is well-suited to this problem for the following reasons. Walk-in patient arrivals are well approximated by a Poisson process, as supported by the broader ED queuing literature \cite{saghafian2015edreview, nambiar2023}. The two-class preemptive priority structure directly reflects the acuity-based triage protocol in which urgent patients (ESI levels 1--2) receive immediate priority over non-urgent patients (ESI levels 3--5). Exponential service times provide the analytical tractability required for exact steady-state computation; Section~\ref{sec:robust-serv} confirms that the optimal threshold $\theta^*$ and the objective value are robust to Erlang-2 and Lognormal service distributions. The Quasi-Birth-Death (QBD) framework, outlined in Section~\ref{sec:methodology}, supports exact steady-state computation for each fixed $\theta$, making the enumeration over $\{0,\ldots,k-1\}$ computationally efficient while also supporting the structural analysis of Section~\ref{sec:struct-props}. 
This differs from a full MDP, which would solve Bellman equations for an unstructured state-dependent control over $(N_u, N_n)$, which may be harder to interpret and implement in an ED. Instead, we optimize the single threshold $\theta$, evaluate it exactly via the QBD, and obtain the monotone comparative statics of Section~\ref{sec:struct-props} that simulation-based optimization cannot provide.
A detailed comparison to these approaches is provided in Section~\ref{sec:lit-review-quant-methods}.

\subsection{Optimization Problem}
The objective is to determine the optimal alternative care threshold $\theta$ that maximizes the net benefit $Z(\theta)$, defined as:
\begin{equation}\label{eq:objective_function}
Z(\theta) =  \omega_R R(\theta) - \omega_B B(\theta) - \omega_W W(\theta),
\end{equation}
where $R(\theta)$, $B(\theta)$, and $W(\theta)$ denote the long-run time averaged revenue, balking cost, and waiting cost, respectively under the alternative care threshold $\theta$. The parameters $\omega_R$, $\omega_B$, and $\omega_W$ weight the relative importance of these components and reflect managerial or institutional priorities. For clarity of exposition, we set $\omega_R = \omega_B = \omega_W=1$ in the numerical analysis, but alternative weights can be used to represent different policy preferences. The objective function $Z(\theta)$ is formulated from the
\textit{hospital's perspective}: the hospital controls the redirection
threshold $\theta$ to maximize net financial benefit, while patients
make independent acceptance decisions captured through the empirically calibrated parameter $p_a$. From the perspective of ED management, the objective is to identify the optimal alternative care threshold $\theta^*$ that maximizes overall net benefit:
\begin{equation}
{\theta^*} = \argmax_{\theta\in\{0,\ldots,k-1\}} Z(\theta).
\label{eq:optimal_policy}
\end{equation}
This optimization problem captures the trade-offs among revenue from ED and alternative care services, congestion-induced patient balking costs, and waiting costs associated with system delays.
The finite feasible set for $\theta$ allows direct enumeration of candidate policies. 
Rather than solving a full MDP problem, the policy is parametrized by the single scalar $\theta$ (see the model justification in Section~\ref{sec:queue-model}), steady-state performance is derived analytically via the QBD framework for each fixed $\theta$, and the optimal threshold $\theta^*$ is identified by enumeration over $\{0,\ldots,k-1\}$. The resulting policy is directly implementable as an occupancy-based protocol without requiring an online optimization engine.

\subsection{Stationary Analysis Framework}

The two-class queuing system described above can be represented as a Quasi-Birth-Death (QBD) chain.
All performance measures in this study are computed from the stationary distribution of the corresponding QBD chain. Since the objective function components are defined as long-run time averages, the ergodic theorem ensures that these quantities converge to steady-state expectations.


Formally, for any system metric $H(t)$ observed at time $t$, we have:
\begin{equation}\label{eq:ergodic_theorem}
\lim_{T \to \infty} \frac{1}{T} \int_0^T H(t)dt = E[H] = \sum_{i=0}^{\infty} \sum_{j=0}^{k-1} H(i,j) \cdot \pi(i,j),
\end{equation}
where $\pi(i,j)$ denotes the stationary probability of $i$ urgent and $j$ non-urgent patients in the system, and $H$ is the steady-state random variable corresponding to $H(t)$. 
By the PASTA (Poisson Arrivals See Time Averages; \cite{wolff1982}) property, since both arrival streams are Poisson, $\pi(i,j)$ simultaneously represents the large-time asymptotic time-average probability and the fraction of arriving patients (of either or both class) who find the system in state $(i,j)$. 

This stationary analysis framework allows us to evaluate all objective function components exactly as steady-state expectations under a given threshold policy~$\theta$, without the need to consider transient system behavior.

\subsection{Objective Function Components}
\label{sec:obj_components}
Using the stationary distribution $\pi(i,j)$, each component of the objective function $Z(\theta)$ can be expressed as a steady-state expectation.

\paragraph*{Average Revenue.}
We decompose revenue into revenue generated by urgent patients, denoted by $R_u$, and those generated by non-urgent patients, denoted by $R_n(\theta)$ where the latter is affected by the particular threshold policy $\theta$. $R_n(\theta)$ can be further decomposed as revenue generated from non-urgent patients treated in the ED, denoted $R_n^{\mathrm{ED}}(\theta)$ and revenue from accepted alternative-care referrals, denoted $R_n^{\mathrm{Alt}}(\theta)$. Let $r_u^{\mathrm{ED}}$ and $r_n^{\mathrm{ED}}$ denote the revenue per urgent and per non-urgent patient treated in the ED, and $r^{\mathrm{Alt}}$ the revenue per accepted alternative-care referral. Define $D_u(t)$, $D_n(t)$ and $X^{{\mathrm{Alt}}}(t)$ to be the cumulative number of urgent ED departures, non-urgent ED departures and accepted alternative-care referrals by time $t$. Then the long-run time-averaged revenue rate is:
\begin{equation}\label{eq:r}
R(\theta) = R_u + R_n(\theta),
\end{equation}
where by the ergodic property of continuous time Markov chains
$$
R_u = \lim_{T \to \infty} \dfrac{1}{T} \left[ r_u^{\mathrm{ED}} \cdot D_u(T) \right] = r_u^{\mathrm{ED}} \cdot  d_u,
$$
and
\begin{align}\label{eq:revenue-comp}
R_n(\theta) = R_n^{\mathrm{ED}}(\theta) + R_n^{{\mathrm{Alt}}}(\theta) &= \lim_{T \to \infty} \frac{1}{T} \left[ r_n^{\mathrm{ED}} \cdot D_n(T) + r^{\mathrm{Alt}} \cdot X^{{\mathrm{Alt}}}(T) \right]\nonumber \\ &= r_{n}^{\mathrm{ED}} \cdot  d_n(\theta) + r^{\mathrm{Alt}} \cdot  x(\theta).
\end{align}
Here $d_u$,  $d_n(\theta)$and $x(\theta)$ are steady-state rates of urgent ED departures, non-urgent ED departures, and accepted alternative care referrals, respectively.
By PASTA
$$
x(\theta)=\lambda_n \cdot p_a \cdot P(\theta\le N<k),
$$
where $P(\cdot)$ is the steady-state probability of the number $N$ in the system. 
We also denote by $p_x(\theta) := p_a P_{\theta} (\theta \leq N < k)$ the unconditional probability that an arriving non-urgent patient is both offered and accepts alternative care, that is, $x(\theta) = \lambda_n p_x(\theta)$.
Note that the steady-state non-urgent ED departure rate is given by
\begin{equation}\label{eq:departure_rate}
d_n(\theta) = \mu_n \cdot E[N_n^s].
\end{equation}
Although preemptive priority determines which patients occupy servers at any instant, the per-server service rate $\mu_n$ is unaffected: each occupied non-urgent server completes service at rate $\mu_n$, so the aggregate departure rate at any state is $\mu_n N_n^s$, and the long-run rate $d_n(\theta) = \mu_n E[N_n^s]$ follows by the ergodic theorem. The memoryless property of exponential service ensures that a non-urgent patient re-entering service after preemption faces a fresh $\mathrm{Exp}(\mu_n)$ service time. From the stationary distribution: $E[N_n] = \sum_{i,j} j\,\pi(i,j)$, and $ \lambda_n^{\mathrm{eff}}(\theta) = \lambda_n [P_\theta(N<\theta) + (1-p_a)\,P_\theta(\theta \leq N < k)]$. 
The steady state urgent departure rate
$d_u=\mu_u \cdot E[N_u^s]$ follows analogously.

\paragraph*{Average Balking Cost.} 
Let $B_n(t)$ denote the cumulative number of balking non-urgent patients until time $t$.
Then the long-run time-averaged balking cost rate is:
\begin{equation}
B(\theta) = \lim_{T \to \infty} \frac{1}{T}  [c^{b} \cdot B_n(T)] = c^{b} \cdot b_n(\theta),
\label{eq:balking-cost_lim}
\end{equation}
where $c^{b}$ is the cost associated with each balking patient and $b_n(\theta)$ is the steady-state balking rate. Let $p_b(\theta) = P_{\theta}(N \geq k)$ be the steady-state probability that the system exceeds the balking threshold $k$. By PASTA, $b_n(\theta)=\lambda_n \cdot  p_b(\theta)$ and therefore
\begin{equation}
B(\theta)= c^{b} \cdot b_n(\theta) = c^{b} \cdot \lambda_n \cdot p_b(\theta).
\label{eq:Btheta}
\end{equation}  
The balking cost $c^b$ represents the hospital's \textit{opportunity
cost} of lost demand: a balking patient would have either been treated in the ED with probability $(1-p_a)$ or accepted alternative care with probability $p_a$, yielding expected foregone revenue.

\paragraph*{Average Waiting Cost.} For non-urgent patients, the average waiting cost is evaluated by the long-run time-averaged waiting cost rate:
\begin{equation}
W_n(\theta) = \lim_{T \to \infty} \frac{1}{T} \int_0^T [c_{n}^w \cdot N_n(t)]dt,
\label{eq:waiting-cost_lim}
\end{equation} where $N_n(t)$ is the instantaneous number of non-urgent patients in the system at time $t$, including those in the queue and those receiving service, while $c_n^w$ denotes the waiting cost rate per patient per unit time. This time-averaged formulation simplifies under the steady-state distribution to a function of the expected individual delay:
\begin{equation}
W_n(\theta) =  \lambda_n^{\mathrm{eff}} \cdot c_n^{w} \cdot {E[S_n]}.
\label{eq:Wtheta}
\end{equation} In this expression, ${E[S_n]}$ is the expected time a non-urgent patient spends in the system under stationarity. This value is related to the expected number of non-urgent patients in the system, $E[N_n]$, via Little's law: 
$$
{E[S_n]} = E[N_n] / \lambda_n^{\mathrm{eff}},
$$ where $\lambda_n^{\mathrm{eff}}$ is the effective arrival rate of non-urgent patients who join the ED queue after accounting for those who balk or accept alternative care redirections. The long-run average waiting cost rate for urgent patients, $W_u$ can be calculated in a similar way.
The non-urgent waiting cost rate $c_n^w$ is calibrated from empirical Left-Without-Being-Seen (LWBS) data \cite{rathlev2020}: longer waits
increase the probability that patients leave without being seen,
representing additional lost revenue. The urgent waiting cost $c_u^w$ reflects clinical liability and quality-of-care costs associated with treatment delays, computed as the product of the mortality risk
increase per hour of delay \cite{liu2017} and the present value of remaining lifetime productivity \cite{grosse2019}.

\section{Structural Properties and Comparative Statics of the Threshold Policy}\label{sec:struct-props}

We next establish several structural properties of the threshold policy. These results clarify how the threshold affects system performance and provide comparative statics for the objective function. 
For a fixed threshold $\theta$, define the admission probability of a non-urgent patient in state $(N_u, N_n) = (i,j)$:
\begin{equation}\label{eq:al_theta}
\alpha_{\theta}(i,j)
= \begin{cases} 
1 & i+j < \theta, \\
1-p_{a}  & \theta \leq i+j < k, \\
0  & i+j \geq k.
\end{cases}
\end{equation}
Thus, a larger value of $\theta$ corresponds to a less aggressive alternative-care referral policy.

Consider two thresholds $\theta_1<\theta_2$, and consider two corresponding systems on a common probability space using the same arrivals, service completions, and alternative-care acceptance decisions. Urgent patients with preemptive priority are unaffected by the threshold policy. Hence,
\begin{equation}\label{eq:urgent}
N_u^{\theta_1}(t)=N_u^{\theta_2}(t), \qquad t\geq0.
\end{equation} However, from \eqref{eq:al_theta} we have
$
\alpha_{\theta_1}(i,j)\leq \alpha_{\theta_2}(i,j)
$
for all $(i,j)$. 
To verify that this ordering implies $N_n^{\theta_1}(t)\leq N_n^{\theta_2}(t)$, let $G(t):=N_n^{\theta_2}(t)-N_n^{\theta_1}(t)$ and consider two cases. When $G(t)=0$ (both systems occupy the same state $(i,j)$), the ordering $\alpha_{\theta_1}(i,j)\leq\alpha_{\theta_2}(i,j)$ ensures that any patient admitted to system~$1$ is also admitted to system~$2$ under a common uniform admission draw, so $G$ cannot decrease. When $G(t)>0$, each Poisson event (arrival or departure) changes $G$ by at most $\pm 1$. Since $G\geq 1$ beforehand, the ordering $G\geq 0$ is preserved after any such event.
That is, for same initial condition, the coupling implies
\begin{equation}\label{eq:non-urgent}
N_n^{\theta_1}(t)\leq N_n^{\theta_2}(t), \qquad t\geq0.
\end{equation}
From \eqref{eq:urgent} and \eqref{eq:non-urgent}
\begin{equation}
N^{\theta_1}(t)\leq N^{\theta_2}(t), \qquad t\geq0,
\end{equation}
and, in steady state, the following stochastic ordering in the threshold holds:
\begin{equation}\label{eq:st-ord-steady}
N_n^{\theta_1}\leq_{\mathrm{st}}N_n^{\theta_2}, \qquad N^{\theta_1}\leq_{\mathrm{st}}N^{\theta_2}.
\end{equation}
It follows that several performance measures are monotone in $\theta$. 
\begin{lemma}\label{lem:mon-1}
In steady state, the expected non-urgent in system $L_n(\theta) = E_{\theta} [N_n]$, the non-urgent balking probability $p_b(\theta) = P_{\theta}(N \geq k)$, the non-urgent departure rate $d_n(\theta) = \mu_n E_{\theta}[N_n^s]$ are non-decreasing in $\theta$, while the alternative care acceptance rate $x(\theta) = \lambda_n p_a P_{\theta}(\theta \leq N < k)$ is non-increasing in $\theta$
\end{lemma}
\begin{proof}
Monotonicity of the first three quantities follows from \eqref{eq:st-ord-steady}. In steady state, $\lambda_n = d_n(\theta) + x(\theta) + \lambda_n p_b(\theta)$, that is, $x(\theta) = \lambda_n (1-p_b(\theta)) -d_n(\theta)$ which is now non-increasing in $\theta$ by considering the monotonicities of $p_b(\theta)$ and $d_n(\theta)$.
\end{proof}
The stochastic ordering \eqref{eq:st-ord-steady} also implies monotonicity of the expected non-urgent sojourn time.
\begin{lemma}\label{lem:sojourn}
The expected sojourn time $S_n(\theta):=E_\theta[S_n]$ of non-urgent patients
is non-decreasing in $\theta$.
\end{lemma}
\begin{proof}
Under preemptive priority, the sojourn time of a non-urgent patient is non-decreasing in the number of non-urgent patients present at arrival. The coupling \eqref{eq:urgent}--\eqref{eq:non-urgent} gives $N_n^{\theta_2}(t)\geq N_n^{\theta_1}(t)$ a.s. while $N_u^{\theta_1}(t)=N_u^{\theta_2}(t)$, so the sojourn functional is a.s. at least as large under $\theta_2>\theta_1$, giving $S_n(\theta_1)\leq S_n(\theta_2)$.
\end{proof}
Now going back to the objective \eqref{eq:objective_function} and re-arranging 
\begin{equation}\label{eq:obj-2}
Z(\theta) = C + \omega_R[r_n^{\mathrm{ED}} d_n(\theta) + r^{\mathrm{Alt}}x(\theta)] - \omega_B c^b \lambda_n p_b(\theta) - \omega_W c_n^w L_n(\theta),
\end{equation}
where $C$ is independent of $\theta$. Using the notation $\Delta_{\theta} f(\theta) = f(\theta+1)-f(\theta)$, this implies 
\begin{equation}\label{eq:Delta_Z}
\Delta_{\theta} Z(\theta) = \omega_R[r_n^{\mathrm{ED}} \Delta_{\theta}d_n(\theta) + r^{\mathrm{Alt}}\Delta_{\theta} x(\theta)] - \omega_B c^b \lambda_n \Delta_{\theta} p_b(\theta) - \omega_W c_n^w \Delta_{\theta} L_n(\theta).
\end{equation}
Raising the threshold from $\theta$ to $\theta+1$ is beneficial if and only if $\Delta_{\theta} Z(\theta) \geq 0$ or equivalently
$$
 \omega_R r_n^{\mathrm{ED}} \Delta_{\theta}d_n(\theta) \geq  
 \omega_R r^{\mathrm{Alt}}[-\Delta_{\theta} x(\theta)] + \omega_B c^b \lambda_n \Delta_{\theta} p_b(\theta) + \omega_W c_n^w \Delta_{\theta} L_n(\theta).
$$
This condition conveys the operational interpretation that the threshold should be raised only when the marginal ED revenue from admitting additional non-urgent patients exceeds the marginal loss in alternative-care revenue along with the additional balking and waiting costs. From Lemma~\ref{lem:mon-1} and \eqref{eq:Delta_Z} we have
\begin{align*}
&\frac{\partial \Delta_{\theta} Z(\theta)}{\partial r_n^{\mathrm{ED}}} = \omega_R\Delta_{\theta} d_n(\theta) \geq0, 
&\frac{\partial \Delta_{\theta} Z(\theta)}{\partial r^{\mathrm{Alt}}}=\omega_R\Delta_{\theta} x(\theta) \leq0,\\
&\frac{\partial \Delta_{\theta} Z(\theta)}{\partial c^b} = -\omega_B\lambda_n\Delta_{\theta} p_b(\theta) \leq0,
&\frac{\partial \Delta_{\theta} Z(\theta)}{\partial c_n^w} = -\omega_W\Delta_{\theta} L_n(\theta) \leq0.
\end{align*}
This provides the following comparative statics on revenue and cost parameters.
\begin{lemma}\label{lem:op-ord}
Fix all parameters except the one being varied. Then the optimal threshold $\theta^*$ satisfies: 
\begin{enumerate}[wide, label = \upshape(\alph*)]
\item $\theta^*$ is non-decreasing in $r_{n}^{\mathrm{ED}}$.
\item $\theta^*$ is non-increasing in $r^{\mathrm{Alt}}$.
\item $\theta^*$ is non-increasing in $c^b$.
\item $\theta^*$ is non-increasing in $c_{n}^{w}$.
\item $\theta^*$ is \emph{invariant} to $c_{u}^{w}$.
\end{enumerate}
\end{lemma}
\begin{proof}
Parts~(a)--(d) follow directly from the signed partial derivatives above, which show that $\Delta_\theta Z(\theta)$ is non-decreasing in $r_n^{\mathrm{ED}}$ and non-increasing in $r^{\mathrm{Alt}}$, $c^b$, and $c_n^w$ for every $\theta$, giving the stated monotonicities of $\theta^*$.  For part~(e): from~\eqref{eq:obj-2}, $c_u^w$ enters the objective only through the constant $C$ via the term $-\omega_W c_u^w E[N_u]$, where $E[N_u]$ is independent of $\theta$ by~\eqref{eq:urgent}.  
\end{proof}
Similarly, analogous stochastic orderings for the operational parameters $\lambda$, $c$, and $p_a$ can be derived. These are contained in the following lemma, which can be proved using the same gap argument as in~\eqref{eq:urgent}--\eqref{eq:non-urgent} applied to each parameter, and is omitted.
\begin{lemma}\label{lem:st-ord-op-params}
While other parameters including the threshold policy $\theta$ is held fixed
\begin{enumerate}[label=\upshape(\alph*)]
\item If $\lambda_1 \leq \lambda_2$, then
        $N_{n}^{(\theta)}(\lambda_1)\leq_{\mathrm{st}} N_{n}^{(\theta)}(\lambda_2)$.
\item If $c_1 \geq c_2$ (with $c_{u}/c$ held fixed), then
        $N_{n}^{(\theta)}(c_1)\leq_{\mathrm{st}} N_{n}^{(\theta)}(c_2)$.
\item If $p_{a}^1 \geq p_{a}^2$, then
        $N_{n}^{(\theta)}(p_{a}^1)\leq_{\mathrm{st}} N_{n}^{(\theta)}(p_{a}^2)$.
\end{enumerate}
\end{lemma}
\begin{remark}\label{rem:op-cs}
Since $S_n$ is non-decreasing in $N_n$ at arrival (Lemma~\ref{lem:sojourn}), the stochastic orderings in Lemma~\ref{lem:st-ord-op-params} carry over directly to the expected sojourn time. That is,  $S_n(\theta;\lambda)$ is non-decreasing in $\lambda$, and $S_n(\theta;c)$ and $S_n(\theta;p_a)$ are non-increasing in $c$ and $p_a$ respectively, for every fixed $\theta$.  Parameters that increase system congestion raise waiting costs at every occupancy level, making earlier redirection more valuable and shifting the zero-crossing of $\Delta_\theta Z$ to a lower $\theta$. Parameters that reduce congestion have the opposite effect.  These orderings therefore suggest the following directional comparative statics:
$$
\text{(a) $\theta^*$ is non-increasing in $\lambda$;
(b) $\theta^*$ is non-decreasing in $c$;
(c) $\theta^*$ is non-decreasing in $p_a$.}
$$
A formal analytical proof would require additional properties of the stationary distribution, which is not established here. The results (a)--(c) are however expected from the stochastic orderings of Lemma~\ref{lem:st-ord-op-params} and {confirmed numerically in Section~\ref{sec:sensitivity-op-params}}.
\end{remark}
\paragraph*{Justification of the threshold structure.}
Threshold policies are the classical optimal structure in admission control, where a single-crossing property of the marginal admission value yields a threshold rule \cite{stidham1985,lippman1975}. The same trade-off is captured by the first difference $\Delta_\theta Z(\theta)$ in \eqref{eq:Delta_Z}. By Lemmas~\ref{lem:mon-1} and~\ref{lem:sojourn} the ED-revenue increment $\omega_R r_n^{\mathrm{ED}}\Delta_\theta d_n(\theta)$ is non-negative while the remaining increments in $\Delta_\theta Z$ are non-positive. If $\Delta_\theta Z$ were non-increasing in $\theta$, it would change sign from positive to negative at most once, making the optimal policy a threshold. The lemmas provide the sign of each increment but not whether $\Delta_\theta Z$ is monotone in $\theta$, so this single-crossing structure, though strongly suggested, is not established. Our two-class priority queuing model with probabilistic redirection and balking lies outside the canonical one-dimensional admission problem and we therefore obtain $\theta^*$ by enumeration over $\{0,\ldots,k-1\}$. The rigorous content lies in the structural results, namely the stochastic orderings and monotone performance measures (Eq.~\eqref{eq:st-ord-steady}, Lemmas~\ref{lem:mon-1}--\ref{lem:sojourn}, Lemma~\ref{lem:st-ord-op-params}) and the comparative statics of $\theta^*$ (Lemma~\ref{lem:op-ord}), which show how the optimum responds to the model parameters. If $\Delta_\theta Z(\theta)=0$ over consecutive $\theta$, the maximizers form a range in which any element is equally optimal, and the comparative statics hold for any selection.

\paragraph*{Managerial interpretation.}
The threshold $\theta^*$ reflects a \emph{marginal balance}: the hospital should admit non-urgent patients until the marginal ED revenue equals the marginal congestion cost they impose on all patients in the system. Lemma~\ref{lem:op-ord} shows how this balance shifts with economic parameters: a more profitable ED or less attractive alternative care raises $\theta^*$, while higher waiting or balking costs lower it. Remark~\ref{rem:op-cs} further shows that operational parameters ($\lambda$, $c$, $p_a$) shift $\theta^*$ in the direction consistent with their effect on system congestion.

Because the components of $Z(\theta)$ exhibit competing monotonicity (throughput and congestion move in opposite directions as $\theta$ increases) the full objective is neither monotone nor convex in general. Therefore, the enumeration of Section~\ref{sec:methodology} computes $\theta^*$ , and the structural results above to interpret it.


\section{Methodology}\label{sec:methodology}
\subsection{Solution Method}

Our multi-class preemptive-priority queuing system can be represented as a level-dependent process where each level \(i\) corresponds to the number of urgent patients currently in the system. Within each level \(i\), the phase \(j\) represents the number of non-urgent patients in the system, where \(j \in \{0, 1, \ldots, k-1\}\). This structure allows us to model the system as a Quasi-Birth-Death (QBD) process following the approach developed by \cite{kao1991} for modeling multiprocessor systems with preemptive priorities. For the sake of completeness, we provide the QBD background modified to our case. The generator matrix $Q$, matrix component definitions, and balance equations are provided in Appendix~\ref{app:qbd_structure}, the detailed
recursive solution steps in Appendix~\ref{app:proofs}, and the solution
algorithm in Appendix~\ref{app:algorithms}.
We validate $\pi(i,j)$ by comparing the QBD marginal probabilities $\pi(i) = \sum_j \pi(i,j)$ to exact $M/M/c$ probabilites, and expected relative errors for both case studies are provided in the section below. Objective function components are computed from $\pi(i,j)$ via the ergodic theorem equation~\eqref{eq:ergodic_theorem}, with derivations in Appendix \ref{app:obj_comp}. The optimal threshold $\theta^*$ is found by enumerating $\theta \in \{0, \ldots,k-1\}$ and selecting the maximizer of \eqref{eq:objective_function}.

\subsection{Parametrization and Computational Results}
Emergency departments operate differently depending on their location and the populations they serve. Based on research on rural-urban healthcare differences \cite{greenwood2019}, we study two types of EDs: \emph{rural} and \emph{urban}. We base our input parameters on several references including national healthcare data \cite{cairns2024}, emergency department performance studies \cite{edba2023}, and research on ED operations and costs \cite{fairhealth2024}.

The computational analysis is performed on these two representative ED configurations to demonstrate the model's applicability across different healthcare settings. Parameter values are derived from published healthcare data; as summarized
in Table~\ref{tab:ed_parameters} and described below.

{\scriptsize
\begin{longtable}{|p{4cm}|p{1.5cm}|c|c|p{5cm}|}
\caption{Emergency Department Input Parameters by Geographic Setting (Based on Literature)}
\label{tab:ed_parameters}\\
\hline
\textbf{Parameter} & \textbf{Unit} & \textbf{Rural ED} & \textbf{Urban ED} & \textbf{Ref.} \\
\hline
\endfirsthead

\hline
\textbf{Parameter} & \textbf{Unit} & \textbf{Rural ED} & \textbf{Urban ED} & \textbf{Ref.} \\
\hline
\endhead

\hline
\endfoot

\hline
\endlastfoot

Arrival rate ($\lambda$) & patients/ hour & $\approx 2$ & $\approx 5$ & \cite{cairns2024,aha2025} \\
Urgent proportion ($p_u$) & - & 0.39 & 0.85  & \cite{alnasser2023,patel2024} \\
Urgent service rate ($\mu_u$) & patients/ hour & 0.15 & 0.15 & \cite{edba2023,cairns2024} \\
Non-urgent service rate ($\mu_n$) & patients/ hour & 0.32 & 0.32 & \cite{edba2023,cairns2024} \\
Total beds ($c$) & beds & $\approx 9$ & 34 & \cite{greenwood2019} \\
Bed allocation ratio ($c_u/c$) & - & $\approx 0.4$ & $\approx 0.4$ & \cite{mermiri2021} \\
Alt. care acceptance rate ($p_a$) & - & 0.52 & 0.52 & \cite{weinick2010} \\
ED revenue - non-urgent ($r_n^{\mathrm{ED}}$) & \$/patient & 675.50  & 675.50  & \cite{fairhealth2024} \\
ED revenue - urgent ($r_u^{\mathrm{ED}}$) & \$/patient & 2,221.00 & 2,221.00 & \cite{fairhealth2024} \\
Alt. option revenue ($r^{\mathrm{Alt}}$) & \$/patient & 436.00 & 436.00 & \cite{fairhealth2024}\\
Balking cost w.r.t. $r^{\mathrm{Tele}}$ ($c^b$) & \$/patient & 384.82 & 384.82 &  \cite{cms2024,weinick2010,fairhealth2024}\\
Balking cost w.r.t. $r^{\mathrm{Alt}}$ ($c^b$) & \$/patient & 550.96 & 550.96 &  \cite{cms2024,weinick2010,fairhealth2024}\\
Urgent waiting cost ($c_u^w$) & \$/hour & 5,531.61  & 5,531.61 & \cite{cairns2024,liu2017,grosse2019} \\
Non-urgent waiting cost($c_n^w$) & \$/hour & 53.21 & 53.21 & \cite{edba2023,rathlev2020} \\
Balking threshold ($k$) & patients & 37 & 39 & \cite{dark2020,fayyaz2013} \\
\end{longtable}
}
\begin{enumerate}[label = (\roman*), labelwidth = !, wide, labelindent = 0pt]
\item \textbf{Arrival Rate.} The arrival rate $\lambda$ is calculated from NHAMCS 2022 national ED visit data by patient residence \cite{cairns2024} and AHA hospital counts \cite{aha2025}. Urban visits (123.246 million visits/3,316 hospitals) and rural visits (27.548 million visits/1,796 hospitals) are converted to hourly rates.

\item \textbf{Urgent Patient Proportion.} The proportion of ED patients triaged as urgent ($p_u$) is calculated using triage severity levels. For urban hospitals, we use the Emergency Severity Index (ESI) from \cite{patel2024}, where ESI 1 (2\%), ESI 2 (22\%), and ESI 3 (61\%) yields $p_u = 0.85$. For rural hospitals, since literature suggests that the proportion of non-urgent patients in rural ED is higher than Urban ED, we use findings from \cite{alnasser2023}, where 61.4\% of ED visits were less-urgent or non-urgent, yielding approximately $p_u = 0.39$.

\item \textbf{Service Rates.} The service rates $\mu_u, \mu_n$ are derived from national ED operations benchmarks \cite{edba2023} showing median length of stay for admitted patients (289 minutes) and discharged patients (147 minutes), adjusted for door-to-clinician time (16 minutes) from NHAMCS 2022 \cite{cairns2024}. Service rates are calculated using the exponential distribution relationship $\mu = \ln(2)/\text{median}$, yielding $\mu_u = 0.15$ and $\mu_n = 0.32$ patients per hour.
\item \textbf{Bed Capacity.} The total bed capacity $c$ is derived from \cite[Table 1]{greenwood2019}, which reports median emergency department bed counts by hospital type. Rural hospitals have a median of approximately 9 beds while urban hospitals have a median of 34 beds.
\item \textbf{Bed Allocation Ratio.} Based on empirical data from Stony Brook University Medical Center \cite{mermiri2021}, which operates a 22-bed resuscitative emergency department ICU (RED-ICU) with 8 beds designated as an Acute Critical Care Subunit for the highest acuity patients, we obtain a bed allocation ratio of $c_u/c = 8/22 = 0.36 \approx 0.4$ for urgent bed allocation.

\item \textbf{Alternative Care Acceptance Rate.} The probability of accepting alternative care or alternative care offers ($p_a$) is derived from \cite{weinick2010}, which found that 52\% of eligible non-urgent ED patients accepted referral to alternative option when offered after triage.

\item \textbf{Revenue Parameters.} ED revenue parameters are derived from FairHealth Consumer medical cost data \cite{fairhealth2024} for ZIP code 16803, combining facility and physician costs for both in-network and out-of-network pricing across CPT codes 99281-99285. Non-urgent ED revenue ($r_n^{\mathrm{ED}} = \$675.50$) represents the simple average of CPT codes 99281 (\$554) and 99282 (\$797). Urgent ED revenue ($r_u^{\mathrm{ED}} = \$2,221.00$) represents the average of CPT codes 99283 (\$1,483), 99284 (\$2,356), and 99285 (\$2,824). Each code average combines facility costs and primary medical procedure costs for both in-network and out-of-network pricing, reflecting the mixed payer composition typical of emergency departments. Alternative option revenue ($r^{\mathrm{Alt}} = \$436.00$) is derived from FairHealth Consumer medical cost data \cite{fairhealth2024} for hospital outpatient/clinic visits code G0463, yielding \$436.00.

\item \textbf{Balking Cost.} We represent balking cost $c^b$ as expected revenue loss when patients balk or do not enter the system. We calculate this for two scenarios: With respect to telemedicine, $c^b = (1-p_a) \cdot r_n^{\mathrm{ED}} + p_a \cdot r^{\mathrm{Tele}}$ yielding $c^b = 0.48 \times 675.50 + 0.52 \times 116.50 = \$384.82$. With respect to alternative care, $c^b = (1-p_a) \cdot r_n^{\mathrm{ED}} + p_a \cdot r^{\mathrm{Alt}}$ yielding $c^b = 0.48 \times 675.50 + 0.52 \times 436.00 = \$550.96$.

\item \textbf{Waiting Costs (Non-urgent).} Based on empirical LWBS (Left Without Being Seen) behavior from \cite{rathlev2020}, showing baseline LWBS rate of 1.41\% for patients seen within 30 minutes, increasing to 3.25\%, 7.00\%, 10.02\%, and 14.46\% for successive 30-minute intervals. The incremental increases per 30-minute period are: 3.75\% (30-60 min), 3.02\% (60-90 min), and 4.44\% (90-120 min). The average linear growth rate is calculated as $(3.75 + 3.02 + 4.44)/3 = 11.21/3 \approx 3.74\%$ per 30-minute period. This yields the model $c_n^w = \frac{(0.014 + 0.0374 \times 2 \times \text{LOS}) \times r_n^{\mathrm{ED}}}{\text{LOS}}$, where the LWBS rate increases linearly with waiting time. Mean non-urgent LOS of 3.53 hours is calculated from the exponential distribution median relationship: $\text{Mean LOS} = \text{Median}/\ln(2) = 147\text{ min}/0.693 = 212.08\text{ min} = 3.53$ hours, using median LOS for discharged patients \cite{edba2023}. With $r_n^{\mathrm{ED}} = \$675.50$ and substituting values: $c_n^w = \frac{(0.014 + 0.0748 \times 3.53) \times 675.50}{3.53} = 53.21\$$ per hour.

\item \textbf{Waiting Costs (Urgent).} Average urgent patient age of 47.0 years is calculated from high-priority patients (triage Levels 1, 2, and 3
) in the 2022 National Hospital Ambulatory Medical Care Survey \cite[Table 6]{cairns2024} using weighted averages: urgent visits by age group are Under 15: 26,463 × (0.2\% + 4.2\% + 22.3\%) = 7,066 visits; 15-24: 20,675 × (0.7\% + 6.9\% + 29.7\%) = 7,712 visits; 25-44: 39,547 × (0.8\% + 8.7\% + 34.5\%) = 17,401 visits; 45-64: 35,778 × (1.2\% + 12.9\% + 35.1\%) = 17,603 visits; 65+: 32,935 × (1.8\% + 15.4\% + 40.3\%) = 18,938 visits, yielding $\frac{\sum(\text{urgent visits}_i \times \text{midpoint age}_i)}{\sum(\text{urgent visits}_i)} = \frac{7{,}066 \times 7.5 + 7{,}712 \times 19.5 + 17{,}401 \times 34.5 + 17{,}603 \times 54.5 + 18{,}938 \times 77.5}{68{,}720} = 47.0$ years. Based on sepsis mortality studies \cite{liu2017}, analysis shows 0.4\% mortality increase per hour of treatment delay. The present value of remaining lifetime productivity at age 47 is calculated by linear interpolation between ages 40 and 50 from productivity estimates \cite{grosse2019} using 1\% annual productivity growth and 3\% discount rate: $\frac{(50-47) \times \$1{,}769{,}257 + (47-40) \times \$1{,}217{,}322}{50-40} = \$1{,}382{,}902.5$.  The urgent waiting cost model is: urgent waiting cost per hour $=$ mortality increase per hour $\times$ present value of lifetime productivity, yielding $c_u^w = 0.004 \times \$1{,}382{,}902.5 = \$5{,}531.61$ per hour.

\item \textbf{Balking Threshold.} The balking threshold $k$ is
calibrated so that the steady-state balking probability
$p_{b}(\theta^*)$ at the optimal policy aligns with empirically
reported left-without-being-seen (LWBS) rates. For the rural
setting, $k=37$ yields $p_{b} \approx 0.01$ (1\%), consistent with the LWBS rates of 0.9\%--1.5\% documented at lower-volume hospital-based emergency departments \cite{dark2020}. For the
urban setting, $k=39$ yields $p_{b} \approx 0.14$ (14\%), consistent with the substantially elevated LWBS rates observed in high-volume metropolitan emergency departments, where 13\% of patients have been reported to leave without being seen, with rates approaching 20\% during peak periods \cite{fayyaz2013}.

\end{enumerate}

Since the alternative-care threshold policy leaves urgent-patient quantities unchanged, all analyses below are conducted on the non-urgent, threshold-dependent component of the objective.

\subsubsection{Rural ED Setting.}
Using parameters from Table~\ref{tab:ed_parameters} (Rural ED column), the state space encompasses $j \in [0, 36]$ representing 37 non-urgent states, and the system remains stable with $\rho_u < 1$.
The QBD solver achieves an expected relative error of $1.825\times 10^{-7}$ against exact $M/M/c$ probabilities. Performance measures for this scenario at the optimal threshold $\theta^* = 5$ show $E[N_n] = 11.08$ as the expected non-urgent patients in system, $\lambda_n^{\mathrm{eff}} = 1.01$ as the effective non-urgent arrival rate, ${E[S_n]} = 10.97$ as the expected non-urgent waiting time, with an average of $3.15$ non-urgent patients in service and a balking probability of $p_{b} = 0.01$.

\begin{figure}[h]
\centering
\includegraphics[width=0.7\textwidth]{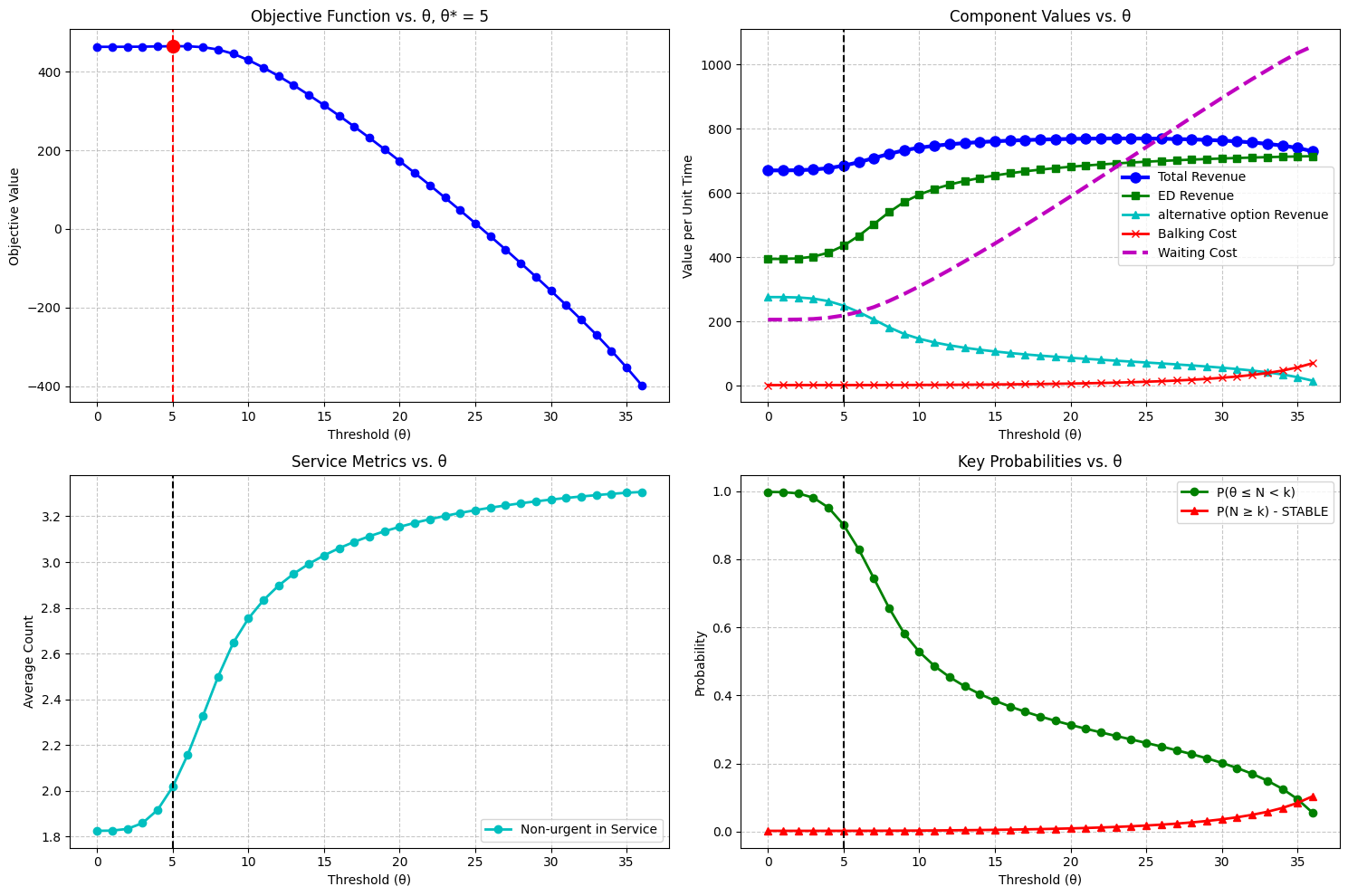}
\caption{Rural ED Setting: Objective function vs. $\theta$ with optimal $\theta^*$ = 5, showing component values, service metrics, and key probabilities -- all panels show
the non-urgent (threshold-dependent) component of the objective and
its constituents.}
\label{fig:set1_overview}
\end{figure}

\subsubsection{Urban ED Setting.}

With parameters specified in Table~\ref{tab:ed_parameters} for the Urban ED setting, the state space encompasses $j \in [0, 38]$ representing 39 non-urgent states, and the system remains stable with $\rho_u < 1$. The QBD solver achieves an expected relative error of $1.063\times 10^{-7}$ against exact $M/M/c$ probabilities.
Performance measures for this scenario at the optimal threshold $\theta^* = 27$ show $E[N_n] = 1.56$ as the expected non-urgent patients in the system, $\lambda_n^{\mathrm{eff}} = 0.32$ as the effective non-urgent arrival rate, ${E[S_n]} = 4.90$ as the expected non-urgent waiting time, with an average of $0.99$ non-urgent patients in service and a balking probability of $p_{b} = 0.14$. 

\begin{figure}[h]
\centering
\includegraphics[width=0.7\textwidth]{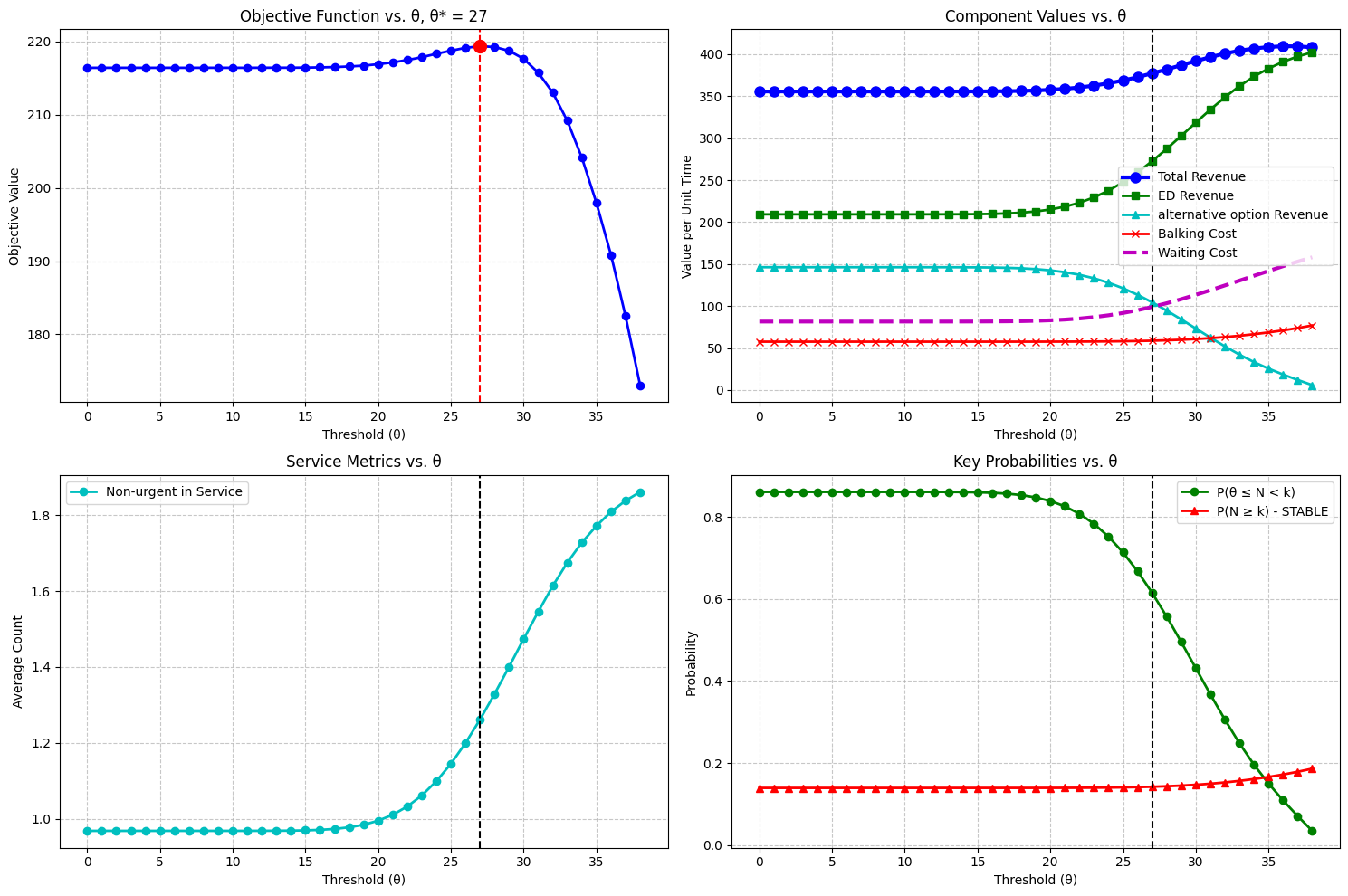}
\caption{Urban ED setting: Objective function vs. $\theta$ with optimal $\theta^*$ = 27, showing component values, service metrics, and key probabilities -- all panels show
the non-urgent (threshold-dependent) component of the objective and
its constituents.}
\label{fig:set2_overview}
\end{figure}

\section{Sensitivity Analysis}\label{sec:sensitivity}
{
\subsection{Sensitivity Analysis: $p_a$, $c$, and $\lambda$}\label{sec:sensitivity-op-params}

To examine how key operational parameters influence the optimal policy,
we conduct three separate sensitivity analyses, varying each parameter
individually while holding all others fixed and re-solving for the
optimal threshold $\theta^*$ and objective value at each level. Since the threshold policy leaves urgent-patient quantities unchanged, all analyses below are conducted on the non-urgent, threshold-dependent component of the objective. We consider the alternative care acceptance probability $p_a$, the total bed capacity $c$ (with $c_u/c=0.40$ held fixed), and the total arrival rate $\lambda$. For both the $c$ and $\lambda$ analyses, configurations
in which the system is unstable ($\rho_u \geq 1$) are excluded. Results
for the rural setting are presented in Table~\ref{tab:sensitivity_rural};
urban results are provided in Appendix~\ref{app:sensitivity_urban}.

\paragraph*{Sensitivity to $p_a$.}
In the \textit{rural} setting, the objective rises from $-168.54$ at
$p_a=0.10$ to $559.31$ at $p_a=0.90$, and $\theta^*$ rises
monotonically from $2$ to $8$: when more patients accept alternative
care, the ED can afford to wait for higher occupancy before offering it.
The urban setting confirms the same directional result with a more
muted response, as detailed in Appendix~\ref{app:sensitivity_urban}.
Both settings confirm Remark~\ref{rem:op-cs}(c): $\theta^*$ is
non-decreasing in $p_a$.

\paragraph*{Sensitivity to $c$.}
In the \textit{rural} setting, $\theta^*$ rises from $0$ at $c=6$ to
$13$ at $c=13$ and the objective improves from $-1580.45$ to $608.98$:
with more beds the system is less congested and the ED can afford to
offer alternative care at higher occupancy. The urban setting shows the same
monotone pattern over a wider capacity range; see
Appendix~\ref{app:sensitivity_urban}. Both settings confirm
Remark~\ref{rem:op-cs}(b): $\theta^*$ is non-decreasing in $c$.

\paragraph*{Sensitivity to $\lambda$.}
In the \textit{rural} setting, $\theta^*$ falls from $10$ at
$\lambda=0.5$ to $0$ at $\lambda=3.0$: as arrivals increase, the system
becomes more congested and the ED must offer alternative care at lower
occupancy. The objective deteriorates from $464.64$ at the base case
$\lambda=2$ to $-1541.77$ at $\lambda=3$. The urban setting mirrors
this pattern; details are in Appendix~\ref{app:sensitivity_urban}.
Both settings confirm Remark~\ref{rem:op-cs}(a): $\theta^*$ is
non-increasing in $\lambda$.

\setlength{\tabcolsep}{26pt}
{\tiny
\begin{longtable}{|l|c|c|r|}
\caption{Separate sensitivity analyses of $p_a$, $c$, and $\lambda$
--- Rural setting. Each block varies one parameter while holding all
others fixed. All reported values are obtained by optimizing the
non-urgent (threshold-dependent) component of the objective.}
\label{tab:sensitivity_rural}\\
\hline
\textbf{Parameter} & \textbf{Value} & \textbf{$\theta^*$} &
\textbf{Objective} \\
\hline
\endfirsthead
\hline
\textbf{Parameter} & \textbf{Value} & \textbf{$\theta^*$} &
\textbf{Objective} \\
\hline
\endhead
\hline
\endfoot
\hline
\endlastfoot

\multirow{9}{*}{\parbox{1.2cm}{\centering Vary $p_a$\\}}
 & 0.10 & 2 & $-168.54$ \\
 & 0.20 & 2 &    $76.67$ \\
 & 0.30 & 4 &   $258.71$ \\
 & 0.40 & 4 &   $378.40$ \\
 & 0.50 & 5 &   $453.65$ \\
 & 0.60 & 6 &   $499.88$ \\
 & 0.70 & 7 &   $528.73$ \\
 & 0.80 & 7 &   $546.65$ \\
 & 0.90 & 8 &   $559.31$ \\
\hline

\multirow{8}{*}{\parbox{1.2cm}{\centering Vary $c$\\}}
 &  6 &  0 & $-1580.45$ \\
 &  7 &  0 &  $-509.33$ \\
 &  8 &  1 &   $239.16$ \\
 &  9 &  5 &   $464.64$ \\
 & 10 &  8 &   $544.99$ \\
 & 11 & 10 &   $579.84$ \\
 & 12 & 12 &   $597.18$ \\
 & 13 & 13 &   $608.98$ \\
\hline

\multirow{6}{*}{\parbox{1.2cm}{\centering Vary $\lambda$\\}}
 & 0.50 & 10 &   $155.26$ \\
 & 1.00 &  9 &   $307.54$ \\
 & 1.50 &  8 &   $436.89$ \\
 & 2.00 &  5 &   $464.64$ \\
 & 2.50 &  3 &  $-140.12$ \\
 & 3.00 &  0 & $-1541.77$ \\
\end{longtable}
}
}
\subsection{Robustness Analysis: Service Time Distributions}\label{sec:robust-serv}
The QBD model assumes exponentially distributed service times. To verify
that the optimal threshold policy $\theta^*$ remains effective under
non-exponential service, we conduct a discrete-event simulation
robustness study using preemptive priority. The warm-up period
$\ell$ is determined via Welch's method \cite{welch1983}; full methodology, equations,
and diagnostic plots are provided in Appendix~\ref{app:warmup}.

We test three distributions with identical means but different
variability, measured by the squared coefficient of variation $c^2=\mathrm{Var}[\tau]/(E[\tau])^2$: \textbf{Exponential}
($c^2=1.00$, QBD baseline), \textbf{Erlang-2} ($c^2=0.50$, modeling two sequential treatment stages), and
\textbf{Lognormal} ($c^2=1.50$, modeling
occasional complex cases with a heavy right tail). We run $m=30$
replications per $\theta$ under exponential service to identify
$\theta^*_{\mathrm{sim}}$, then fix $\theta^*_{\text{sim}}$ and run 30
replications under each distribution. The policy is declared
\emph{robust} if the 95\% CI overlaps with the exponential baseline.
Full simulation formulas and the robustness criterion are given in
Appendix~\ref{app:warmup}.

{\tiny
\begin{longtable}{|l|c|r|r|r|r|r|r|r|r|}
\caption{Robustness results -- Rural setting
($\theta^*=7$, $\ell=1003$, $T_{\text{sim}}=5000$, $m=30$). All
reported values are for the non-urgent (threshold-dependent)
component of the objective. $\hat{p}_b(\theta^*)$ and
$\hat{p}_{x}(\theta^*)$ are empirical estimates of the balking
probability $p_b(\theta^*)$ and the alternative-care acceptance
probability $p_x(\theta^*)$ defined in
Section~\ref{sec:obj_components}.}
\label{tab:robust_rural}\\
\hline
\textbf{Distribution} & \textbf{$c^2$} & \textbf{Objective} &
\textbf{95\% CI} & \textbf{Revenue} & \textbf{Balk\$} &
\textbf{Wait\$} & \textbf{$\hat{p}_b(\theta^*)$} &
\textbf{$\hat{p}_{x}(\theta^*)$} & \textbf{$\widehat{E}[N_n]$} \\
\hline
\endfirsthead
\hline
\textbf{Distribution} & \textbf{$c^2$} & \textbf{Objective} &
\textbf{95\% CI} & \textbf{Revenue} & \textbf{Balk\$} &
\textbf{Wait\$} & \textbf{$\hat{p}_b(\theta^*)$} &
\textbf{$\hat{p}_{x}(\theta^*)$} & \textbf{$\widehat{E}[N_n]$} \\
\hline
\endhead
\hline
\endfoot
\hline
\endlastfoot
Exponential & 1.00 & 475.26 & [463.57,\;486.95] &
  712.65 & 1.14 & 236.26 & 0.0017 & 0.384 & 5.92 \\
\hline
Erlang-2    & 0.50 & 489.60 & [480.63,\;498.57] &
  711.42 & 0.38 & 221.44 & 0.0006 & 0.389 & 5.59 \\
\hline
Lognormal   & 1.50 & 465.91 & [453.96,\;477.86] &
  711.03 & 2.46 & 242.66 & 0.0036 & 0.387 & 6.12 \\
\end{longtable}}

\subsubsection{Results: Rural Setting ($\theta^*_{\mathrm{sim}}=7$, $\ell=1003$)}
Both distributions yield overlapping CIs with the exponential baseline, with objective differences of $+3.02\%$ (Erlang-2) and $-1.97\%$ (Lognormal) relative to the exponential.
Revenue is nearly identical across all three ($\approx711$--$713$),
confirming throughput is insensitive to distributional shape. Differences
are driven by waiting cost: Erlang-2 reduces it from $236.26$ to
$221.44$ due to lower queue variability, while Lognormal raises it to
$242.66$ from occasional very long service times. Balking probability is negligible in all cases.


\subsubsection{Results: Urban Setting
($\theta^*_{\mathrm{sim}}=30$, $\ell=1942$)}

Both distributions yield overlapping CIs with objective differences
below $1.5\%$, confirming robustness to non-exponential service in the
urban setting as well. Waiting cost is nearly identical across
distributions ($\approx110$--$112$), indicating that the dominant
urgent stream ($p_u=0.85$) constrains behavior more than service time
variability. Full tables are provided in Appendix~\ref{app:robust_urban}.

\subsection{Simulation Analysis: Priority Discipline and Referral Policy}\label{sec:sim-priority-referral}
The QBD model assumes preemptive priority, where an urgent arrival can
interrupt a non-urgent patient in service. To assess whether this
assumption materially affects the optimal policy, we compare four
configurations combining two priority disciplines, preemptive and
non-preemptive, with two referral policies, with and without the
alternative care option, via discrete-event simulation. All simulations use exponential service times, $T_{\text{sim}}=5000$, and $m=500$ replications per $\theta$ value. The warm-up period $\ell$ is determined separately for each configuration via Welch's method. For the referral cases, $\theta^*$ is found by enumerating over all $\theta \in \{0,\ldots,k-1\}$ and selecting the maximizer of the mean objective. For the no-referral cases, no threshold exists and the
objective is evaluated directly.

\subsubsection{Rural Setting}

Both disciplines identify $\theta^*=6$ as the optimal threshold when
referral is available. Non-preemptive outperforms preemptive by
$492.93 - 473.34 = 19.59$ units ($+4.14\%$), driven entirely by lower
waiting cost ($201.80$ vs.\ $221.53$) since non-urgent patients are never
interrupted mid-service. Without referral, the rural system deteriorates
catastrophically: the objective falls to $-365.35$ (non-preemptive) and
$-449.88$ (preemptive), with $E[N_n]$ rising from $\approx 4$ to
$\approx 19$--$20$ and waiting cost exploding from $\approx 202$--$222$
to $\approx 1010$--$1077$. The gain from referral exceeds 858 units
($+235\%$) for non-preemptive and 923 units ($+205\%$) for preemptive.

{
\begin{table}[h!]
\tiny
\caption{Rural setting --- all four configurations ($T_{\text{sim}}=5000$, $m=500$, $k=37$). All reported values are for the non-urgent (threshold-dependent) component of the objective. $\hat{p}_b(\theta^*)$ and $\hat{p}_x(\theta^*)$ are empirical estimates of the balking probability $p_b(\theta^*)$ and the alternative-care acceptance probability $p_x(\theta^*)$ defined in Section~\ref{sec:obj_components}. $\widehat{E}[N_n]$ and $\widehat{E}[S_n]$ denote the estimated mean non-urgent census and mean non-urgent sojourn time, respectively. N/A applies to no-referral configurations, where neither $\theta^*$ nor $p_x(\theta^*)$ is defined.}
\label{tab:rural_all_sim}
\centering
\begin{tabular}{|l|l|r|r|r|r|r|r|r|r|}
\hline
\textbf{Priority} & \textbf{Referral} & \textbf{$\theta^*$} &
\textbf{$\ell$} & \textbf{Objective} & \textbf{95\% CI} &
\textbf{Wait\$} & \textbf{Balk\$} &
\textbf{$\widehat{E}[N_n]$} & \textbf{$\hat{p}_b(\theta^*)$} \\
\hline
Non-preemptive & Yes & 6   & 212  & 492.93    &
  [490.69,\;495.18]       & 201.80  & 1.30  & 3.79  & 0.0019 \\
\hline
Preemptive     & Yes & 6   & 1003 & 473.34    &
  [470.81,\;475.87]       & 221.53  & 1.41  & 4.16  & 0.0021 \\
\hline
Non-preemptive & No  & N/A & 1615 & $-$365.35 &
  [$-$374.66,\;$-$356.05] & 1010.47 & 80.40 & 18.99 & 0.1196 \\
\hline
Preemptive     & No  & N/A & 1738 & $-$449.88 &
  [$-$460.27,\;$-$439.49] & 1077.41 & 87.68 & 20.25 & 0.1305 \\
\hline
\end{tabular}

\vspace{0.4cm}

\begin{tabular}{|l|l|r|r|r|r|r|r|}
\hline
\textbf{Priority} & \textbf{Referral} & \textbf{Revenue} &
\textbf{ED Rev} & \textbf{Alt Rev} &
\textbf{$\widehat{E}[S_n]$} & \textbf{$\widehat{\lambda_n^{\mathrm{eff}}}$} &
\textbf{$\hat{p}_{x}(\theta^*)$} \\
\hline
Non-preemptive & Yes & 696.04 & 466.64 & 229.39 & 5.49  & 0.6909 & 0.4314 \\
\hline
Preemptive     & Yes & 696.28 & 466.52 & 229.76 & 6.04  & 0.6905 & 0.4319 \\
\hline
Non-preemptive & No  & 725.52 & 725.52 & N/A    & 17.72 & 1.0738 & N/A    \\
\hline
Preemptive     & No  & 715.20 & 715.20 & N/A    & 19.17 & 1.0590 & N/A    \\
\hline
\end{tabular}

\end{table}
}

\subsubsection{Urban Setting}

With referral, the two disciplines produce slightly different optimal
thresholds ($\theta^*=31$ for non-preemptive, $\theta^*=29$ for
preemptive). Non-preemptive outperforms preemptive by $26.30$ units
($+11.65\%$), a larger gap than in the rural case due to higher system
load ($\rho_u=0.833$). Without referral, the urban system remains
profitable under both disciplines, in contrast to the rural case, since
only 15\% of arrivals are non-urgent. {The gain from referral is
$+36.68$ units ($+17\%$) for non-preemptive and $+53.16$ units ($+31\%$)
for preemptive}. Full tables are provided in
Appendix~\ref{app:urban_sim}.

\subsubsection{Summary}
The results yield two key findings. First, the alternative care referral
policy improves performance under both priority disciplines and both
settings: in the rural case the gain is $+858$ ($+235\%$) units for
non-preemptive and $+923$ ($+205\%$) units for preemptive, while in the
urban case the gain is more modest at $+37$ units ($+17\%$) and $+53$
units ($+31\%$) respectively. Notably, the optimal threshold $\theta^*$
is identical across both disciplines in the rural setting ($\theta^*=6$)
and nearly identical in the urban setting ($\theta^*=31$ vs.\
$\theta^*=29$), indicating that the choice of priority discipline does
not materially affect the optimal redirection policy. Second, as expected
non-preemptive priority consistently outperforms preemptive across all
four configurations, since non-urgent patients are never interrupted
mid-service. Preemptive priority is the conservative case and the one that keeps urgent dynamics $\theta$-independent and QBD-tractable. Taken together, these results confirm that the referral
policy provides substantial operational benefit regardless of the
priority discipline assumed, and that the optimal threshold is robust
to the preemptive versus non-preemptive modeling choice.

{{\tiny
\begin{longtable}{|l|l|r|r|r|r|}
\caption{\tiny Objective gains across settings and configurations. All reported values are for the non-urgent (threshold-dependent) component of the objective.}
\label{tab:sim_summary}\\
\hline
\textbf{Setting} & \textbf{Priority} &
\textbf{$Z_{\text{referral}}$} & \textbf{$Z_{\text{no referral}}$} &
\textbf{Gain from referral} & \textbf{Gain: NP over P} \\
\hline
\endfirsthead
\hline
\textbf{Setting} & \textbf{Priority} &
\textbf{$Z_{\text{referral}}$} & \textbf{$Z_{\text{no referral}}$} &
\textbf{Gain from referral} & \textbf{Gain: NP over P} \\
\hline
\endhead
\hline
\endfoot
\hline
\endlastfoot
Rural & Non-preemptive & 492.93 & $-$365.35 &
  $+$858.28 (235\%)  & \multirow{2}{*}{$+$19.59 (4.1\%)} \\
\cline{1-5}
Rural & Preemptive     & 473.34 & $-$449.88 &
  $+$923.22 (205\%) & \\
\hline
Urban & Non-preemptive & 252.21 & 215.53 &
  $+$36.68 (17\%)   & \multirow{2}{*}{$+$26.30 (11.7\%)} \\
\cline{1-5}
Urban & Preemptive     & 225.91 & 172.75 &
  $+$53.16 (31\%)   & \\
\end{longtable}
}}

\subsection{Tornado Diagram Analysis}
To assess the robustness of the optimal alternative care threshold policy, we conduct a comprehensive sensitivity analysis using tornado diagrams. This analysis evaluates how parameter uncertainties affect the objective function $Z(\theta)$ at the optimal operating point $\theta^* = \arg\max_{\theta \in \{0,1,\ldots,k-1\}} Z(\theta)$ from equation~\eqref{eq:objective_function}. Unlike the sensitivity analyses of Sections~\ref{sec:sensitivity-op-params}-\ref{sec:sim-priority-referral}, which focus on non-urgent patient quantities only, all analyses from this point forward use the complete objective function $Z(\theta)$ including all performance metrics and components for both patient types: revenue from urgent and non-urgent patients (ED and alternative care), waiting costs for urgent and non-urgent patients, and balking cost of non-urgents. 

\subsubsection{Methodology}
Our tornado analysis evaluates sensitivity using operational ratios rather than individual parameters, comparing our alternative care threshold model against a no-referral baseline case. The analysis employs a comprehensive set of proportional relationships including bed allocation ratio $(c_u/c)$, service rate ratio $(\mu_u/\mu_n)$, revenue ratio $(r_u^{\mathrm{ED}}/r_n^{\mathrm{ED}})$, waiting cost ratio $(c_u^w/c_n^w)$, alternative care revenue ratio $(r^{\mathrm{Alt}}/r_n^{\mathrm{ED}})$, balking cost ratio $(c^{b}/r_n^{\mathrm{ED}})$, and threshold proportion $(\theta/k)$. 
For each operational ratio $\Tilde{r}_i$, the sensitivity impact is measured by:
\begin{equation}
I^{(i)} = |Z(\theta^*; \Tilde{r}_{i,\text{high}}) - Z(\theta^*; \Tilde{r}_{i,\text{low}})|,
\end{equation}
where $\Tilde{r}_{i,\text{low}} = \Tilde{r}_i \cdot 0.95$ and $\Tilde{r}_{i,\text{high}} = \Tilde{r}_i \cdot 1.05$.

The analysis generates comprehensive tornado plots for two operational modes. First, for our alternative care-enabled ED model operating at optimal threshold $\theta^*$, we provide tornado plots beginning with the base case (rural/urban parameter set), then systematically varying each operational ratio by $20\%$ in their feasible region to assess high and low sensitivity scenarios while fixing other ratios, incorporating both urgent and non-urgent revenue streams and associated costs. Second, for the alternative care-disabled case where all non-urgent patients enter the ED if $N(t) < k$ without alternative care options, we apply the same procedure. This comparative approach reveals differential sensitivity patterns between our alternative care policy and traditional ED operations, identifying which operational ratios exert the greatest influence on system performance and demonstrating the robustness of our threshold policy across varying proportional configurations.
We analyze rural and urban ED settings separately.

\subsection{Rural ED} This section presents the sensitivity analysis results examining how parameter variations affect the Rural ED model's performance.

\subsubsection{Rural Alternative Care-enabled ED: Base Case.}
We use parameter values and key operational ratios from Table~\ref{tab:ed_parameters}: $\lambda = 2$ patients/hour, $p_u = 0.39$, total capacity $c = 9$ beds, $k = 37$ patients, $p_a = 0.52$, bed allocation ratio $(c_u/c = 0.4)$, service rate ratio $(\mu_u/\mu_n =  0.15/0.32)$, revenue ratio $(r_u^{\mathrm{ED}}/r_n^{\mathrm{ED}} = 2,221/675.50)$, waiting cost ratio $(c_u^w/c_n^w = 5,531.61/53.21)$, alternative care revenue ratio $(r^{\mathrm{Alt}}/r_n^{\mathrm{ED}} = 436/675.50)$, and balking cost ratio $(c^{b}/r_n^{\mathrm{ED}} = 550.96/675.50)$. 

\begin{figure}[h]
\includegraphics[width=0.6\textwidth]{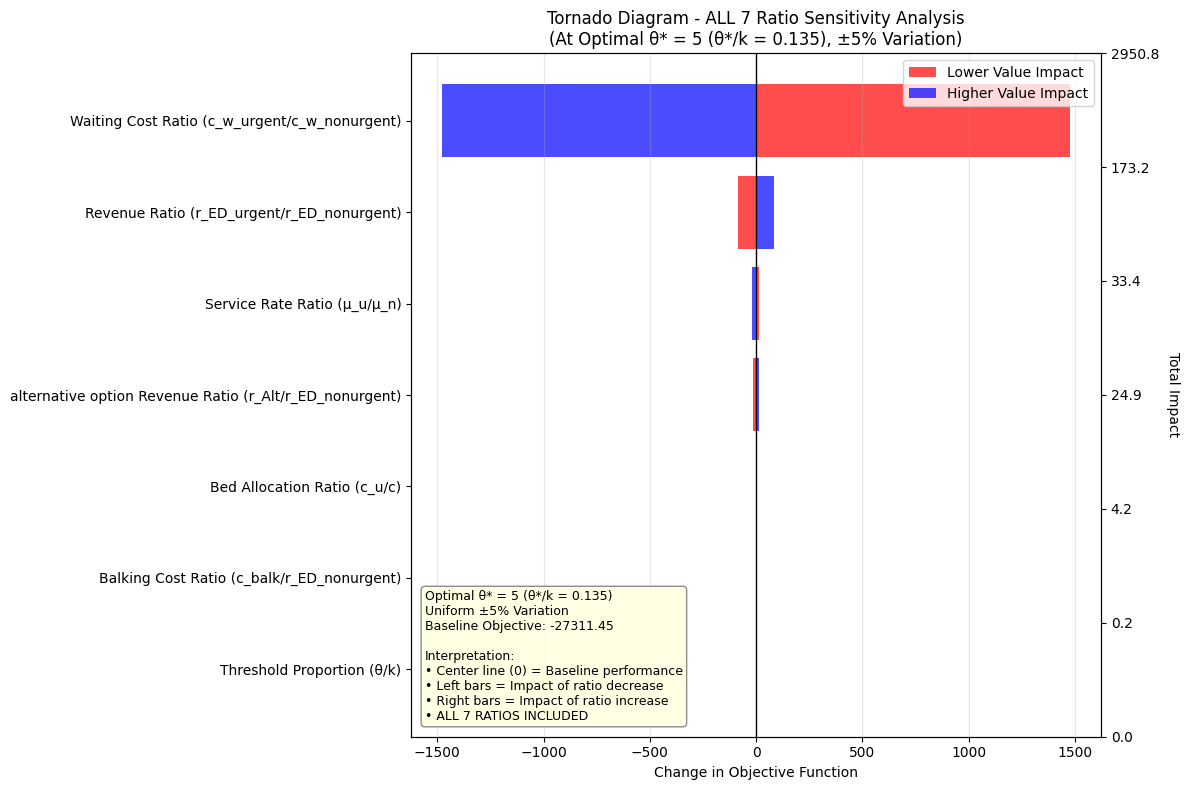}
\caption{Tornado Diagram for Rural Alternative Care-enabled ED- (±5\% Variation)}
\label{fig:tornado_baseline_optimal}
\end{figure}

{\tiny
\begin{longtable}{|c|l|c|c|}
\caption{Sensitivity Analysis Summary - Rural Alternative Care-enabled ED}
\label{tab:tornado_sensitivity_analysis-1}\\
\hline
\textbf{Rank} & \textbf{Ratio} & \textbf{Relative Impact (\%)} & \textbf{Base Value} \\
\hline
\endfirsthead

\hline
\textbf{Rank} & \textbf{Ratio} & \textbf{Relative Impact (\%)} & \textbf{Base Value} \\
\hline
\endhead

\hline
\endfoot

\hline
\endlastfoot

1 & Waiting Cost Ratio ($c_u^w$/$c_n^w$) & 10.81 & 103.958 \\
\hline
2 & Revenue Ratio ($r_u^{\mathrm{ED}}$/$r_n^{\mathrm{ED}}$) & 0.63 & 3.288 \\
\hline
3 & Service Rate Ratio ($\mu_u$/$\mu_n$) & 0.12 & 0.469 \\
\hline
4 & Alternative Option Revenue Ratio ($r^{\mathrm{Alt}}$/$r_n^{\mathrm{ED}}$) & 0.09 & 0.645 \\
\hline
5 & Bed Allocation Ratio ($c_u$/$c$) & 0.02 & 0.400 \\
\hline
6 & Balking Cost Ratio ($c^{b}$/$r_n^{\mathrm{ED}}$) & 0.00 & 0.816 \\
\hline
7 & Threshold Proportion ($\theta$/$k$) & 0.00 & 0.135 \\

\end{longtable}
}

\noindent
\subsubsection{Rural Alternative Care-enabled ED: Case Comparison.}
To understand how system characteristics affect sensitivity patterns, we analyze tornado diagrams across 15 different operational scenarios, ranging from high-demand to capacity-constrained systems.

{\tiny
\begin{longtable}{|c|l|l|r|c|c|c|r|}

\caption{Case Comparison Summary for Rural Alternative Care-enabled ED}
\label{tab:case_comparison-1}\\


\hline
\textbf{Rank} & \textbf{Case} & \textbf{Description} & \textbf{Baseline Obj} & \textbf{$\theta^*$} & \textbf{$\theta^*/k$} & \textbf{{Top} Ratio} & \textbf{Rel. Impact (\%)} \\
\hline
\endfirsthead

\hline
\textbf{Rank} & \textbf{Case} & \textbf{Description} & \textbf{Baseline Obj} & \textbf{$\theta^*$} & \textbf{$\theta^*/k$} & \textbf{Top Ratio} & \textbf{Rel. Impact (\%)} \\
\hline
\endhead

\hline
\endfoot

\hline
\endlastfoot

1 & Low Urgent Proportion & $p_u -20\%$& -21,154.99 & 6 & 0.162 & $c_u^w/c_n^w$ & 10.95 \\
\hline
2 & Low Arrival Rate & $\lambda-20\%$ & -21,311.17 & 8 & 0.216 & $c_u^w/c_n^w$ & 10.87 \\
\hline
3 & High Urgent Service Rate & $\mu_u$ +20\% & -21,837.87 & 8 & 0.216 & $c_u^w/c_n^w$ & 11.07 \\
\hline
4 & High Capacity & $c +20\%$ & -26,759.90 & 8 & 0.216 & $c_u^w/c_n^w$ & 10.85 \\
\hline
5 & Low Capacity & $c -20\%$& -26,759.90 & 8 & 0.216 & $c_u^w/c_n^w$ & 10.85 \\
\hline
6 & High Acceptance Rate & $p_{a}$ +20\% & -27,268.20 & 6 & 0.162 & $c_u^w/c_n^w$ & 10.82 \\
\hline
7 & Low Balking Threshold & $k -20\%$ & -27,309.15 & 5 & 0.135 & $c_u^w/c_n^w$ & 10.81 \\
\hline
8 & Baseline & Original parameters & -27,311.45 & 5 & 0.135 & $c_u^w/c_n^w$ & 10.81 \\
\hline
9 & High Theta & $\theta$ +20\% & -27,311.45 & 5 & 0.135 & $c_u^w/c_n^w$ & 10.81 \\
\hline
10 & Low Theta & $\theta -20\%$ & -27,311.45 & 5 & 0.135 & $c_u^w/c_n^w$ & 10.81 \\
\hline
11 & High Balking Threshold & $k$ +20\% & -27,317.46 & 3 & 0.081 & $c_u^w/c_n^w$ & 10.80 \\
\hline
12 & Low Acceptance Rate & $p_{a} -20\%$ & -27,383.28 & 1 & 0.027 & $c_u^w/c_n^w$ & 10.78 \\
\hline
13 & Low Urgent Service Rate & $\mu_u -20\%$& -28,560.91 & 4 & 0.108 & $c_u^w/c_n^w$ & 10.75 \\
\hline
14 & High Urgent Proportion & $p_u$ +20\% & -35,156.48 & 3 & 0.081 & $c_u^w/c_n^w$ & 10.65 \\
\hline
15 & High Arrival Rate & $\lambda +20\%$ &  {-35,287.23} &  {0} & {0.000} & $c_u^w/c_n^w$ & {10.61} \\


\end{longtable}
}

The sensitivity analysis across 15 scenarios reveals that the waiting cost ratio (urgent vs. non-urgent) dominates system performance across all cases. The patient mix emerges as the most critical operational factor: reducing urgent patients by 20\% improves performance by 22.5\%, while increasing them by 20\% deteriorates performance by 28.7\%. Similarly, a 20\% increase in arrival rate results in the largest 
deterioration across all scenarios (29.2\%), with the optimal 
threshold dropping to $\theta^* = 0$. The optimal threshold $\theta^*$ varies substantially across scenarios, demonstrating the need for context-dependent policies. Revenue ratio provides consistent secondary impact, while service rate considerations remain tertiary. These findings highlight that managing patient mix and understanding relative waiting costs are the primary drivers.

\begin{figure}[h]
\centering
\includegraphics[width=0.9\textwidth]{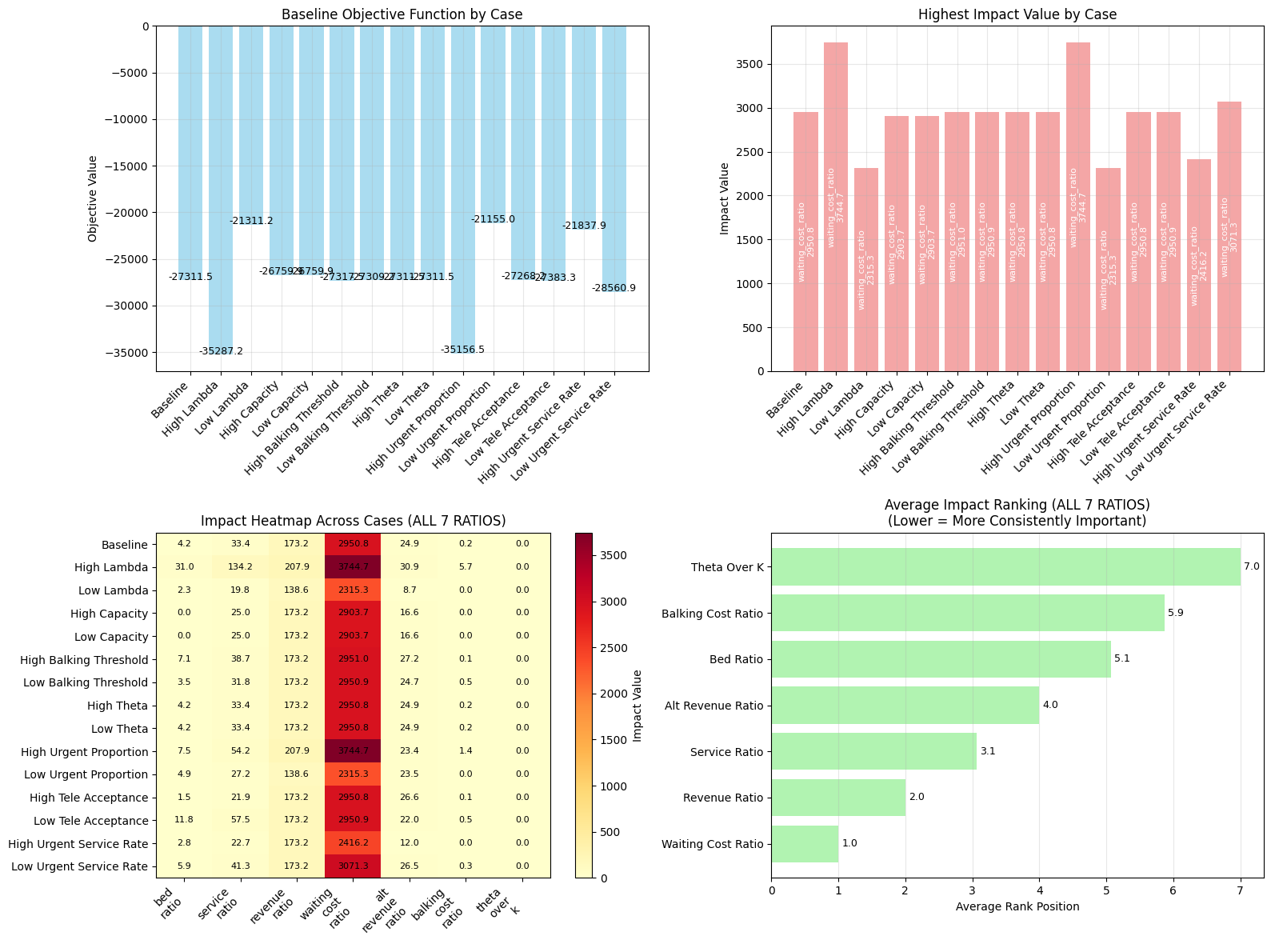}
\caption{Case Comparison Summary of Rural Alternative Care-enabled ED: Baseline Objectives and Impact Rankings Across All Scenarios (±5\% Variation)}
\label{fig:case_comparison_summary}
\end{figure}

\subsubsection{Rural Alternative Care-disabled ED: Base Case.}

\sloppy
Here we consider a simplified emergency department model, that eliminates alternative care referrals entirely, creating a binary decision system where non-urgent patients either enter the ED (when total patients $< k$) or balk (when total patients $\geq k$). This streamlined approach again focuses exclusively on five core operational ratios: bed allocation ratio, service rate ratio, revenue ratio, waiting cost ratio, and balking cost ratio.

The base case configuration maintains the same fundamental parameters from Table~\ref{tab:ed_parameters}: $\lambda = 2$ patients/hour, $p_u = 0.39$, total capacity $c = 9$ beds, $k = 37$ patients, with identical operational ratios as the alternative option-enabled model. However, the simplified balking function eliminates the intermediate alternative option. The probability that a non-urgent patient decides to join the ED upon arrival when the system state is $(i,j)$, denoted $\alpha_{\theta}(i,j)$, equals 1 when the total number of patients $(i+j)$ is below the threshold $k$, meaning non-urgent patients join the system. When the total reaches or exceeds $k$ patients, $\alpha_{\theta}(i,j)$ equals 0, causing all non-urgent patients to balk and leave without receiving care.

This simplified baseline model serves as a useful comparison point to demonstrate the operational benefits that the alternative option provides in ED management.

{\tiny
\begin{longtable}{|c|l|c|c|}
\caption{Sensitivity Analysis Summary - Rural Alternative Care-disabled ED}
\label{tab:tornado_sensitivity_analysis-2}\\
\hline
\textbf{Rank} & \textbf{Ratio} & \textbf{Relative Impact (\%)} & \textbf{Base Value} \\
\hline
\endfirsthead
\hline
\textbf{Rank} & \textbf{Ratio} & \textbf{Relative Impact (\%)} & \textbf{Base Value} \\
\hline
\endhead
\hline
\endfoot
\hline
\endlastfoot
1 & Waiting Cost Ratio ($c_u^w$/$c_n^w$) & 10.45 & 103.958 \\
\hline
2 & Service Rate Ratio ($\mu_u$/$\mu_n$) & 1.10 & 0.469 \\
\hline
3 & Bed Allocation Ratio ($c_u$/$c$) & 0.62 & 0.400 \\
\hline
4 & Revenue Ratio ($r_u^{\mathrm{ED}}$/$r_n^{\mathrm{ED}}$) & 0.61 & 3.288 \\
\hline
5 & Balking Cost Ratio ($c^{b}$/$r_n^{\mathrm{ED}}$) & 0.04 & 1.000 \\
\hline
\end{longtable}
}

\subsubsection{Rural Alternative Care-disabled ED: Case Comparison.}
To understand sensitivity patterns in the simplified model, we analyze tornado diagrams across 11 different operational scenarios, examining how system characteristics affect parameter importance without alternative care complexity.

{\tiny
\begin{longtable}{|c|l|l|r|c|r|}
\caption{Case Comparison Summary for Rural Alternative Care-disabled ED}
\label{tab:case_comparison-2}\\

\hline
\textbf{Rank} & \textbf{Case} & \textbf{Description} & \textbf{Baseline Obj} & \textbf{Top Ratio} & \textbf{Rel. Impact (\%)} \\
\hline
\endfirsthead

\hline
\textbf{Rank} & \textbf{Case} & \textbf{Description} & \textbf{Baseline Obj} & \textbf{Top Ratio} & \textbf{Rel. Impact (\%)} \\
\hline
\endhead

\hline
\endfoot

\hline
\endlastfoot

1 & Low Arrival Rate & $\lambda -20\% $& -21,451.87 & $c_u^w/c_n^w$ & 10.79 \\
\hline
2 & Low Urgent Proportion & $p_u -20\% $& -21,978.96 & $c_u^w/c_n^w$ & 10.53 \\
\hline
3 & High Urgent Service Rate & $\mu_u$ +20\% & -22,017.27 & $c_u^w/c_n^w$ & 10.97 \\
\hline
4 & High Capacity & $c$ +20\% & -27,117.04 & $c_u^w/c_n^w$ & 10.71 \\
\hline
5 & Low Capacity & $c -20\% $& -27,117.04 & $c_u^w/c_n^w$ & 10.71 \\
\hline
6 & Low Balking Threshold & $k -20\% $& -27,974.34 & $c_u^w/c_n^w$ & 10.55 \\
\hline
7 & Baseline & Original parameters & -28,245.93 & $c_u^w/c_n^w$ & 10.45 \\
\hline
8 & High Balking Threshold & $k$ +20\% & -28,706.18 & $c_u^w/c_n^w$ & 10.28 \\
\hline
9 & Low Urgent Service Rate & $\mu_u -20\% $& -29,656.36 & $c_u^w/c_n^w$ & 10.36 \\
\hline
10 & High Urgent Proportion & $p_u$ +20\% & -36,185.67 & $c_u^w/c_n^w$ & 10.35 \\
\hline
11 & High Arrival Rate & $\lambda +20\%$ & {-36,739.30} & $c_u^w/c_n^w$ & {10.19} \\
\hline

\end{longtable}
}

In the absence of alternative care options, the waiting cost ratio remains the dominant factor across all 11 scenarios. Service rate ratio emerges as the secondary factor, followed by revenue ratio as tertiary. Patient mix and arrival rate drive dramatic performance variation, with the worst-case scenario (high arrival rate) performing {71.3\%} worse than the best-case (low arrival rate). Without alternative care pathways, service efficiency becomes more critical for managing system bottlenecks, though waiting cost differentials remain the primary optimization driver.

\begin{figure}[H]
\centering
\includegraphics[width=0.9\textwidth]{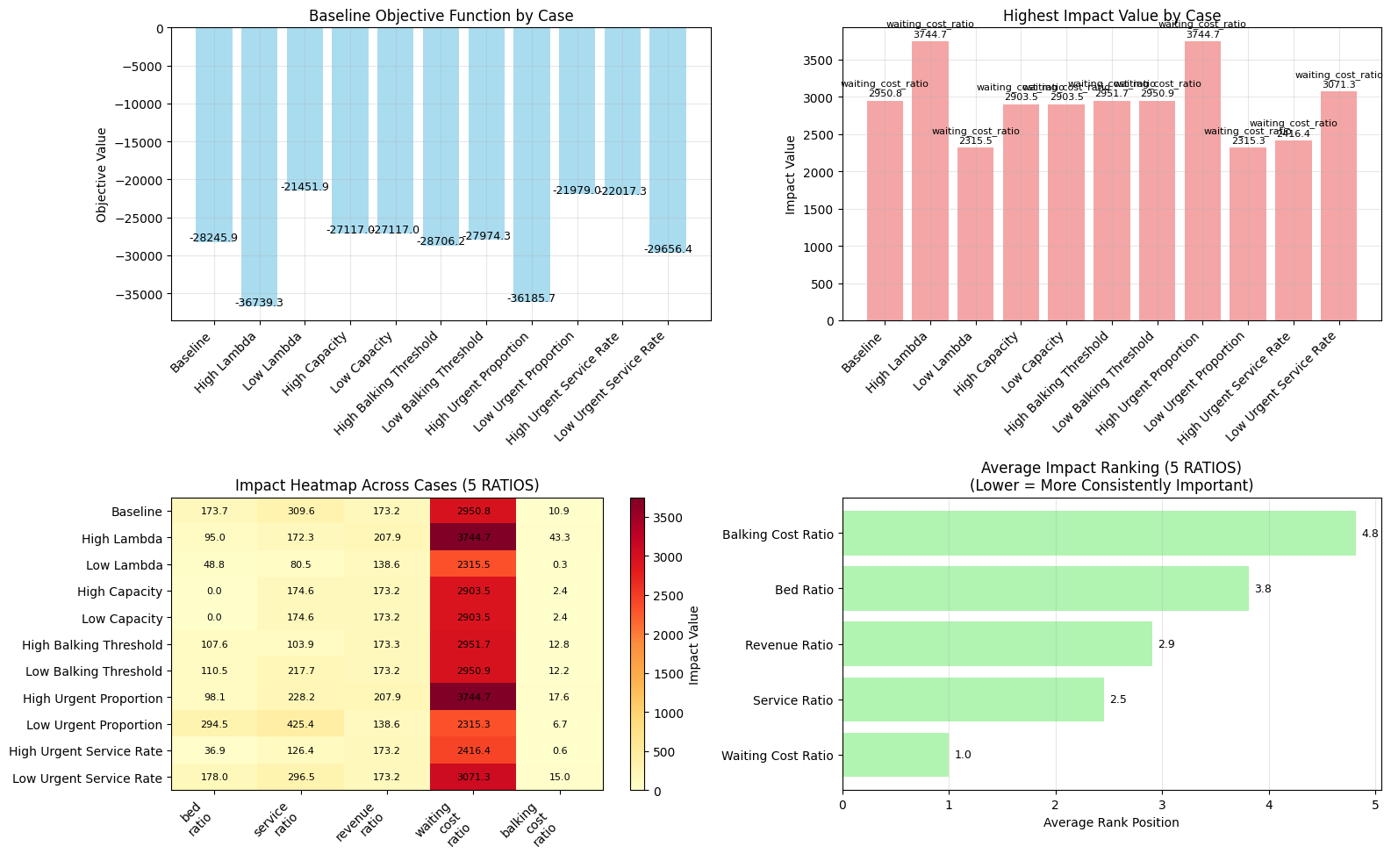}

\caption{Case Comparison Summary of Rural Alternative Care-disabled ED: Baseline Objectives and Impact Rankings Across All Scenarios (±5\% Variation)}
\label{fig:case_comparison_summary_no_tele}
\end{figure}

\subsubsection{Rural ED: Alternative Care Enabled vs. Disabled.}
Alternative care improves performance across all 11 scenarios, with gains ranging from 0.66\% (low arrival rate) to 4.84\% (high balking threshold). Under baseline conditions, alternative care provides a 3.31\% improvement, while maintaining consistent benefits even under high demand conditions (2.85\% improvement in high urgent proportion scenario). The waiting cost ratio remains the dominant factor in both configurations, indicating that alternative care enhances operational efficiency without altering the fundamental performance drivers.

{\tiny
\begin{longtable}{|l|c|c|c|c|c|c|}

\caption{Alternative Care Impact: Enabled vs. Disabled with Relative Impact Comparison}
\label{tab:alternative_option_impact}\\

\hline
\textbf{Scenario} & \textbf{Enabled} & \textbf{Enabled} & \textbf{Disabled} & \textbf{Disabled} & \textbf{Benefit} & \textbf{Gain} \\
 & \textbf{(\$/hr)} & \textbf{Rel. Impact (\%)} & \textbf{(\$/hr)} & \textbf{Rel. Impact (\%)} & \textbf{(\$/hr)} & \textbf{(\%)} \\
\hline
\endfirsthead

\hline
\textbf{Scenario} & \textbf{Enabled} & \textbf{Enabled} & \textbf{Disabled} & \textbf{Disabled} & \textbf{Benefit} & \textbf{Gain} \\
 & \textbf{(\$/hr)} & \textbf{Rel. Impact (\%)} & \textbf{(\$/hr)} & \textbf{Rel. Impact (\%)} & \textbf{(\$/hr)} & \textbf{(\%)} \\
\hline
\endhead

\hline
\endfoot

\hline
\endlastfoot

Low Urgent Proportion & -21,155 & 10.95 & -21,979 & 10.53 & +824 & +3.75 \\
\hline
Low Arrival Rate & -21,311 & 10.87 & -21,452 & 10.79 & +141 & +0.66 \\
\hline
High Urgent Service Rate & -21,838 & 11.07 & -22,017 & 10.97 & +179 & +0.81 \\
\hline
High Capacity & -26,760 & 10.85 & -27,117 & 10.71 & +357 & +1.32 \\
\hline
Low Capacity & -26,760 & 10.85 & -27,117 & 10.71 & +357 & +1.32 \\
\hline
Low Balking Threshold & -27,309 & 10.81 & -27,974 & 10.55 & +665 & +2.38 \\
\hline
Baseline & -27,311 & 10.81 & -28,246 & 10.45 & +935 & +3.31 \\
\hline
High Balking Threshold & -27,317 & 10.80 & -28,706 & 10.28 & +1,389 & +4.84 \\
\hline
Low Urgent Service Rate & -28,561 & 10.75 & -29,656 & 10.36 & +1,095 & +3.69 \\
\hline
High Urgent Proportion & -35,156 & 10.65 & -36,186 & 10.35 & +1,030 & +2.85 \\
\hline
High Arrival Rate &  {-35,287} & {10.61} &  {-36,739} & {10.19} & {+1,452} & {3.95\%} \\

\end{longtable}
}


\subsection{Urban ED}
For the urban setting, we conducted the same comprehensive analysis: we examined the base case and compared scenarios with alternative care both enabled and disabled, following the same methodology as the rural setting. Details are provided in Appendix \ref{app:urban setting analysis}. Under baseline conditions, alternative care yields a modest improvement of 0.046\%. However, the most substantial benefit emerges under high urgent proportion conditions, where alternative care achieves a 5.90\% improvement. Consistent with the rural analysis, the waiting cost ratio remains the dominant factor influencing system performance in both configurations. Notably, the service rate ratio demonstrates exceptional importance in low urgent proportion scenarios.

\subsection{Proportional Analysis: Optimal Objective Function Response}
To complement the tornado diagram sensitivity analysis, we conduct a comprehensive proportional analysis that examines how systematic variations in each operational ratio affect the optimal objective function. Unlike the tornado analysis which evaluates parameter sensitivity at a fixed threshold policy, the proportional analysis optimizes the alternative care threshold $\theta^*$ for each ratio configuration, revealing the true performance envelope under varying operational conditions. Complete results are provided in Appendix \ref{app:proportional_plots}. The most important key findings reveal that in both settings, the waiting cost ratio exhibits the most dramatic performance impact, while the threshold proportion $\theta/k$ reveals different behavior for each case.

\section{Capacity Allocation}\label{sec:capacity}
\subsection{Capacity Allocation Problem}

While the previous sections focused on optimizing the alternative care threshold $\theta$ for a fixed capacity allocation $(c_u, c_n)$, this section addresses the complementary problem of optimizing the capacity allocation itself. Given a total ED capacity $c = c_u + c_n$, we seek to determine the optimal allocation of urgent beds $c_u^*$ and non-urgent beds $c_n^*$ that maximizes the system's net benefit $Z(c_u, c_n, \theta^*)$, where for each candidate split the threshold is re-optimized: $\theta^*(c_u, c_n) = \argmax_{\theta} Z(c_u, c_n, \theta)$. 
The optimal split solves
\begin{equation}\label{eq:capacity_optimization}
(c_u^*, c_n^*) = \arg\max_{c_u+c_n = c, c_u, c_n \in \mathbb{Z}^{+}} Z(c_u, c_n, \theta^*(c_u,c_n)),
\end{equation}
the threshold re-optimized for each split.
Because stability ($\rho_u < 1$) leaves only a few feasible integer splits, we employ direct optimization through complete enumeration to obtain under the nested policy an optimal $(c_u^*, c_n^*) = (6,3)$ for the rural ED and $(c_u^*, c_n^*) = (29,5)$ for the urban ED. The algorithm is provided in Appendix~\ref{app:algorithms}.

\subsection{Nested vs. Fixed-partitioned Bed Configuration}
To further evaluate the impact of capacity allocation strategies, we compared two bed configuration approaches. The comparison examines nested bed allocation (where urgent patients can overflow to non-urgent beds with preemptive priority) versus fixed bed allocation (where urgent and non-urgent patients are strictly assigned to their designated bed types).\\

The fixed-partitioned approach models the emergency department as two independent queues that interact only through the alternative care decision mechanism based on total system occupancy $(i + j)$.  \begin{enumerate}[label = (\roman*), labelwidth = !, wide, labelindent = 0pt]
\item The urgent queue operates as a standard $M/M/c_u/\infty$ system with arrival rate $\lambda_u = \lambda \cdot p_u$, service rate $\mu_u$, and requiring stability condition $\lambda_u/(c_u \mu_u) < 1$. The steady-state probabilities follow $\pi_u(0) = [\sum_{n=0}^{c_u-1} \rho_u^n/n! + \rho_u^{c_u}/(c_u!(c_u - \rho_u))]^{-1}$, with $\pi_u(i) = \pi_u(0) \cdot \rho_u^i/i!$ for $0 \leq i \leq c_u$ and $\pi_u(i) = \pi_u(0) \cdot \rho_u^i/(c_u! \cdot c_u^{i-c_u})$ for $i > c_u$ \cite{shortle2018}.
\item The non-urgent queue follows a birth-death process with state-dependent arrival rates $\lambda_j = \lambda_n \sum_{i=0}^{\infty} \pi_u(i) \cdot \alpha_{\theta}(i,j)$ where $\alpha_{\theta}(i,j)$ is defined in \ref{eq:al_theta} and death rates $\mu_j = \mu_n \cdot \min(j, c_n)$. 
The steady-state probabilities are $\pi_n(j) = \pi_n(0) \prod_{m=0}^{j-1} \lambda_m/\mu_{m+1}$ with normalization $\pi_n(0) = [1 + \sum_{j=1}^{k-1} \prod_{m=0}^{j-1} \lambda_m/\mu_{m+1}]^{-1}$ \cite{crawford2018}.
\end{enumerate}
The solution algorithm begins by verifying urgent queue stability and solving the $M/M/c_u/\infty$ system to obtain $\{\pi_u(i)\}$, then calculates birth rates $\lambda_j$ and death rates $\mu_j$ for the non-urgent birth-death process. 

\subsubsection{Computational Results}
Stability analysis reveals limited feasible bed allocations in both settings under $\rho_u < 1$. The rural ED (9 total beds) achieves stability in only 3 of 8 allocations (37.5\%), requiring a minimum of 6 urgent beds. The urban ED (34 total beds) shows even tighter constraints with 5 of 33 stable allocations (15.2\%), requiring a minimum of 29 urgent beds. Optimizing over these feasible allocations under the nested policy yields $(c_u^*, c_n^*) = (6,3)$ for the rural ED and $(c_u^*, c_n^*) = (29,5)$ for the urban ED. Due to these limited stable configurations, we introduce an alternative parameter set that yields a broader range of stable allocations for comprehensive comparison between fixed and nested threshold policies.This set of parameters and our computational results are detailed in Appendix \ref{app:capacity_allocation}. The key findings reveal that: (1) the nested configuration consistently outperformed fixed-partitioned across all threshold comparisons with the same bed allocation, and (2) in the bed combination analysis with fixed total capacity, nested allocation dominated most configurations, though fixed-partitioned showed advantages in specific scenarios with very low or very high urgent-capacity. The managerial message is that pooling protects performance against bed-allocation error, while partitioning is competitive only under precise tuning.

\section{Conclusion}\label{sec:conclusion}
We propose and analyze an easily implementable occupancy-triggered redirection policy for managing ED congestion by routing selected low-acuity arrivals to alternative care when the department is crowded. Using a tractable two-class priority-queue framework, we characterize steady-state performance and determine an optimal redirection threshold under a long-run objective that balances revenue and congestion-related costs from the hospital perspective. 
We establish structural properties of the optimal threshold, including monotone comparative statics with respect to all model parameters, that analytically characterize how $\theta^*$ responds to changes in the operational environment without resolving the numerical optimization.
Numerical experiments calibrated to representative rural and urban settings indicate that the optimal threshold is highly context dependent and that enabling alternative care yields consistent improvements relative to a no-redirection baseline across the parameter ranges tested. These improvements were up to 4.84\% in rural settings and 5.90\% in urban settings. Sensitivity analysis highlights that across both configurations, the waiting cost ratio demonstrates the strongest influence on performance, whereas the threshold proportion $\theta/k$
displays distinct patterns in each scenario. Robustness analysis confirms that the optimal threshold and 
objective value are insensitive to non-exponential service time 
distributions. 
Simulation experiments further show that the optimal threshold is stable across preemptive and non-preemptive priority disciplines, and that the referral policy yields substantial gains under both disciplines. 
In addition, our capacity-allocation experiments suggest that allowing urgent patients to overflow into non-urgent capacity under preemptive priority often outperforms strict partitioning. Future work should incorporate richer operational features, including nonstationary arrivals, multiple acuity classes, and alternative-care capacity constraints, and should validate key economic and behavioral inputs using site-specific data. Additionally, the fixed acceptance probability $p_a$ is a modeling simplification;
incorporating state-dependent acceptance behavior where patients respond to real-time system occupancy represents an important direction for future work. Extending the model to explicitly account for boarding dynamics, fast-tracking of low-acuity patients to dedicated treatment areas, and triage-based patient classification would bring the framework closer to full operational fidelity and is a natural direction for future research. Extending the objective to a social planner's perspective, incorporating patient welfare, equity of access, and system-wide externalities, would also allow the approach to be considered from alternative perspectives such as from government policymakers.

\bibliographystyle{plain}
\bibliography{references}

\newpage
\appendix
\section{QBD Structure and Component Definitions}
\label{app:qbd_structure}

We will solve for the steady state distribution of our multi-class queuing system using a QBD process with the generator matrix \(Q\):
\begin{equation}
Q = \begin{bmatrix}
A_0 & B     &        &        &        &        &        &        &        \\
C_1 & A_1   & B      &        &        &        &        &        &        \\
     & C_2  & A_2    & B      &        &        &        &        &        \\
     &      & \ddots & \ddots & \ddots &        &        &        &        \\
     &      &        & C_{h-1} & A_{h-1} & B      &        &        &        \\
     &      &        &        & C      & A      & B      &        &        \\
     &      &        &        &        & C      & A      & B      &        \\
     &      &        &        &        &        & \ddots & \ddots & \ddots
\end{bmatrix},
\label{eq:qbd_matrix}
\end{equation}
where all blocked matrices are $k \times k$  and \(h = \max(k, c)\).\\

\subsubsection{Matrix Component Definitions}
\label{sec:matrix_components}
The block matrices in the QBD structure represent different types of transitions:

\begin{enumerate}[wide, labelindent=0pt, label=(\roman*)]
\item \(B = \lambda_u I_k\) represents urgent arrivals that increase the level from \(i\) to \(i+1\). This matrix is level-independent since urgent arrivals occur at rate \(\lambda_u\) regardless of the current system state.

\item \label{eq:d} \(C_i = \mu_u \min\{i, c\} \cdot I_k\) 
(with \(C = \mu_u \cdot c \cdot I_k\) for \(i \geq c\)) denotes urgent departures that decrease the level from \(i\) to \(i-1\). The service rate depends on the minimum of urgent patients present and available servers.

\item the matrix \(A_i\) is tridiagonal for \(i < h\) and captures transitions within level \(i\), including non-urgent arrivals (when system capacity allows), non-urgent departures (when servers are available), and diagonal elements ensuring row sums equal zero, with:
\begin{align*}
&a_{j,j+1}^{(i)} = \lambda_n \alpha_{\theta}(i,j), \\
&a_{j,j-1}^{(i)} = \mu_n \min\{\max\{0, c - i\}, j, c_n\}, \\ 
&a_{j,j}^{(i)} = -\left(\lambda_u + \lambda_n \alpha_{\theta}(i,j) + \mu_u \min\{i, c\} + \mu_n \min\{\max\{0, c - i\}, j, c_n\}\right),
\end{align*}
where $\alpha_{\theta}(i,j)$ is as defined in \eqref{eq:al_theta}.
For $i \geq h$, $A_i = A$ reflects only urgent dynamics. 
\end{enumerate}

We define the steady-state distribution as
\[
\mathbf{X} = [\mathbf{x}_0, \mathbf{x}_1,\mathbf{x}_2, \ldots],
\]where each \(\mathbf{ x}_i \) represents the steady-state probability of being in level \( i \):
\[
\mathbf{x}_i = \begin{bmatrix} x_{i,0}, x_{i,1}, \cdots, x_{i,k-1} \end{bmatrix}_{1 \times k}.
\]

\subsubsection{Repeating Levels.} To find the steady-state distribution for \( i \geq h \), we can express them as:
\begin{equation}\label{eq:x-geom}
\mathbf{x}_i = \mathbf{x}_h \, \rho_u^{i - h}, \text{where }  \rho_u=\frac{\lambda_u}{c\cdot\mu_u},
\end{equation}
and where \( \mathbf{x}_h \) is the steady-state probability at level \( h \).
\\

\subsubsection{Boundary Levels.} For $i=0,...,h-1$, we use a recursive approach.

For level 0:
\begin{equation}
\mathbf{x}_0 A_0 + \mathbf{x}_1 C_1 = \mathbf{0}.
\label{eq:e}
\end{equation} This equation represents the balance condition at level 0, where the first term captures within-level transitions (non-urgent arrivals and departures when no urgent patients are present), and the second term represents urgent departures from level 1 that bring the system back to level 0.

For levels \( i = 1, \ldots, h-1 \):
\[
\mathbf{x}_{i-1} B_i + \mathbf{x}_i A_i + \mathbf{x}_{i+1} C_{i+1} = \mathbf{0},
\]
where the first term represents arrivals from level \(i-1\), the second term captures within-level transitions at level \(i\), and the third term accounts for departures from level \(i+1\).

For level $h$:
\[
\mathbf{x}_{h-1} B + \mathbf{x}_h A + \mathbf{x}_{h+1} C = \mathbf{0},
\]
\noindent
substituting $\mathbf{x}_{h+1} = \mathbf{x}_h \rho_u$, we get:
\begin{equation}
\mathbf{x}_{h-1} B + \mathbf{x}_h (A + \rho_u C) = \mathbf{0}.
\label{eq:c}
\end{equation}
\noindent
Looking at the above equations, we can solve for each \(\mathbf{x}_i \) in terms of \(\mathbf{ x}_{i+1} \). The steady state distribution is obtained using the methodology as described by \cite{kao1991}. The detailed recursive solution steps are provided in Appendix \ref{app:proofs}. The algorithm for determining steady-state and evaluating queue metrics is provided in Appendix~\ref{app:algorithms}.

\section{Additional Proofs}
\label{app:proofs}

\emph{Backward Recursive Formula:}
\[
\mathbf{x}_h U_{i-1} B + \mathbf{x}_h U_i A_i + \mathbf{x}_h U_{i+1} C_{i+1} = \mathbf{0}.
\]
\noindent
Since \(\mathbf{x}_h\) has strictly positive entries, solving for $U_{i-1}$ we get
\[
U_{i-1} = -(U_i A_i + U_{i+1} C_{i+1})B^{-1} =- \frac{1}{\lambda_u}(U_i A_i + U_{i+1} C_{i+1}),\]
where we have used the fact that \(B^{-1}=\frac{1}{\lambda_u}I_k\).
\noindent
Substituting \(C_{i+1}\):
\begin{equation}
U_{i-1} = -\frac{1}{\lambda_u} \left(U_i A_i + \left(\mu_u \min\{i+1, c\}\right) U_{i+1} \right).
\end{equation}
\noindent
Now, we convert this equation using the backward indexing notation $U_{h-i}$. That is, for levels \(i=2,\dots,h:\)
\[
U_{h-i} = -\frac{1}{\lambda_u} \left(U_{h-i+1} A_{h-i+1} + \mu_u \min\{h-i+2, c\} U_{h-i+2}\right).
\]
\noindent
From the equation for level $h$ in (\ref{eq:c}) we have similarly
\[
\mathbf{x}_{h}( U_{h-1}B + A + \rho_u C) = \mathbf{0}; 
\] therefore,
\[
U_{h-1} = -\frac{1}{\lambda_u} \left( A + \rho_u C \right).
\]
\noindent
Substituting the value of \(\rho_u\) from (\ref{eq:x-geom}) and \(C\) from \ref{sec:matrix_components}.\ref{eq:d} we get
\[
U_{h-1} = -\frac{1}{\lambda_u} \left( A + \frac{\lambda_u}{\mu_u c} \cdot \mu_u c \right) = -\frac{1}{\lambda_u} \left( A + \lambda_u I_k\right) = - \frac{1}{\lambda_u}A-I_k.
\]
Our complete recursive approach can now be summarized below:
\begin{align*}
&U_h = I, \\
&U_{h-1} = -\frac{1}{\lambda_u} A - I, \quad \text{{and}}\\
&U_{h-i} = -\frac{1}{\lambda_u} \left(U_{h-i+1} A_{h-i+1} + \mu_u \min\{h-i+2, c\} U_{h-i+2}\right).
\end{align*}
\noindent
We need to find  \(\mathbf{x}_h\) in order to get the steady state distribution from the backward recursive approach.
From (\ref{eq:e}) and using \(\mathbf{x}_i=\mathbf{x}_{h}U_i\),
\[
\mathbf{x}_h (U_0 A_0 + U_1 C_1) = \mathbf{0}.
\]
\noindent
Furthermore notice
\begin{equation}
\sum_{i=0}^{h-1} \mathbf{x}_i \mathbf{e} + \sum_{i=h}^{\infty} \mathbf{x}_i \mathbf{e} = 1.
\label{eq:f}
\end{equation}
\noindent
For \( i = 0, \ldots, h-1 \):
\begin{equation}
\sum_{i=0}^{h-1} \mathbf{x}_i \mathbf{e} = \sum_{i=0}^{h-1} (\mathbf{x}_h U_i) \mathbf{e} = \mathbf{x}_h \sum_{i=0}^{h-1} U_i \mathbf{e} = \mathbf{x}_h \left( \sum_{i=0}^{h-1} U_i \right) \mathbf{e}.
\label{eq:h}
\end{equation}
\noindent
Now using \eqref{eq:x-geom}
\begin{equation}
\sum_{i=h}^{\infty} \mathbf{x}_i \mathbf{e} 
= \sum_{n=0}^{\infty} \left(\mathbf{x}_h \rho_u^n \right) \mathbf{e}
= (1-\rho_u)^{-1}\mathbf{x}_h\cdot\mathbf{e},
\label{eq:g}
\end{equation}
\noindent
substituting (\ref{eq:h}) and (\ref{eq:g}) to (\ref{eq:f}) we get
\begin{equation}
1 = \mathbf{x}_h \left( \sum_{i=0}^{h-1} U_i \right) \mathbf{e} + \left(1 -\rho_u \right)^{-1} \mathbf{x}_h \cdot\mathbf{e} = \mathbf{x}_h \left[ \sum_{i=0}^{h-1} U_i + \left(1 - \rho_u \right)^{-1} I \right] \mathbf{e}.
\end{equation}
\noindent
Therefore,\[
\mathbf{x}_h \left[ \sum_{i=0}^{h-1} U_i + \left(1 - \rho_u \right)^{-1} \right] \mathbf{e} = 1
.
\]
\section{Algorithms}
\label{app:algorithms}
\vspace{0.5em}
\subsubsection{QBD Algorithm}The following algorithm outlines the steps to determine the steady-state and evaluate queue metrics.
\\
\vspace{0.5em}

\begin{breakablealgorithm}
\caption{QBD Steady-State Solution for Emergency Department Model - Part 1}
\label{alg:qbd_steady_state_part1}
\begin{algorithmic}[1]
\STATE \textbf{Input:} \(\lambda, p_u, \mu_u, \mu_n, c_u, c_n, k, \theta, p_{\text{accept}}\)
\STATE \textbf{Output:} Steady-state probabilities \(\pi(i,j)\) and performance measures

\STATE \textbf{Step 1: Initialize Parameters}
\STATE \(p_n \leftarrow 1 - p_u\), \(\lambda_u \leftarrow p_u \cdot \lambda\), \(c \leftarrow c_u + c_n\)
\STATE \(h \leftarrow \max(k, c)\), \(\rho_u \leftarrow \frac{p_u \lambda}{\mu_u c}\)

\STATE \textbf{Step 2: Verify Stability}
\IF{\(\rho_u \geq 1\)}
    \STATE \textbf{return} "System is unstable"
\ENDIF

\STATE \textbf{Step 3: Define System Functions}
\[
\alpha(i, j) = \begin{cases}
1 & \text{if } i + j < \theta \\
1 - p_{\text{accept}} & \text{if } \theta \leq i + j < k \\
0 & \text{if } i + j \geq k
\end{cases}
\]
\STATE \(s_u(i) = \min\{c, i\}\), \(s_n(i,j) = \min\{\max\{0, c - i\}, j, c_n\}\)

\STATE \textbf{Step 4: Construct QBD Matrices}
\FOR{\(i = 0\) to \(h-1\)}
    \STATE \(B \leftarrow p_u \lambda I_{k}\)
    \STATE \(\mathbf{C}_i \leftarrow \mu_u \min\{i, c\} \cdot I_{k}\)
    \FOR{\(j = 0\) to \(k-1\)}
        \IF{\(j < k-1\)} 
            \STATE \(a_{j,j+1}^{(i)} \leftarrow \lambda_n \alpha_{\theta}(i,j)\) 
        \ENDIF
        \IF{\(j > 0\)} 
            \STATE \(a_{j,j-1}^{(i)} \leftarrow \mu_n s_n(i,j)\) 
        \ENDIF
        \STATE \(a_{j,j}^{(i)} \leftarrow -(\lambda_u + \lambda_n \alpha_{\theta}(i,j) + \mu_u s_u(i) + \mu_n s_n(i,j))\)
    \ENDFOR
\ENDFOR
\STATE \(A \leftarrow A_h\), \(\mathbf{C} \leftarrow \mu_u c \cdot I_{k}\)

\STATE \textbf{Step 5: Backward Recursive Computation}
\STATE \(U_h \leftarrow I_{k}\), \(U_{h-1} \leftarrow -\frac{1}{p_u \lambda} A - I_{k}\)
\FOR{\(i = 2\) to \(h\)}
    \STATE \(U_{h-i} \leftarrow -\frac{1}{p_u \lambda} \left(U_{h-i+1} A_{h-i+1} + \mu_u \min\{h-i+2, c\} U_{h-i+2}\right)\)
\ENDFOR
\end{algorithmic}
\end{breakablealgorithm}

\begin{breakablealgorithm}
\caption{QBD Steady-State Solution for Emergency Department Model - Part 2}
\label{alg:qbd_steady_state_part2}
\begin{algorithmic}[1]
\setcounter{ALC@line}{35}

\STATE \textbf{Step 6: Solve for \(\mathbf{x}_h\)}
\STATE Solve
\[
\mathbf{x}_h \, (U_0 A_0 + U_1 C_1) = \mathbf{0},
\]
\STATE and use
\[
\mathbf{x}_h \left[ \sum_{i=0}^{h-1} U_i + \left(1 - \frac{p_u \lambda}{\mu_u (c_u + c_n)} \right)^{-1} I_k \right] \mathbf{e} = 1
\]as the last row of the above system to determine \(\mathbf{x}_h^*\).

\STATE \textbf{Step 7: Compute Complete Steady-State Distribution}
\FOR{\(i = 0\) to \(h-1\) }
    \STATE \(\mathbf{x}_i \leftarrow \mathbf{x}_h U_i\)
\ENDFOR
\FOR{\(i = h\) to \(\infty\)}
    \STATE \(\mathbf{x}_i \leftarrow \mathbf{x}_h \rho_u^{i-h}\)
\ENDFOR

\STATE \textbf{Step 8: Extract Individual State Probabilities}
\FOR{\(i = 0\) to \(\infty\) }
    \FOR{\(j = 0\) to \(k-1\)}
        \STATE \(\pi(i,j) \leftarrow \mathbf{x}_i[j]\) \COMMENT{\(j\)-th element of vector \(\mathbf{x}_i\)}
    \ENDFOR
\ENDFOR

\STATE \textbf{Step 9: Compute Performance Measures}
\STATE \(E[N] \leftarrow \sum_{i=0}^{\infty} \sum_{j=0}^{k-1} (i + j) \pi(i,j)\)
\STATE \(E[N_n] \leftarrow \sum_{i=0}^{\infty} \sum_{j=0}^{k-1} j \cdot \pi(i,j)\)
\STATE \(\lambda_n^{\mathrm{eff}}\leftarrow \sum_{i=0}^{\infty} \sum_{j=0}^{k-1} \lambda_n \alpha_{\theta}(i,j) \pi(i,j)\)

\STATE \textbf{Step 10: Calculate Waiting Time for Non-Urgent Patients Using Little's Law}
\STATE \({E[S_n]} \leftarrow \frac{E[N_n]}{\lambda_n^{\mathrm{eff}}}\)
\STATE where $\lambda_{n}^{\mathrm{eff}}$ shows the effective rate of non-urgent patients entering the ED.

\STATE \textbf{return} \(\{\pi(i,j), E[N], E[N_n], {E[S_n]}\}\)

\end{algorithmic}
\end{breakablealgorithm}

\begin{breakablealgorithm}
\caption{Direct Capacity Optimization for Emergency Department}
\label{alg:direct_capacity}
\begin{algorithmic}[1]
\REQUIRE Total capacity $c$, system parameters
\ENSURE Optimal capacity allocation $(c_u^*, c_n^*)$

\STATE Initialize best\_objective $= -\infty$, $(c_u^*, c_n^*) = (1, c-1)$

\FOR{$c_u = 1$ \TO $c-1$}
    \STATE Set $c_n = c - c_u$
    \STATE Find optimal threshold: $\theta^* = \arg\max_{\theta} Z(c_u, c_n, \theta)$
    \STATE Solve QBD system with $(c_u, c_n, \theta^*)$
    \STATE Calculate objective: $Z(c_u, c_n, \theta^*)$
    
    \IF{$Z(c_u, c_n, \theta^*) >$ best\_objective}
        \STATE best\_objective $= Z(c_u, c_n, \theta^*)$
        \STATE $(c_u^*, c_n^*) = (c_u, c_n)$
    \ENDIF
\ENDFOR

\STATE \textbf{Output:} Optimal capacity allocation
\end{algorithmic}
\end{breakablealgorithm}
\section{Implementation of Objective Function Components}
\label{app:obj_comp}
\vspace{0.5em}

The objective function components from Section~\ref{sec:obj_components} are implemented using the steady-state distribution $\pi(i,j)$ obtained from the QBD model:

\subsubsection{Average Revenue Implementation}
\sloppy
The average revenue $R(\theta)$ from equation~\eqref{eq:r} is evaluated by computing $R_n^{\mathrm{ED}}(\theta)$, and $R_n^{\mathrm{Tele}}(\theta)$ separately using the formulae in ~\eqref{eq:revenue-comp}. Note that 
$$
E[N_n^s] = \sum_{i=0}^{\infty} \sum_{j=0}^{k-1} s_n(i,j) \cdot \pi(i,j),
$$ 
where $s_n(i,j) = \min\{j, \max\{0, c - i\}, c_n\}$, and 
$$
P(\theta \leq N < k) = \sum_{i=0}^{k-1} \sum_{j=\max(0, ( \theta - i ))}^{k-i-1} \pi(i,j).
$$
\subsubsection{Average Balking Cost Implementation}
The average balking cost $B(\theta)$ is computed using \eqref{eq:balking-cost_lim} where-in the steady-state balking probability 
$$
P(N \geq k) = \sum_{i=0}^{\infty} \sum_{j=\max(0, k-i)}^{k-1} \pi(i,j) = \pi_{f} + \pi_{g},
$$
where 
$$
\pi_{f} = \sum_{i= 0} ^ {h-1}\sum_{j= \max(0,k-i)}^{k-1} \pi(i,j) \text{ and } \pi_{g}= \sum_{i= h}^{\infty}\sum_{j= \max(0,k-i)}^{k-1} \pi(i,j).
$$
In order to compute $\pi_g$ we use the geometric structure \eqref{eq:x-geom} for the tail of the steady state probabilities. Since $\pi(i,j) = x_{h,j} \cdot \rho_u^{i-h}$ for $i \geq h$ and $h = \max(k,c) \geq k$, we have:

\begin{equation}
\pi_g = \sum_{i=h}^{\infty} \sum_{j=0}^{k-1} x_{h,j} \cdot \rho_u^{i-h} = \sum_{j=0}^{k-1} x_{h,j} \sum_{i=h}^{\infty} \rho_u^{i-h} = \left(\sum_{j=0}^{k-1} x_{h,j}\right) \cdot \frac{1}{1-\rho_u}
\label{eq:geometric_balking_sum}
\end{equation}
\noindent
where the geometric series $\sum_{i=h}^{\infty} \rho_u^{i-h} = \frac{1}{1-\rho_u}$ converges since $\rho_u < 1$ for system stability.

\subsubsection{Average Waiting Cost Implementation}
The average waiting cost $W(\theta)$ is calculated using equation~\eqref{eq:Wtheta}, where $E[N_n] = \sum_{i=0}^{\infty} \sum_{j=0}^{k-1} j \cdot \pi(i,j)$.
{
\section{Urban Sensitivity Analysis: $p_a$, $c$, and $\lambda$}
\label{app:sensitivity_urban}

Table~\ref{tab:sensitivity_urban} reports the urban sensitivity results. The key findings mirror the rural case while reflecting the higher system load ($\rho_u = 0.833$) and dominant urgent proportion ($p_u = 0.85$). Specifically: (i) higher $p_a$ raises the objective and $\theta^*$ monotonically {(except at $p_a = 0.3$, where $\theta^* = 28$ rather than $27$ due to an extremely flat objective surface where the objective values at $\theta = 27$ and $\theta = 28$ differ by less than $0.01$)}, but the gain is more muted than in the rural setting
since non-urgent diversion has limited impact when urgents dominate;
(ii) increasing $c$ raises both $\theta^*$ and the objective, with
gains diminishing as capacity grows large; and (iii) increasing
$\lambda$ lowers $\theta^*$ and deteriorates the objective, turning
negative at $\lambda=5.5$ ($\rho_u=0.917$). All three results confirm
Remark~\ref{rem:op-cs}(a)--(c).

\setlength{\tabcolsep}{26pt}
{\tiny
\begin{longtable}{|l|c|c|r|}
\caption{Separate sensitivity analyses of $p_a$, $c$, and $\lambda$
--- Urban setting. Each block varies one parameter while holding all
others fixed. All reported values are obtained by optimizing the
non-urgent (threshold-dependent) component of the objective.}
\label{tab:sensitivity_urban}\\
\hline
\textbf{Parameter} & \textbf{Value} & \textbf{$\theta^*$} &
\textbf{Objective} \\
\hline
\endfirsthead
\hline
\textbf{Parameter} & \textbf{Value} & \textbf{$\theta^*$} &
\textbf{Objective} \\
\hline
\endhead
\hline
\endfoot
\hline
\endlastfoot

\multirow{9}{*}{\parbox{1.2cm}{\centering Vary $p_a$\\}}
 & 0.10 & 26 & $174.73$ \\
 & 0.20 & 27 & $185.25$ \\
 & 0.30 & {28} & $195.90$ \\
 & 0.40 & 27 & $206.63$ \\
 & 0.50 & 27 & $217.21$ \\
 & 0.60 & 27 & $227.79$ \\
 & 0.70 & 28 & $238.28$ \\
 & 0.80 & 28 & $248.73$ \\
 & 0.90 & 28 & $259.07$ \\
\hline

\multirow{16}{*}{\parbox{1.2cm}{\centering Vary $c$\\}}
 & 30 & 14 & $-212.55$ \\
 & 31 & 19 &  $-32.50$ \\
 & 32 & 22 &   $86.56$ \\
 & 33 & 25 &  $165.78$ \\
 & 34 & 27 &  $219.36$ \\
 & 35 & 29 &  $256.24$ \\
 & 36 & 31 &  $281.94$ \\
 & 37 & 32 &  $300.04$ \\
 & 38 & 33 &  $312.61$ \\
\hline

\multirow{5}{*}{\parbox{1.2cm}{\centering Vary $\lambda$\\}}
 & 3.00 & 38 &   $230.44$ \\
 & 4.00 & 31 &  $297.27$ \\
 & 4.50 & 31 &  $299.42$ \\
 & 5.00 & 27 &  $219.36$ \\
 & 5.50 & 21 &  $-49.31$ \\
\end{longtable}
}}
\section{Urban Robustness Analysis: Service Time Distributions}
\label{app:robust_urban}

Table~\ref{tab:robust_urban} reports the full robustness results for
the urban setting ($\theta^*_{\text{sim}}=30$, $\ell=1942$). Both
Erlang-2 and Lognormal distributions yield overlapping CIs with the
exponential baseline, with objective differences below $1.5\%$.
Revenue is stable across all three distributions
($\approx394$--$397$), and waiting cost varies only between $110$
and $112$, confirming that in a high-load system the dominant urgent
stream constrains behavior more than service time variability.

{\tiny
\begin{longtable}{|l|c|r|r|r|r|r|r|r|r|}
\caption{Robustness results --- Urban setting
($\theta^*=30$, $\ell=1942$, $T_{\text{sim}}=5000$, $m=30$)
All reported values are for the non-urgent (threshold-dependent)
component of the objective. $\hat{p}_b(\theta^*)$ and
$\hat{p}_{x}(\theta^*)$ are empirical estimates of the balking
probability $p_b(\theta^*)$ and the alternative-care acceptance
probability $p_x(\theta^*)$ defined in
Section~\ref{sec:obj_components}.}
\label{tab:robust_urban}\\
\hline
\textbf{Distribution} & \textbf{$c^2$} & \textbf{Objective} &
\textbf{95\% CI} & \textbf{Revenue} & \textbf{Balk\$} &
\textbf{Wait\$} & \textbf{$\hat{p}_b(\theta^*)$} &
\textbf{$\hat{p}_{x}(\theta^*)$} & \textbf{$\widehat{E}[S_n]$} \\
\hline
\endfirsthead
\hline
\textbf{Distribution} & \textbf{$c^2$} & \textbf{Objective} &
\textbf{95\% CI} & \textbf{Revenue} & \textbf{Balk\$} &
\textbf{Wait\$} & \textbf{$\hat{p}_b(\theta^*)$} &
\textbf{$\hat{p}_{x}(\theta^*)$} & \textbf{$\widehat{E}[S_n]$} \\
\hline
\endhead
\hline
\endfoot
\hline
\endlastfoot
Exponential & 1.00 & 230.18 & [222.78,\;237.57] &
  396.91 & 56.25 & 110.48 & 0.1364 & 0.223 & 4.33 \\
\hline
Erlang-2    & 0.50 & 228.60 & [218.05,\;239.14] &
  393.52 & 54.25 & 110.67 & 0.1325 & 0.237 & 4.45 \\
\hline
Lognormal   & 1.50 & 226.87 & [215.05,\;238.68] &
  395.94 & 56.97 & 112.10 & 0.1379 & 0.230 & 4.45 \\
\end{longtable}
}

\section{Simulation Robustness: Warm-up Methodology and Plots}
\label{app:warmup}
 
\subsection{Warm-Up Period Determination}
 
\begin{figure}[H]
\centering
\includegraphics[width=0.90\textwidth]{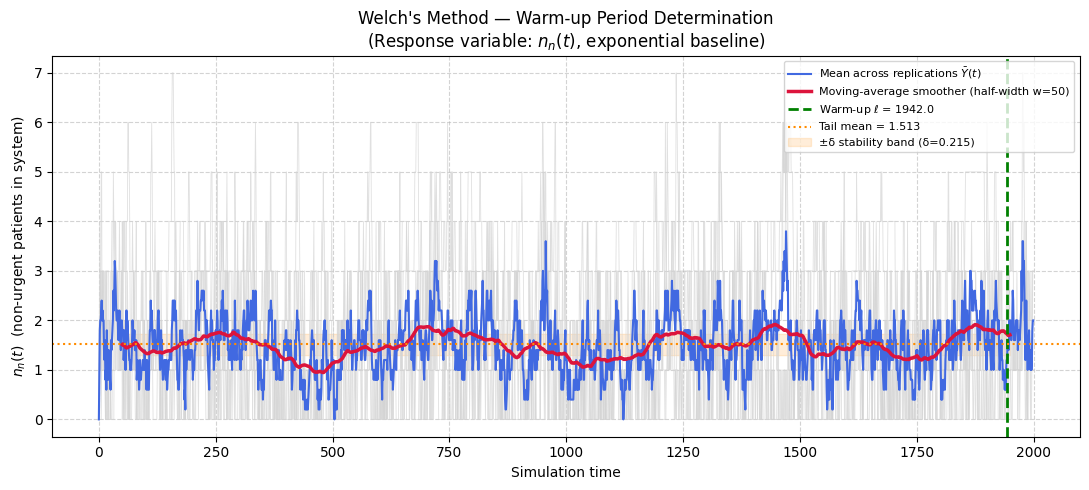}
\caption{Welch warm-up plot --- Urban setting
  ($\ell=1942$, tail mean $=1.513$, $\delta=0.215$,
  $M=5$, $T_{\text{pilot}}=2000$)}
\label{fig:welch_urban}
\end{figure}

We use $N_n(t)$ as the response variable since it is the most sensitive
indicator of congestion relevant to our objective. We run $M=5$ pilot
replications of length $T_{\text{pilot}}=2000$, each starting empty, and record $N_n(t)$ at every integer time point, where $m=1,\ldots,M$ indexes replications. Averaging across
replications gives:
\begin{equation}
\bar{Y}(t) = \frac{1}{M}\sum_{m=1}^{M} N_n^{(m)}(t),
\end{equation}
where $\bar{Y}(t)$ denotes the replication-averaged non-urgent census at time $t$. A centred moving-average smoother of half-width $w=50$ is applied:
\begin{equation}
\bar{Y}^{(w)}(t) = \frac{1}{101}\sum_{s=-50}^{50}\bar{Y}(t+s),
\end{equation}
where $\bar{Y}^{(w)}(t)$ denotes the smoothed average. From the last 25\% of the smoothed series we estimate the steady-state
level $\bar{Y}^*_{\text{tail}}$ and define the stability bandwidth
\begin{equation}
\delta = \max\!\bigl(0.05\times|\bar{Y}^*_{\text{tail}}|,\;
\sigma_{\text{tail}},\;0.01\bigr),
\end{equation}
where $\sigma_{\text{tail}}$ is the standard deviation of the smoothed
tail. The warm-up period length $\ell$ is then defined as the
earliest time at which the smoothed series remains within $\delta$ of
$\bar{Y}^*_{\text{tail}}$ for at least 80\% of subsequent
observations:
\begin{equation}
\ell = \min\!\left\{t:\;
\frac{1}{|\{s\geq t\}|}
\sum_{s\geq t}
\mathbf{1}\!\left[|\bar{Y}^{(w)}(s)-\bar{Y}^*_{\text{tail}}|
\leq\delta\right]\geq 0.80\right\}.
\end{equation}
Although Welch's original method relies on visual inspection
\cite{welch1983}, we automate this using the 80\% rule. Graphical
output is produced for verification (Figures~\ref{fig:welch_rural}
and~\ref{fig:welch_urban}).
 
\subsection{Simulation Objective Estimator}
Let $T_c = T_{\text{sim}} - \ell$ denote the post-warmup collection 
period. For each replication $m$, let $D_n^{(m)}$, $X^{\mathrm{Alt}(m)}$, 
and $B_n^{(m)}$ denote the number of non-urgent patients served in the 
ED, the number of accepted alternative-care referrals, and the number 
of balking patients, respectively, observed during $[\ell, T_{\text{sim}}]$,
corresponding to the cumulative quantities $D_n(t)$, $X^{\mathrm{Alt}}(t)$,
and $B_n(t)$ defined in Section~\ref{sec:descr-prob}.
For replication $m$, the post-warmup objective is
\begin{equation}
\widehat{Z}^{(m)}(\theta)=
  m_n^{\mathrm{ED}}\cdot\frac{D_n^{(m)}}{T_c}
+ m^{\mathrm{Alt}}\cdot\frac{X^{\mathrm{Alt}(m)}}{T_c}
- c^{b}\cdot\frac{B_n^{(m)}}{T_c}
- c_n^{w}\cdot\widehat{E}^{(m)}[N_n],
\end{equation}
where $\widehat{E}^{(m)}[N_n]$ is computed exactly via the rectangle
rule since $N_n(t)$ is a step function constant between events, with 
$t_e$ and $t_{\text{last}}$ denoting the current and previous event 
times, respectively:
\begin{equation}
\widehat{E}^{(m)}[N_n]
= \frac{1}{T_c}\sum_{\text{events}} N_n\times(t_e - t_{\text{last}}).
\end{equation}
 
\subsection{Robustness Criterion}
 
Let $\bar{Z}_d$ and $s_d$ denote the sample mean and standard
deviation of the objective across $n=30$ replications for distribution $d$, and let $\mathrm{lo}_d$ and $\mathrm{hi}_d$ denote the lower and upper bounds of its 95\% confidence interval, with
$\mathrm{lo}_{\mathrm{Exp}}$ and $\mathrm{hi}_{\mathrm{Exp}}$ denoting the corresponding bounds for the exponential baseline. The 95\% CI for distribution $d$ is given by
\begin{equation}
\bar{Z}_d \;\pm\; t_{0.025,29}\cdot\frac{s_d}{\sqrt{30}}, 
\qquad t_{0.025,29}=2.045.
\end{equation}
Robustness requires both (i) CI overlap with the exponential baseline,
\begin{equation}
\max(\mathrm{lo}_{\mathrm{Exp}},\mathrm{lo}_d)\;\leq\;
\min(\mathrm{hi}_{\mathrm{Exp}},\mathrm{hi}_d),
\end{equation}
and (ii) failure of Welch's two-sample $t$-test to reject
$H_0\colon\mu_d=\mu_{\mathrm{Exp}}$ at the significance level $\alpha=0.05$, assuming unequal variances.

\subsection{Warm-Up Plots} The plots are shown as below:
 
\begin{figure}[H]
\centering
\includegraphics[width=0.90\textwidth]{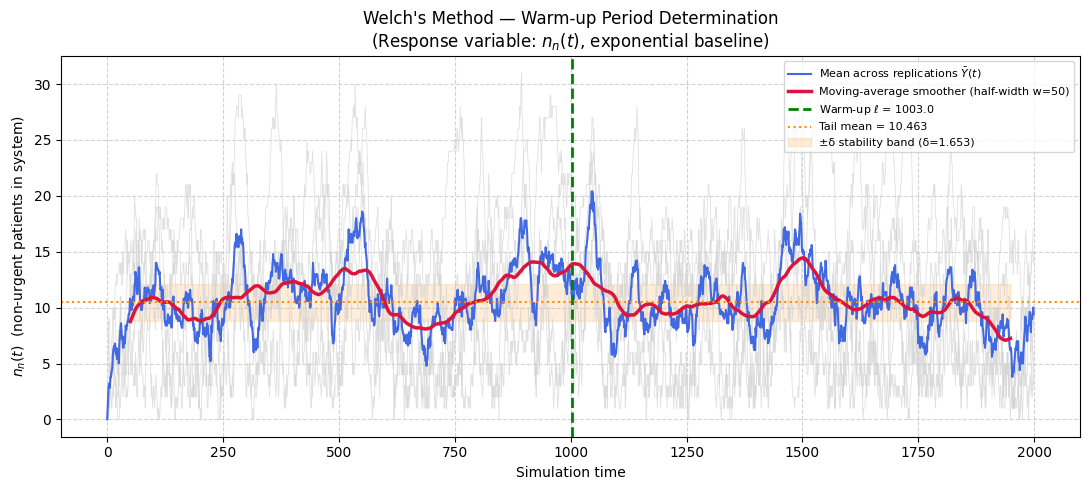}
\caption{Welch warm-up plot --- Rural setting
  ($\ell=1003$, tail mean $=10.463$, $\delta=1.653$,
  $M=5$, $T_{\text{pilot}}=2000$)}
\label{fig:welch_rural}
\end{figure}
 
\section{Urban Simulation Analysis: Priority Discipline and Referral Policy}
\label{app:urban_sim}

Table~\ref{tab:urban_all_sim} report
the full simulation results for the urban setting. With referral, the
two disciplines yield $\theta^*=31$ (non-preemptive) and $\theta^*=29$
(preemptive). Non-preemptive outperforms preemptive by $26.30$ units
($+11.65\%$), a larger gap than in the rural case because the high
system load ($\rho_u=0.833$) makes preemption events more frequent,
substantially increasing non-urgent waiting time
($\widehat{E}[S_n]=4.43$ vs.\ $3.51$). Without referral, the urban
system remains profitable under both disciplines ($215.53$ and
$172.75$), since only 15\% of arrivals are non-urgent. The gain from
referral is $+36.68$ units ($+17\%$) for non-preemptive and $+53.16$
units ($+31\%$) for preemptive.

{
\begin{table}[h]
\tiny
\caption{Urban setting --- all four configurations
  ($T_{\text{sim}}=5000$, $m=500$, $k=39$). All reported values are for the non-urgent (threshold-dependent) component of the objective. $\hat{p}_b(\theta^*)$ and $\hat{p}_x(\theta^*)$ are empirical estimates of the balking probability $p_b(\theta^*)$ and the alternative-care acceptance probability $p_x(\theta^*)$ defined in Section~\ref{sec:obj_components}. $\widehat{E}[N_n]$ and $\widehat{E}[S_n]$ denote the estimated mean non-urgent census and mean non-urgent sojourn time, respectively. N/A applies to no-referral configurations, where neither $\theta^*$ nor $p_x(\theta^*)$ is defined.}
\label{tab:urban_all_sim}
\centering
\begin{tabular}{|l|l|r|r|r|r|r|r|r|r|}
\hline
\textbf{Priority} & \textbf{Referral} & \textbf{$\theta^*$} &
\textbf{$\ell$} & \textbf{Objective} & \textbf{95\% CI} &
\textbf{Wait\$} & \textbf{Balk\$} &
\textbf{$\widehat{E}[N_n]$} & \textbf{Balk. Prob.} ($\hat{p}_b(\theta^*)$) \\
\hline
Non-preemptive & Yes & 31  & 1943 & 252.21 &
  [250.01,\;254.41] & 93.07  & 56.44 & 1.75 & 0.1366 \\
\hline
Preemptive     & Yes & 29  & 1942 & 225.91 &
  [223.45,\;228.36] & 106.17 & 57.65 & 2.00 & 0.1394 \\
\hline
Non-preemptive & No  & N/A & 1869 & 215.53 &
  [213.18,\;217.88] & 125.35 & 74.41 & 2.36 & 0.1801 \\
\hline
Preemptive     & No  & N/A & 1835 & 172.75 &
  [169.97,\;175.54] & 160.88 & 77.95 & 3.02 & 0.1884 \\
\hline
\end{tabular}

\vspace{0.4cm}

\begin{tabular}{|l|l|r|r|r|r|r|r|}
\hline
\textbf{Priority} & \textbf{Referral} & \textbf{Revenue} &
\textbf{ED Rev} & \textbf{Alt Rev} &
\textbf{$\widehat{E}[S_n]$} & \textbf{$\widehat{{\lambda}^{\mathrm{eff}}}$} &
\textbf{$\hat{P}_{{\mathrm{Alt}}}$} \\
\hline
Non-preemptive & Yes & 401.72 & 336.85 & 64.87 & 3.51 & 0.4987 & 0.1984 \\
\hline
Preemptive     & Yes & 389.73 & 304.70 & 85.03 & 4.43 & 0.4511 & 0.2597 \\
\hline
Non-preemptive & No  & 415.28 & 415.28 & N/A   & 3.84 & 0.6148 & N/A    \\
\hline
Preemptive     & No  & 411.59 & 411.59 & N/A   & 4.97 & 0.6093 & N/A    \\
\hline
\end{tabular}
\end{table}
}

\section{Urban Setting Tornado Diagram Analysis}
\label{app:urban setting analysis}
\subsubsection{Urban Alternative Care-enabled ED: Base Case.}
We use parameter values and key operational ratios from Table~\ref{tab:ed_parameters} (Urban ED column) with the following key operational ratios: bed allocation ratio $(c_u/c = 0.4)$, service rate ratio $(\mu_u/\mu_n = 0.15/0.32)$, revenue ratio $(r_u^{\mathrm{ED}}/r_n^{\mathrm{ED}} = 2,221/675.50)$, waiting cost ratio $(c_u^w/c_n^w = 5,531.61/53.21)$, alternative care revenue ratio $(r^{\mathrm{Alt}}/r_n^{\mathrm{ED}} = 436/675.50)$, and balking cost ratio $(c^{b}/r_n^{\mathrm{ED}} = 550.96/675.50)$.

\begin{figure}[H]
\centering
\includegraphics[width=0.6\textwidth]{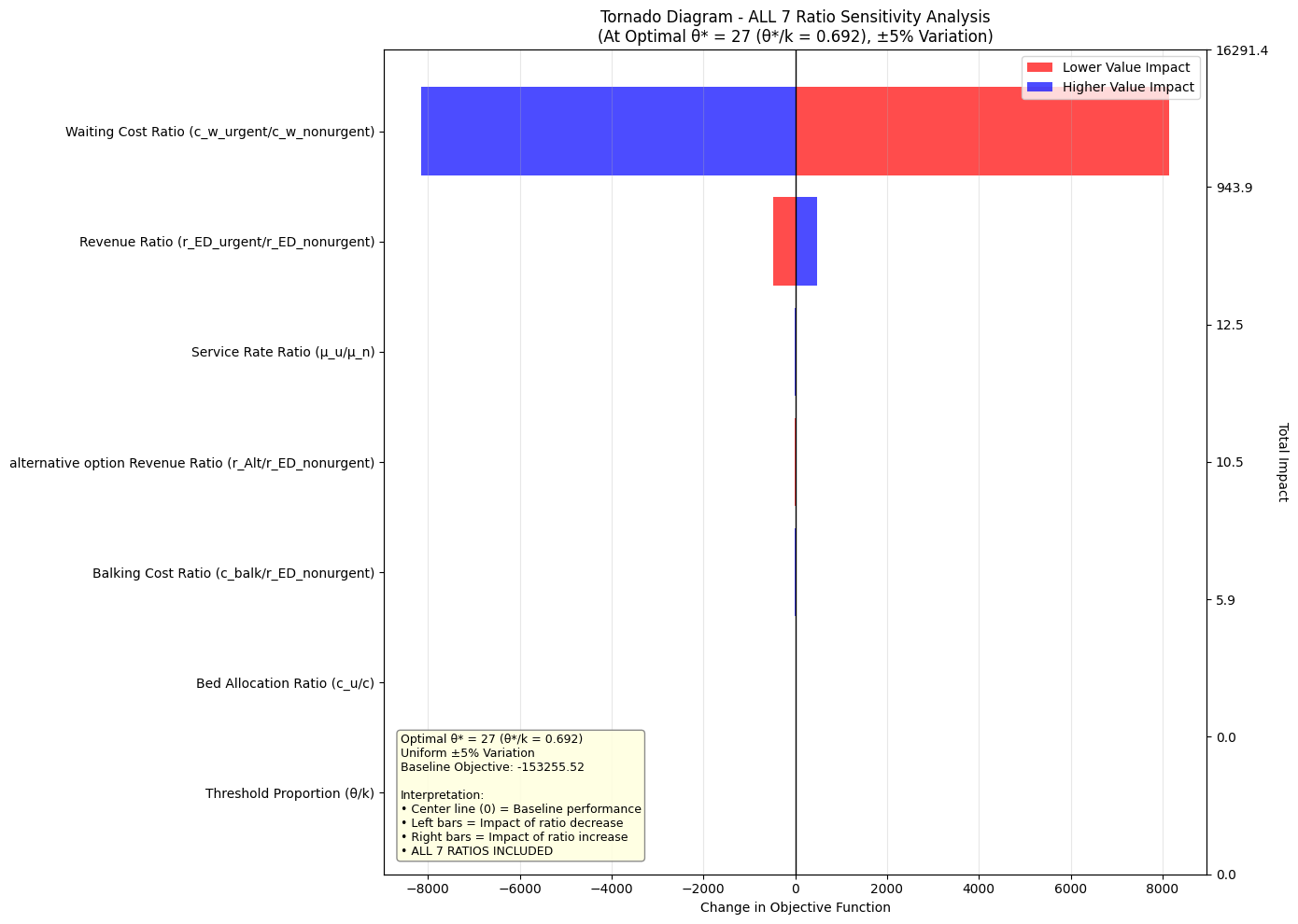}
\caption{Tornado Diagram for Urban Alternative Care-enabled ED- (±5\% Variation)}
\label{fig:tornado_baseline_optimal-2}
\end{figure}

{\tiny
\begin{longtable}{|c|l|c|c|}
\caption{Sensitivity Analysis Summary - Urban Alternative Care-enabled ED}
\label{tab:tornado_sensitivity_analysis-3}\\
\hline
\textbf{Rank} & \textbf{Ratio} & \textbf{Relative Impact (\%)} & \textbf{Base Value} \\
\hline
\endfirsthead

\hline
\textbf{Rank} & \textbf{Ratio} & \textbf{Relative Impact (\%)} & \textbf{Base Value} \\
\hline
\endhead

\hline
\endfoot

\hline
\endlastfoot

1 & Waiting Cost Ratio ($c_u^w$/$c_n^w$) & 10.63 & 103.958 \\
\hline
2 & Revenue Ratio ($r_u^{\mathrm{ED}}$/$r_n^{\mathrm{ED}}$) & 0.62 & 3.288 \\
\hline
3 & Service Rate Ratio ($\mu_u$/$\mu_n$) & 0.01 & 0.469 \\
\hline
4 & Alternative Option Revenue Ratio ($r^{\mathrm{Alt}}$/$r_n^{\mathrm{ED}}$) & 0.01 & 0.645 \\
\hline
5 & Balking Cost Ratio ($c^{b}$/$r_n^{\mathrm{ED}}$) & 0.004 & 0.816 \\
\hline
6 & Bed Allocation Ratio ($c_u$/$c$) & 0.00 & 0.400 \\
\hline
7 & Threshold Proportion ($\theta$/$k$) & 0.00 & 0.692 \\
\hline

\end{longtable}
}
\subsubsection{Urban Alternative Care-enabled ED: Case Comparison.}
To understand how system characteristics affect sensitivity patterns, we analyze tornado diagrams across 15 different operational scenarios, ranging from high-demand to capacity-constrained systems.

{\tiny
\begin{longtable}{|c|l|l|r|c|r|}
\caption{Case Comparison Summary for Urban Alternative Care-enabled ED}
\label{tab:case_comparison-3}\\
\hline
\textbf{Rank} & \textbf{Case} & \textbf{Description} & \textbf{Baseline Obj} & \textbf{Top Ratio} & \textbf{Rel. Impact (\%)} \\
\hline
\endfirsthead
\hline
\textbf{Rank} & \textbf{Case} & \textbf{Description} & \textbf{Baseline Obj} & \textbf{Top Ratio} & \textbf{Rel. Impact (\%)} \\
\hline
\endhead
\hline
\endfoot
\hline
\endlastfoot
1 & Low Urgent Proportion & $p_u -20\%$ & 295,179.19 & $\mu_u/\mu_n$ & 107.89 \\
\hline
2 & High Arrival Rate & $\lambda +20\%$ & -58,761.64 & $c_u^w/c_n^w$ & 10.67 \\
\hline
3 & Low Arrival Rate & $\lambda -20\%$ & -118,019.40 & $c_u^w/c_n^w$ & 10.64 \\
\hline
4 & High Urgent Service Rate & $\mu_u +20\%$ & -121,156.43 & $c_u^w/c_n^w$ & 10.81 \\
\hline
5 & High Urgent Proportion & $p_u +20\%$ & -127,689.06 & $c_u^w/c_n^w$ & 10.77 \\
\hline
6 & High Capacity & $c +20\%$ & -147,308.09 & $c_u^w/c_n^w$ & 10.66 \\
\hline
7 & High Balking Threshold & $k +20\%$ & -152,674.22 & $c_u^w/c_n^w$ & 10.64 \\
\hline
8 & High Acceptance Rate & $p_a +20\%$ & -153,244.53 & $c_u^w/c_n^w$ & 10.63 \\
\hline
9 & Baseline & Original parameters & -153,255.52 & $c_u^w/c_n^w$ & 10.63 \\
\hline
10 & High Theta & $\theta +20\%$ & -153,255.52 & $c_u^w/c_n^w$ & 10.63 \\
\hline
11 & Low Theta & $\theta -20\%$ & -153,255.52 & $c_u^w/c_n^w$ & 10.63 \\
\hline
12 & Low Acceptance Rate & $p_a -20\%$ & -153,266.31 & $c_u^w/c_n^w$ & 10.63 \\
\hline
13 & Low Balking Threshold & $k -20\%$ & -153,455.28 & $c_u^w/c_n^w$ & 10.62 \\
\hline
14 & Low Urgent Service Rate & $\mu_u -20\%$ & -163,139.02 & $c_u^w/c_n^w$ & 10.59 \\
\hline
15 & Low Capacity & $c -20\%$ & -349,628.58 & $c_u^w/c_n^w$ & 10.26 \\
\hline
\end{longtable}
}

\begin{figure}[H]
\centering
\includegraphics[width=0.9\textwidth]{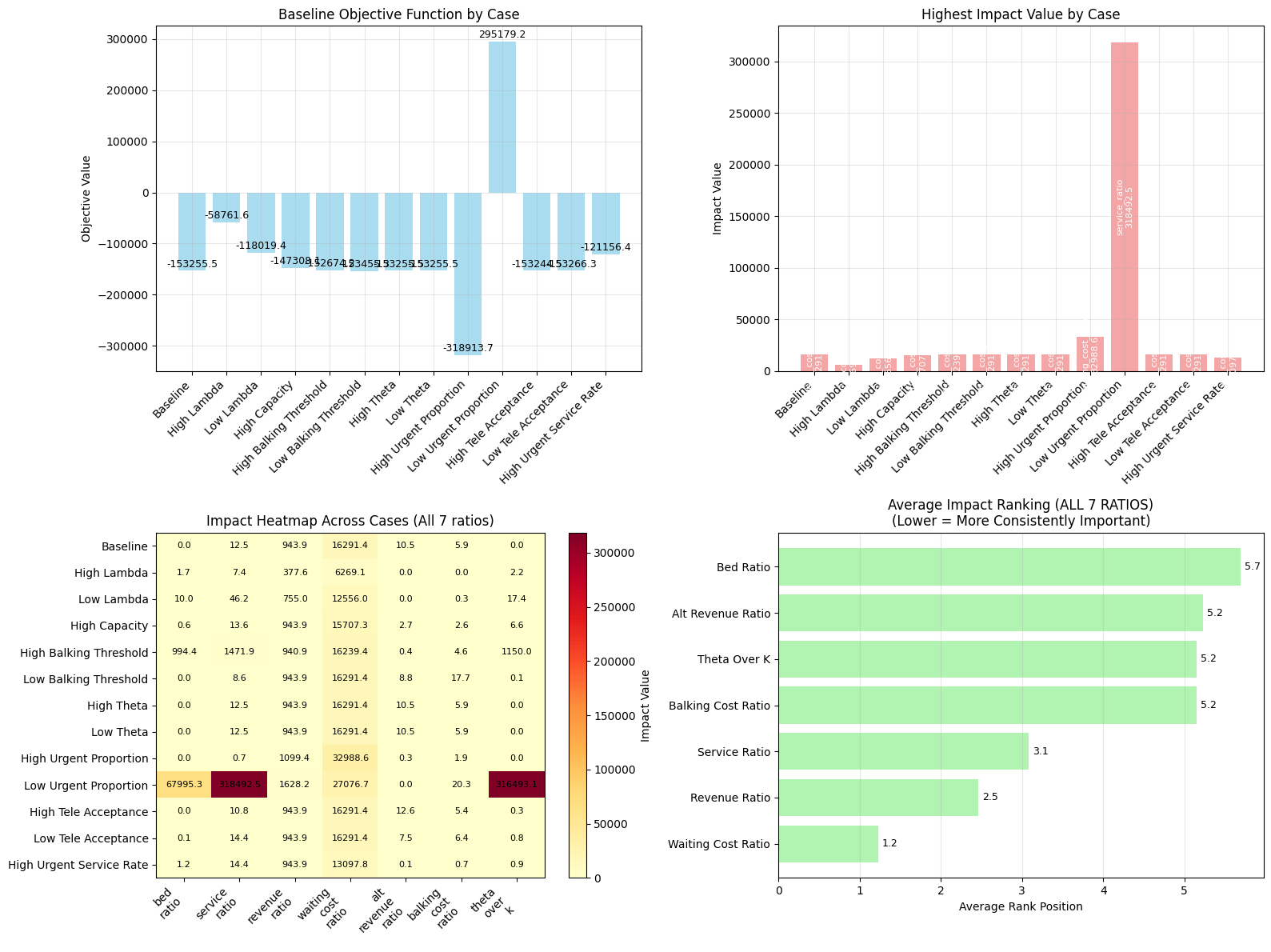}
\caption{Case Comparison Summary of Urban Alternative Care-enabled ED: Baseline Objectives and Impact Rankings Across All Scenarios (±5\% Variation)}
\label{fig:case_comparison_summary_par2}
\end{figure}

The waiting cost ratio dominates 14 of 15 scenarios. However, the service rate ratio demonstrates exceptional sensitivity in the low urgent proportion scenario, far exceeding all other parameters, revealing that service efficiency becomes critical when urgent patient volumes are low and the system becomes revenue-positive. Revenue ratio provides consistent secondary impact (average rank 2.5), while threshold proportion and alternative revenue ratios show moderate importance. Performance varies dramatically across scenarios, with the worst case (low capacity) performing 218\% worse than the best case (low urgent proportion), underscoring the critical importance of capacity management in urban EDs.

\subsubsection{Urban Alternative Care-disabled ED: Base Case.}

\sloppy
Following the alternative care-disabled methodology established for the rural case, we conduct the same analysis for urban settings using base case parameters from Table~\ref{tab:ed_parameters} (Urban ED column). This parallel analysis enables direct comparison of policy sensitivity across operational contexts.


{\tiny
\begin{longtable}{|c|l|c|c|}
\caption{Sensitivity Analysis Summary - Urban Alternative Care-disabled ED}
\label{tab:tornado_sensitivity_analysis_urban_disabled}\\
\hline
\textbf{Rank} & \textbf{Ratio} & \textbf{Relative Impact (\%)} & \textbf{Base Value} \\
\hline
\endfirsthead
\hline
\textbf{Rank} & \textbf{Ratio} & \textbf{Relative Impact (\%)} & \textbf{Base Value} \\
\hline
\endhead
\hline
\endfoot
\hline
\endlastfoot
1 & Waiting Cost Ratio ($c_u^w$/$c_n^w$) & 10.62 & 103.958 \\
\hline
2 & Revenue Ratio ($r_u^{\mathrm{ED}}$/$r_n^{\mathrm{ED}}$) & 0.62 & 3.288 \\
\hline
3 & Service Rate Ratio ($\mu_u$/$\mu_n$) & 0.02 & 0.469 \\
\hline
4 & Balking Cost Ratio ($c^{b}$/$r_n^{\mathrm{ED}}$) & 0.006 & 1.000 \\
\hline
5 & Bed Allocation Ratio ($c_u$/$c$) & 0.002 & 0.400 \\
\hline
\end{longtable}
}

\subsubsection{Urban Alternative Care-disabled ED: Case Comparison.}
To understand sensitivity patterns in the simplified model, we analyze tornado diagrams across 11 different operational scenarios, as with the rural case.

{\tiny
\begin{longtable}{|c|l|l|r|c|r|}
\caption{Case Comparison Summary for Urban Alternative Care-disabled ED}
\label{tab:case_comparison_urban_disabled}\\
\hline
\textbf{Rank} & \textbf{Case} & \textbf{Description} & \textbf{Baseline Obj} & \textbf{Top Ratio} & \textbf{Rel. Impact (\%)} \\
\hline
\endfirsthead
\hline
\textbf{Rank} & \textbf{Case} & \textbf{Description} & \textbf{Baseline Obj} & \textbf{Top Ratio} & \textbf{Rel. Impact (\%)} \\
\hline
\endhead
\hline
\endfoot
\hline
\endlastfoot
1 & Low Urgent Proportion & $p_u -20\%$ & 295,133.20 & $\mu_u/\mu_n$ & 107.81 \\
\hline
2 & High Arrival Rate & $\lambda +20\%$ & -58,761.64 & $c_u^w/c_n^w$ & 10.67 \\
\hline
3 & Low Arrival Rate & $\lambda -20\%$ & -118,020.27 & $c_u^w/c_n^w$ & 10.64 \\
\hline
4 & High Urgent Service Rate & $\mu_u +20\%$ & -121,160.00 & $c_u^w/c_n^w$ & 10.81 \\
\hline
5 & High Urgent Proportion & $p_u +20\%$ & -135,695.39 & $c_u^w/c_n^w$ & 10.63 \\
\hline
6 & High Capacity & $c +20\%$ & -147,327.84 & $c_u^w/c_n^w$ & 10.66 \\
\hline
7 & Baseline & Original parameters & -153,326.45 & $c_u^w/c_n^w$ & 10.62 \\
\hline
8 & Low Balking Threshold & $k -20\%$ & -153,528.20 & $c_u^w/c_n^w$ & 10.61 \\
\hline
9 & High Balking Threshold & $k +20\%$ & -154,898.62 & $c_u^w/c_n^w$ & 10.62 \\
\hline
10 & Low Urgent Service Rate & $\mu_u -20\%$ & -163,239.64 & $c_u^w/c_n^w$ & 10.58 \\
\hline
11 & Low Capacity & $c -20\%$ & -349,884.14 & $c_u^w/c_n^w$ & 10.25 \\
\end{longtable}
}

\begin{figure}[H]
\centering
\includegraphics[width=0.9\textwidth]{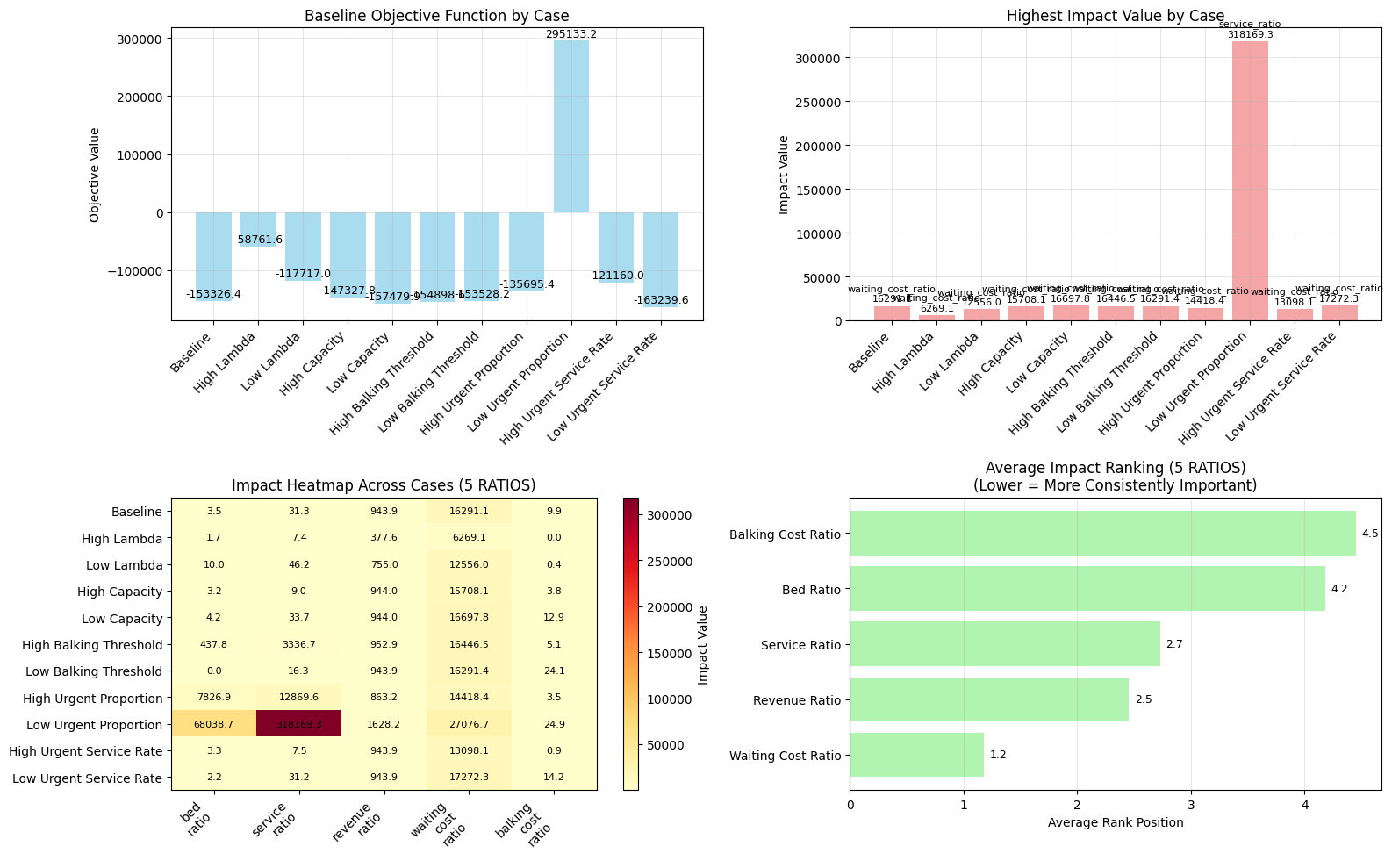}

\caption{Case Comparison Summary of Urban Alternative Care-disabled ED: Baseline Objectives and Impact Rankings Across All Scenarios (±5\% Variation)}
\label{fig:case_comparison_summary_no_tele_par2}
\end{figure}

The waiting cost ratio dominates 10 of 11 scenarios. The service rate ratio demonstrates exceptional sensitivity in the low urgent proportion scenario, the best-performing case where the system achieves positive revenue, far exceeding all other parameters. Revenue ratio provides a consistent secondary impact (average rank 2.5), while bed allocation and balking cost ratios show minimal influence. Performance varies dramatically: the system performs 218\% worse in the worst case (low capacity) compared to the best case (low urgent proportion), underscoring that patient mix and capacity management remain critical drivers even without alternative care pathways.

\subsubsection{Urban ED: Alternative Care Enabled vs. Disabled.}

Alternative care improves or maintains performance across all 11 scenarios. Under baseline conditions, alternative care provides a 0.046\% improvement. The largest benefit occurs under high urgent proportion conditions (5.90\%), while high balking threshold scenarios show 1.44\% improvement. The waiting cost ratio remains dominant in both configurations, while service rate ratio demonstrates exceptional importance in low urgent proportion scenarios, regardless of alternative care availability, indicating fundamental operational leverage points persist across configurations.

{\tiny
\begin{longtable}{|l|c|c|c|c|c|c|}
\caption{Alternative Care Impact: Enabled vs. Disabled with Relative Impact Comparison}
\label{tab:alternative_option_urban_impact}\\
\hline
\textbf{Scenario} & \textbf{Enabled} & \textbf{Enabled} & \textbf{Disabled} & \textbf{Disabled} & \textbf{Benefit} & \textbf{Gain} \\
 & \textbf{(\$/hr)} & \textbf{Rel. Impact (\%)} & \textbf{(\$/hr)} & \textbf{Rel. Impact (\%)} & \textbf{(\$/hr)} & \textbf{(\%)} \\
\hline
\endfirsthead
\hline
\textbf{Scenario} & \textbf{Enabled} & \textbf{Enabled} & \textbf{Disabled} & \textbf{Disabled} & \textbf{Benefit} & \textbf{Gain} \\
 & \textbf{(\$/hr)} & \textbf{Rel. Impact (\%)} & \textbf{(\$/hr)} & \textbf{Rel. Impact (\%)} & \textbf{(\$/hr)} & \textbf{(\%)} \\
\hline
\endhead
\hline
\multicolumn{7}{l}{\textsuperscript{*}Arrival rate capped at $\lambda$ = 5 (baseline value) due to system stability constraints} \\
\hline
\endfoot
\hline
\multicolumn{7}{l}{\textsuperscript{*}Arrival rate capped at $\lambda$ = 5 (baseline value) due to system stability constraints} \\
\hline
\endlastfoot
Low Urgent Proportion & +295,179 & 107.89 & +295,133 & 107.81 & +46 & +0.016 \\
\hline
High Arrival Rate\textsuperscript{*} & -58,762 & 10.67 & -58,762 & 10.67 & +0 & +0.000 \\
\hline
Low Arrival Rate & -118,019 & 10.64 & -118,020 & 10.64 & +1 & +0.001 \\
\hline
High Urgent Service Rate & -121,156 & 10.81 & -121,160 & 10.81 & +4 & +0.003 \\
\hline
High Capacity & -147,308 & 10.66 & -147,328 & 10.66 & +20 & +0.013 \\
\hline
Low Balking Threshold & -153,455 & 10.62 & -153,528 & 10.61 & +73 & +0.048 \\
\hline
Baseline & -153,256 & 10.63 & -153,326 & 10.62 & +71 & +0.046 \\
\hline
High Balking Threshold & -152,674 & 10.64 & -154,899 & 10.62 & +2,224 & +1.44 \\
\hline
Low Urgent Service Rate & -163,139 & 10.59 & -163,240 & 10.58 & +101 & +0.062 \\
\hline
Low Capacity & -349,629 & 10.26 & -349,884 & 10.25 & +256 & +0.073 \\
\hline
High Urgent Proportion & -127,689 & 10.77 & -135,695 & 10.63 & +8,006 & +5.90 \\
\hline
\end{longtable}
}

\section{Proportional Analysis}
\label{app:proportional_plots}
\subsubsection{Methodology:} For each operational ratio in our comprehensive set
$\mathcal{R} = 
\{$\text{bed ratio},  \text{service ratio}, \text{revenue ratio}, \text{waiting cost ratio}, 
\text{alternative care revenue ratio}, \text{balking cost ratio}, \text{and} $\theta/ k \}$,
we systematically vary each ratio across its feasible operating range while allowing the alternative care threshold policy to optimize freely for each configuration. 

The analysis begins by establishing the practical bounds for each ratio based on operational constraints, then systematically exploring how changes in these fundamental operational relationships affect system performance when the threshold policy is allowed to adapt optimally. For each ratio configuration tested, we determine the threshold value that maximizes the objective function, record both the optimal threshold and the resulting performance level, and analyze the patterns that emerge across the ratio's feasible range. This approach reveals how each operational ratio affects not just sensitivity (as shown in tornado analysis) but the actual optimal performance envelope when the system is allowed to adapt its alternative care policy dynamically. 

\subsubsection{Rural Alternative Care-enabled ED: Base Case.}
\begin{figure}[H]
\centering
\subfloat[Waiting Cost Ratio\label{fig:waiting_cost_ratio}]{%
    \includegraphics[width=0.45\textwidth]{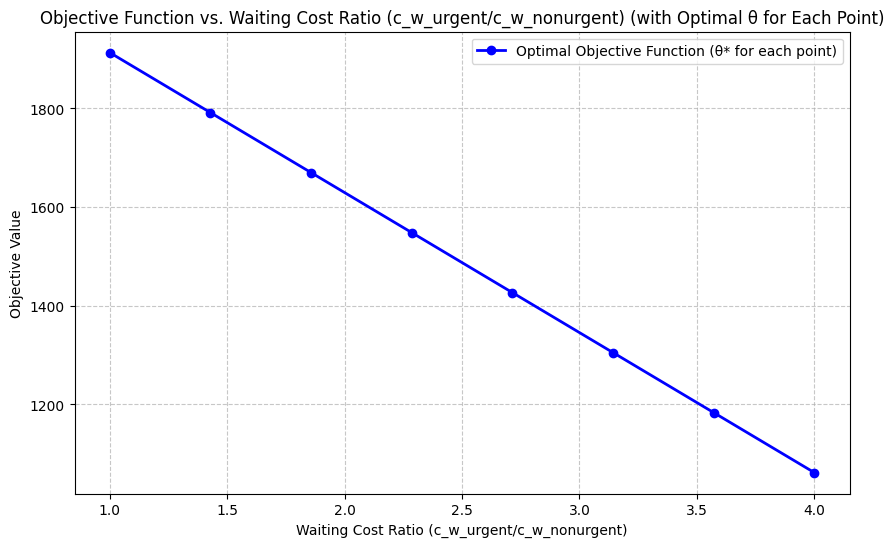}%
}\hfill
\subfloat[Threshold Proportion\label{fig:theta_optimization}]{%
    \includegraphics[width=0.45\textwidth]{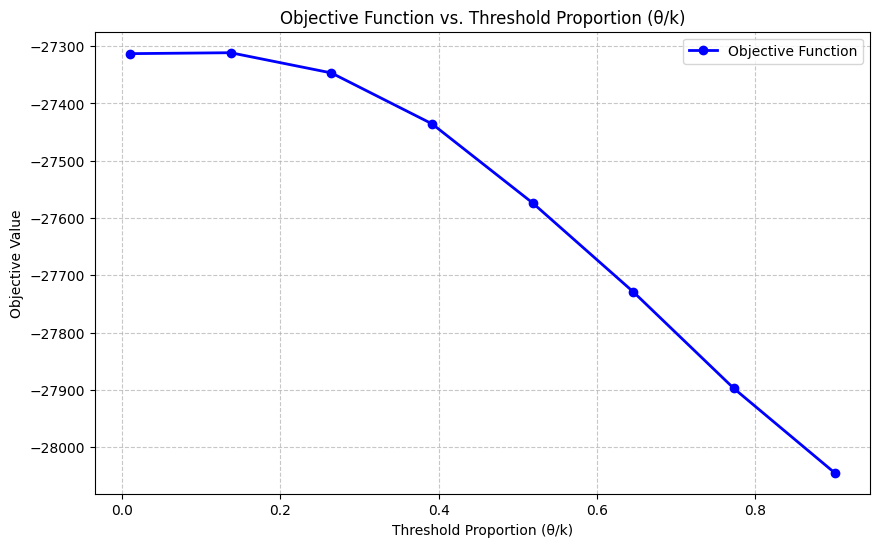}%
}
\caption{Baseline Case Analysis: Critical Sensitivity Ratios}
\label{fig:baseline_analysis_rural}
\end{figure}

\begin{figure}[H]
\centering
\subfloat[Bed Allocation Ratio]{%
  \includegraphics[width=0.32\textwidth]{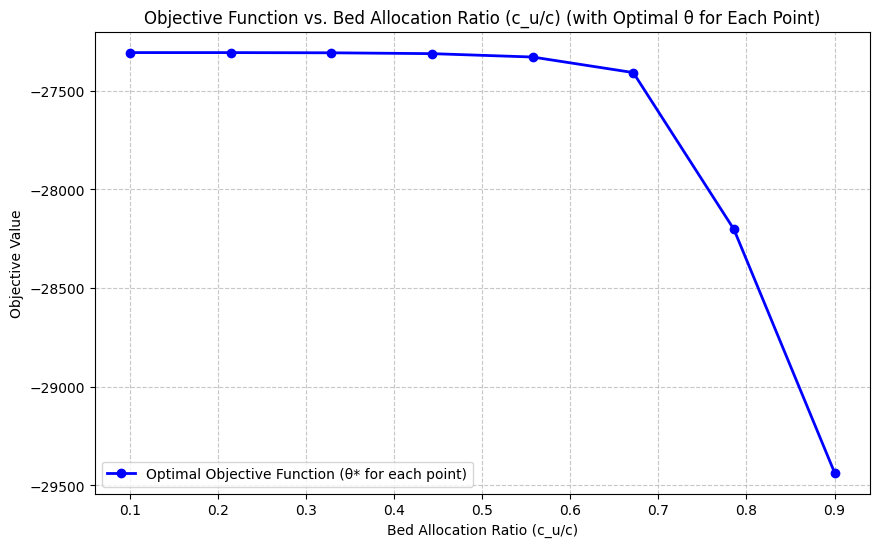}%
}\hfill
\subfloat[Service Rate Ratio]{%
  \includegraphics[width=0.32\textwidth]{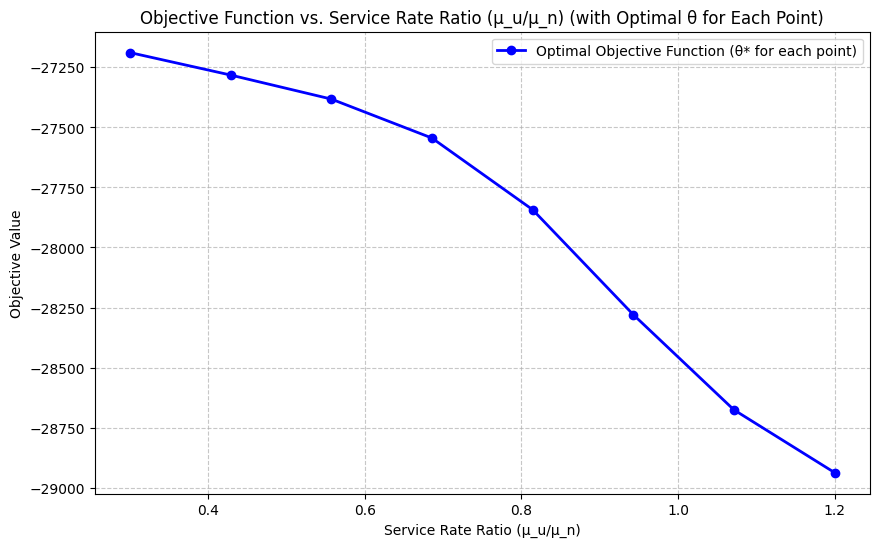}%
}\hfill
\subfloat[Revenue Ratio]{%
  \includegraphics[width=0.32\textwidth]{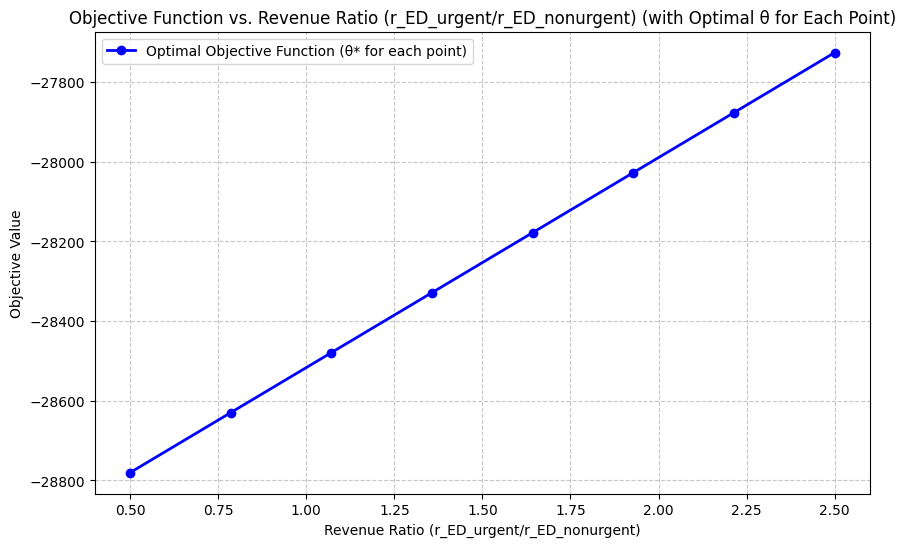}%
}\\[2mm]
\subfloat[Alt Care Revenue Ratio]{%
  \includegraphics[width=0.32\textwidth]{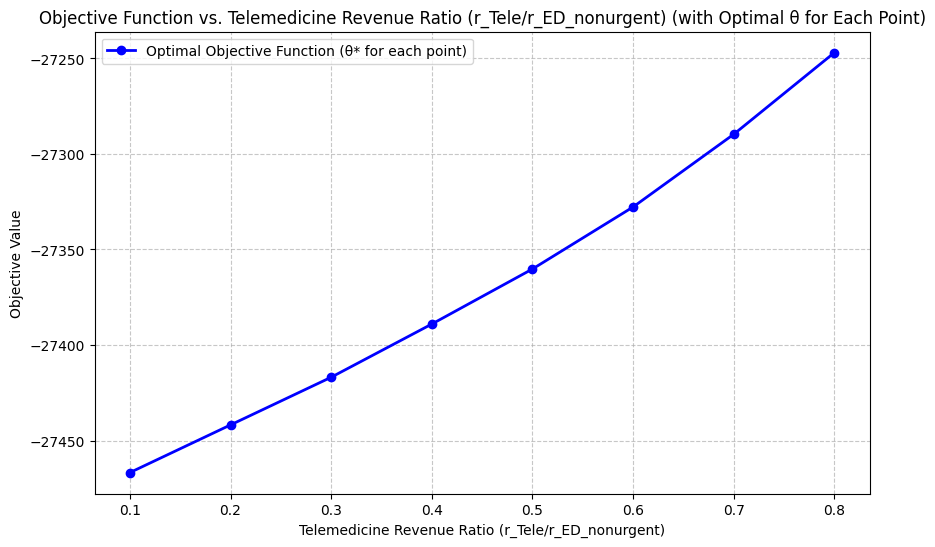}%
}\hfill
\subfloat[Balking Cost Ratio]{%
  \includegraphics[width=0.32\textwidth]{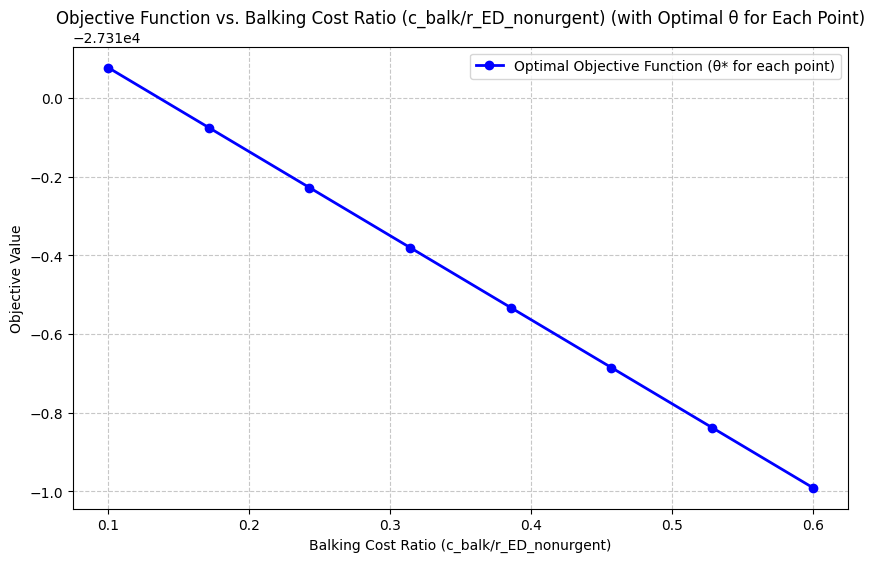}%
}
\caption{Baseline Case Analysis: Rural Alternative Care-enabled ED}
\label{fig:baseline_analysis_rural_others}
\end{figure}

Under standard parameters (optimal $\theta^* = 5$), the system demonstrates balanced optimization with clear sensitivity patterns. Figure~\ref{fig:baseline_analysis_rural} shows the two most critical ratios; additional proportional sensitivity analyses are provided in Figure \ref{fig:baseline_analysis_rural_others}.

\renewcommand\thesubfigure{\Alph{subfigure}}

The waiting cost ratio exhibits the most dramatic performance impact, with the objective function declining steeply and nearly linearly as the ratio increases. This indicates that as urgent patients incur proportionally higher waiting costs relative to non-urgent patients, the system performance deteriorates significantly, creating extreme cost penalties during delays. The threshold proportion shows performance remains relatively stable at low values of $\theta/k$, then declines with increasing steepness at higher proportions. Higher thresholds force patients to wait longer before alternative care referral, accumulating higher waiting costs and increased balking, with performance dropping from approximately $-27{,}300$ to $-28{,}050$ as the proportion increases from 0 to 0.9.

\subsubsection{Urban Alternative Care-enabled ED: Base Case.}
\begin{figure}[H]
\centering
\subfloat[Waiting Cost Ratio]{%
    \includegraphics[width=0.45\textwidth]{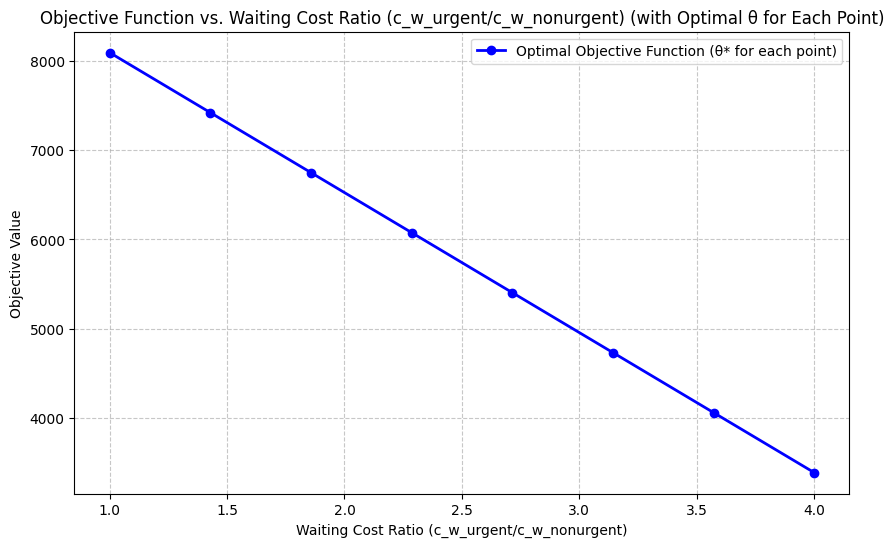}%
}\hfill
\subfloat[Threshold Proportion]{%
    \includegraphics[width=0.45\textwidth]{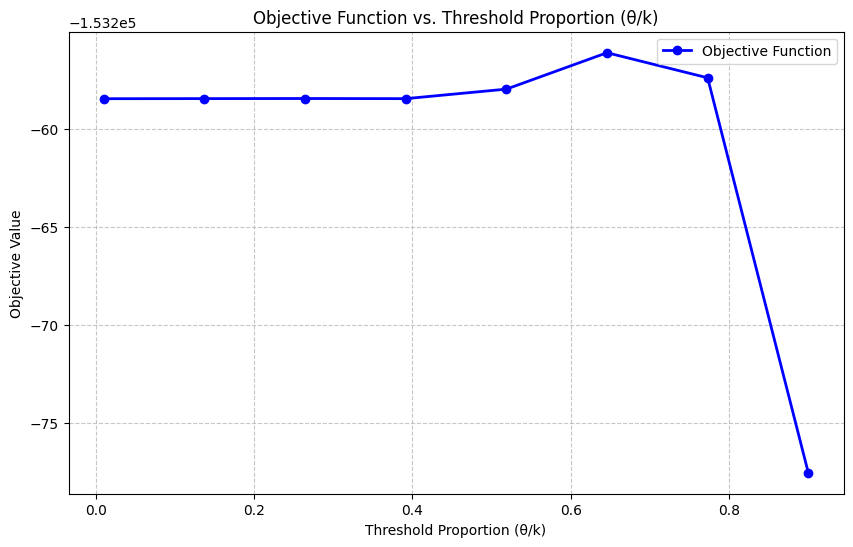}%
}
\caption{Baseline Case Analysis: Critical Sensitivity Ratios}
\label{fig:baseline_analysis_urban}
\end{figure}

\begin{figure}[H]
\centering
\subfloat[Bed Allocation Ratio]{%
  \includegraphics[width=0.32\textwidth]{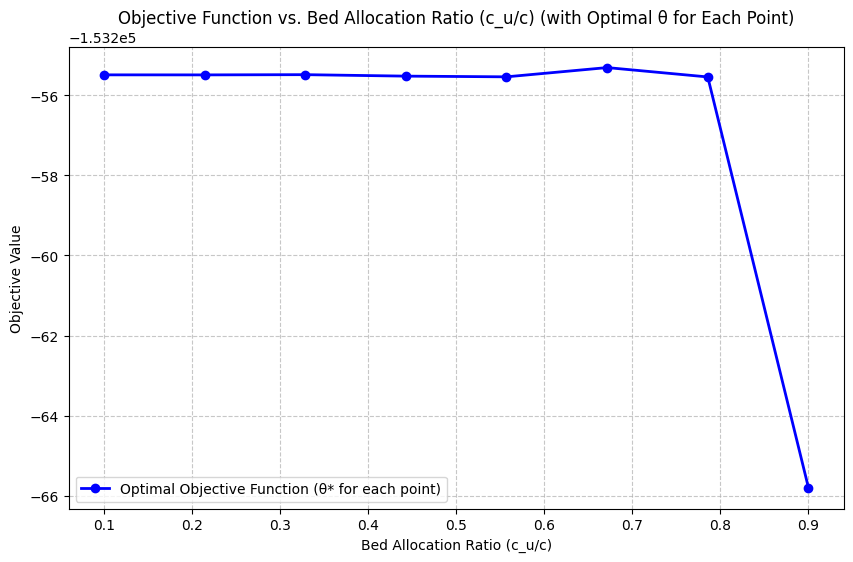}%
}\hfill
\subfloat[Service Rate Ratio]{%
  \includegraphics[width=0.32\textwidth]{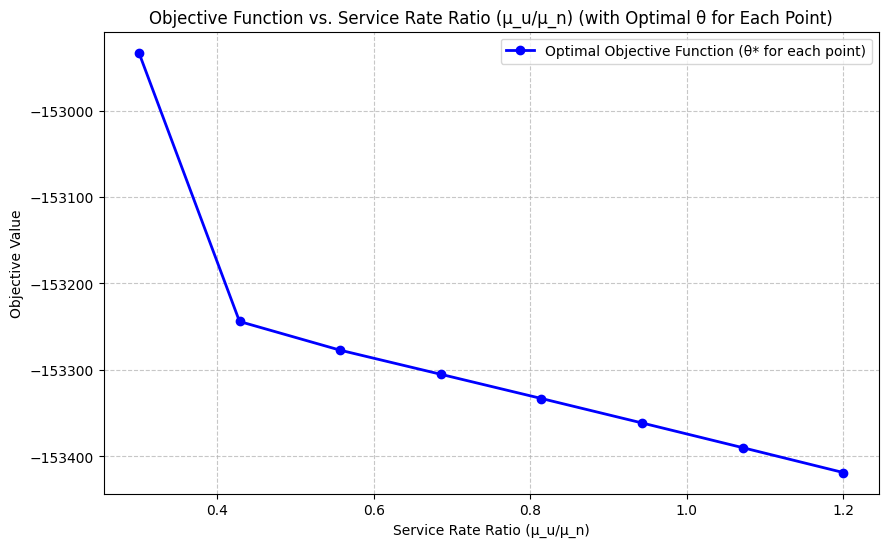}%
}\hfill
\subfloat[Revenue Ratio]{%
  \includegraphics[width=0.32\textwidth]{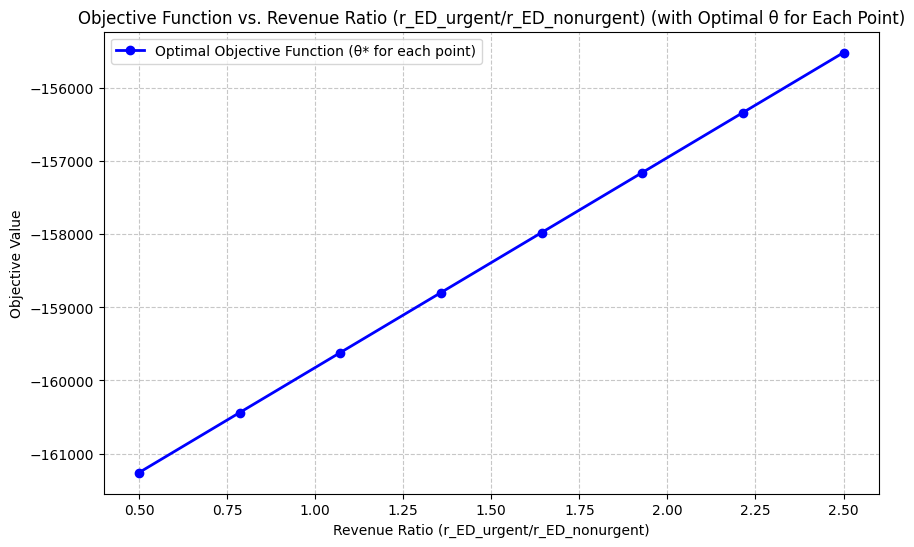}%
}\\[2mm]
\subfloat[Alt Care Revenue Ratio]{%
  \includegraphics[width=0.32\textwidth]{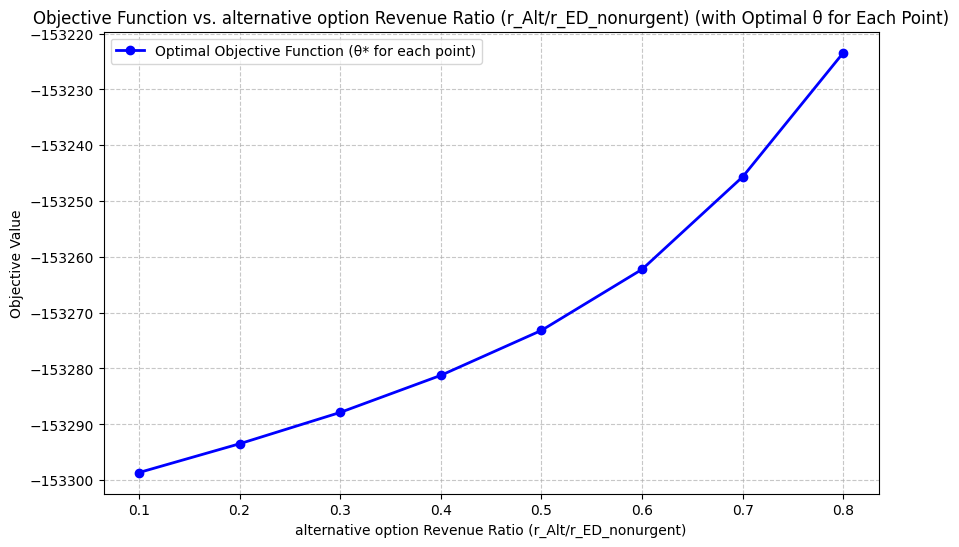}%
}\hfill
\subfloat[Balking Cost Ratio]{%
  \includegraphics[width=0.32\textwidth]{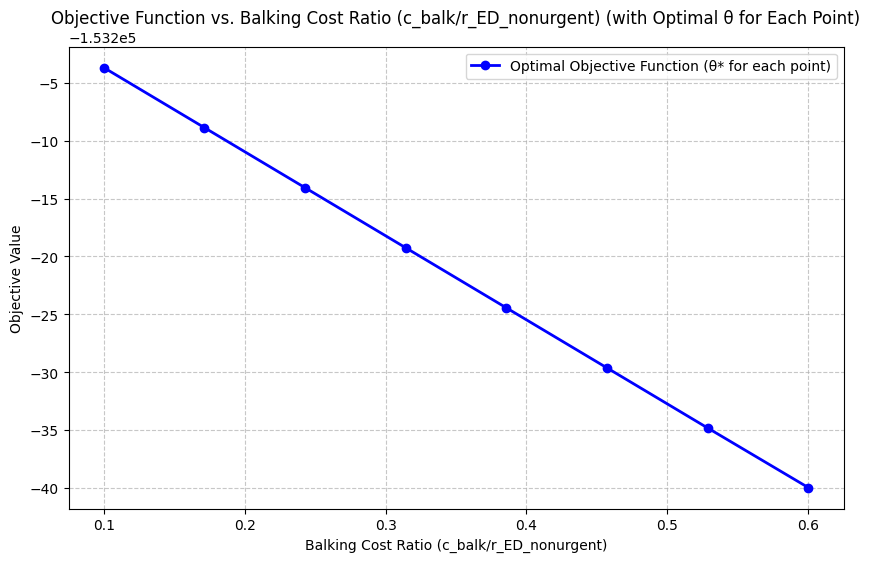}%
}
\caption{Baseline Case Analysis: Urban Alternative Care-enabled ED}
\label{fig:baseline_analysis_urban_other}
\end{figure}

Under standard parameters (optimal $\theta^* = 27$), the system demonstrates balanced optimization with clear sensitivity patterns. Figure~\ref{fig:baseline_analysis_urban} shows the two most critical ratios; additional proportional sensitivity analyses are provided in Figure \ref{fig:baseline_analysis_urban_other}.

\renewcommand\thesubfigure{\Alph{subfigure}}

The waiting cost ratio exhibits extreme performance impact, declining from approximately 8,091 at ratio 1.0 to 3,390 at ratio 4.0. As urgent patients incur proportionally higher waiting costs relative to non-urgent patients, the system performance deteriorates significantly, creating severe cost penalties during delays. The threshold proportion reveals a non-monotonic pattern distinct from rural EDs: performance remains stable for $\theta/k < 0.5$, slightly improves to peak at $\theta/k \approx 0.7$, then collapses dramatically beyond $\theta/k \approx 0.75$. This pattern suggests that moderate-to-high thresholds can be optimal in urban settings before excessive waiting causes severe cost accumulation and balking.

\section{Capacity Allocation}
\label{app:capacity_allocation}

We use this parameter set to compare fixed-partitioned versus nested approaches, as the rural and urban models exhibit instability under various bed configurations, limiting valid comparisons.

$\lambda = 20$, $p_u = 0.8$, $\mu_u = 4$, $\mu_n = 6$, $c_u = 8$, $c_n = 10$, $k = 25$, $\theta = 20$, $p_a = 0.5$, $r_n^{\mathrm{ED}} = 100$, $r^{\mathrm{Tele}} = 40$, $c^{b} = 30$, $c_n^w = 20$, $r_u^{\mathrm{ED}} = 200$, $c_u^w = 30$.

{\tiny
\begin{longtable}{|c|c|c|c|c|}

\caption{Detailed Comparison: NESTED vs FIXED-PARTITIONED Bed Configuration ($c_u = 8$, $c_n = 10$)}
\label{tab:nested_vs_fixed_8_10}\\
\hline
\multicolumn{5}{|c|}{\textbf{Bed Configuration: $c_u = 8$ urgent beds, $c_n = 10$ non-urgent beds}} \\
\multicolumn{5}{|c|}{\textbf{Detailed Comparison Table (NESTED vs FIXED-PARTITIONED)}} \\
\hline
$\boldsymbol{\theta}$ & \textbf{NESTED} & \textbf{FIXED} & \textbf{Diff} & \textbf{Better} \\
\hline
\endfirsthead

\hline
$\boldsymbol{\theta}$ & \textbf{NESTED} & \textbf{FIXED} & \textbf{Diff} & \textbf{Better} \\
\hline
\endhead

\hline
\endfoot

\hline
\multicolumn{5}{|c|}{\textbf{Summary: NESTED wins 25/25, FIXED-PARTITIONED wins 0/25, Ties 0/25}} \\
\hline
\endlastfoot

0 & 3353.33 & 3278.24 & 75.09 & NESTED \\
1 & 3354.69 & 3278.60 & 76.09 & NESTED \\
2 & 3360.53 & 3280.16 & 80.37 & NESTED \\
3 & 3373.22 & 3283.63 & 89.59 & NESTED \\
4 & 3391.87 & 3288.91 & 102.96 & NESTED \\
5 & 3412.73 & 3295.12 & 117.61 & NESTED \\
6 & 3431.68 & 3301.11 & 130.57 & NESTED \\
7 & 3446.21 & 3305.96 & 140.25 & NESTED \\
8 & 3455.84 & 3309.31 & 146.53 & NESTED \\
9 & 3461.45 & 3311.32 & 150.13 & NESTED \\
10 & 3464.37 & 3312.41 & 151.96 & NESTED \\
11 & 3465.74 & 3312.97 & 152.76 & NESTED \\
12 & 3466.32 & 3313.26 & 153.06 & NESTED \\
13 & 3466.55 & 3313.40 & 153.14 & NESTED \\
14 & 3466.63 & 3313.48 & 153.15 & NESTED \\
15 & 3466.65 & 3313.51 & 153.14 & NESTED \\
16 & 3466.66 & 3313.53 & 153.13 & NESTED \\
17 & 3466.67 & 3313.54 & 153.13 & NESTED \\
18 & 3466.67 & 3313.54 & 153.12 & NESTED \\
19 & 3466.67 & 3313.55 & 153.12 & NESTED \\
20 & 3466.67 & 3313.55 & 153.12 & NESTED \\
21 & 3466.67 & 3313.55 & 153.12 & NESTED \\
22 & 3466.67 & 3313.55 & 153.12 & NESTED \\
23 & 3466.67 & 3313.55 & 153.12 & NESTED \\
24 & 3466.67 & 3313.55 & 153.12 & NESTED \\


\end{longtable}
}

{\tiny
\begin{longtable}{|c|c|c|c|c|c|c|c|c|}

\caption{Bed Combination Analysis: NESTED vs FIXED-PARTITIONED (Total Capacity: 18 beds)}
\label{tab:bed_combination_analysis_18}\\

\hline
\multicolumn{9}{|c|}{\textbf{Bed Combination Analysis (Total beds: 18)}} \\
\hline
$\boldsymbol{c_u}$ & $\boldsymbol{c_n}$ & \textbf{Total} & \textbf{NESTED} $\boldsymbol{\theta^*}$ & \textbf{NESTED Obj} & \textbf{FIXED} $\boldsymbol{\theta^*}$ & \textbf{FIXED Obj} & \textbf{Difference} & \textbf{Better} \\
\hline
\endfirsthead

\hline
$\boldsymbol{c_u}$ & $\boldsymbol{c_n}$ & \textbf{Total} & \textbf{NESTED} $\boldsymbol{\theta^*}$ & \textbf{NESTED Obj} & \textbf{FIXED} $\boldsymbol{\theta^*}$ & \textbf{FIXED Obj} & \textbf{Difference} & \textbf{Better} \\
\hline
\endhead

\hline
\endfoot

\hline
\multicolumn{9}{|c|}{\textbf{Summary: NESTED wins: 11, FIXED-PARTITIONED wins: 2, Ties: 0}} \\
\multicolumn{9}{|c|}{\textbf{Best FIXED-PARTITIONED advantage: $c_u=5$, $c_n=13$, advantage: 637.47}} \\
\hline
\endlastfoot

1 & 17 & 18 & -- & -- & -- & -- & -- & FIXED UNSTABLE \\
2 & 16 & 18 & -- & -- & -- & -- & -- & FIXED UNSTABLE \\
3 & 15 & 18 & -- & -- & -- & -- & -- & FIXED UNSTABLE \\
4 & 14 & 18 & -- & -- & -- & -- & -- & FIXED UNSTABLE \\
5 & 13 & 18 & 23 & 3466.67 & 24 & 4104.14 & -637.47 & FIXED \\
6 & 12 & 18 & 23 & 3466.67 & 24 & 3465.72 & 0.95 & NESTED \\
7 & 11 & 18 & 23 & 3466.67 & 24 & 3348.28 & 118.38 & NESTED \\
8 & 10 & 18 & 24 & 3466.67 & 24 & 3313.55 & 153.12 & NESTED \\
9 & 9 & 18 & 23 & 3466.67 & 24 & 3301.41 & 165.25 & NESTED \\
10 & 8 & 18 & 23 & 3466.67 & 24 & 3297.00 & 169.67 & NESTED \\
11 & 7 & 18 & 23 & 3466.67 & 24 & 3295.43 & 171.24 & NESTED \\
12 & 6 & 18 & 24 & 3466.67 & 24 & 3294.91 & 171.76 & NESTED \\
13 & 5 & 18 & 24 & 3466.66 & 24 & 3294.83 & 171.83 & NESTED \\
14 & 4 & 18 & 22 & 3466.65 & 24 & 3295.71 & 170.94 & NESTED \\
15 & 3 & 18 & 24 & 3466.48 & 24 & 3303.72 & 162.76 & NESTED \\
16 & 2 & 18 & 24 & 3465.00 & 24 & 3370.00 & 95.00 & NESTED \\
17 & 1 & 18 & 13 & 3441.02 & 6 & 3685.36 & -244.34 & FIXED \\


\end{longtable}
}

 The nested configuration won all 25 threshold comparisons with consistent advantages ranging from \$75.09 to \$153.12 per hour. In the bed combination analysis with total capacity $c = 18$, nested allocation won 11 configurations while fixed-partitioned won only 2, with the best fixed-partitioned advantage being \$637.47 per hour at $(c_u = 5, c_n = 13)$. This demonstrates that nested allocation still dominates in high-capacity systems with urgent-heavy patient mix, providing superior performance across most bed allocation strategies.
\end{document}